\documentclass[preprint,11pt]{elsarticle}
\usepackage{verbatim}
\usepackage{graphics}
\usepackage{epsfig}
\usepackage{amsmath,amsfonts}
\usepackage[ruled,norelsize,vlined,linesnumbered]{algorithm2e}
\makeatletter
\newcommand{\removelatexerror}{\let\@latex@error\@gobble}
\makeatother
\usepackage{url}
\usepackage[a4paper,left=1in,right=1in]{geometry}
\usepackage{setspace}
\usepackage{multirow}
\usepackage{booktabs}
\usepackage{array}
\newcolumntype{P}[1]{>{\centering\arraybackslash}m{#1}}
\usepackage{color}
\usepackage[section]{placeins}
\FloatBarrier

\biboptions{sort&compress}
\usepackage{booktabs}
\usepackage{makecell}
\usepackage{array}
\usepackage{caption}
\usepackage{booktabs}
\usepackage{threeparttable}

\usepackage{pdflscape}

\usepackage{graphicx}
\usepackage{subcaption}
\usepackage{amssymb}

\begin{document}
\linespread{1.5}
\journal{ }
	\begin{frontmatter}

		\title{EOS-Bench: A Comprehensive Benchmark for Earth Observation Satellite Scheduling}

\address[lab1]{School of Traffic and Transportation Engineering, Central South University, Changsha 410083, China}
\address[lab2]{School of Engineering and Materials Science, Queen Mary University of London, London E1 4NS, UK}
\address[lab24]{College of Automation, Central South University, Changsha, 410083, China}

\address[lab3]{Mechanical and Aerospace Engineering, University of Strathclyde, Glasgow G1 1XQ, UK}
\address[lab4]{Department of Computer Science, University of Exeter, Exeter EX4 4QJ, UK} 
\address[lab6]{Department of Aerospace Engineering, Universidad de Sevilla, Camino de los Descubrimientos s.n., Sevilla, 41092, Spain}
\address[lab7]{Department of Electronics and Telecommunications, Politecnico di Torino, Corso Duca degli Abruzzi, 24, Turin, 10129, Italy}

\address[lab80]{ERATOSTHENES Centre of Excellence, Limassol, 3012, Cyprus}

\address[lab8]{Department of Civil Engineering and Geomatics, Cyprus University of Technology, Limassol, 3036, Cyprus}

\address[lab9]{Department of Computer Science and Engineering, College of Engineering, Qatar University, Doha, 2713, Qatar}
\address[lab10]{Department of Aerospace Engineering, Korea Advanced Institute of Science and Technology, Daejeon, 34141, South Korea} 
\address[lab11]{School of Management, Hefei University of Technology, Hefei, 230009, China}
\address[lab12]{Key Laboratory of Collaborative Intelligence Systems, Ministry of Education, Xidian University, Xi’an 710071, China}
\address[lab13]{School of Astronautics, Beihang University, 102206 Beijing, China}
\address[lab14]{Advanced Space Technology Laboratory, College of Astronautics, Nanjing University of Aeronautics and Astronautics, Nanjing 211106, China}
\address[lab15]{National Key Laboratory of Aerospace Flight Dynamics, Northwestern Polytechnical University, Xi’an, 710072, China}
\address[lab16]{State Key Laboratory of Information Engineering in Surveying, Mapping and Remote Sensing, Wuhan University, Wuhan, 430079, China}
\address[lab17]{School of Computer Science, China University of Geosciences, Wuhan, 430074, China}
\address[lab18]{School of Information Science and Technology, Dalian Maritime University, Dalian, 116026, China}
\address[lab19]{Department of Electrical \& Computer Engineering, University of Alberta, Edmonton, AB T6R 2V4, Canada}
\address[lab20]{Division of Geological and Planetary Science, California Institute of Technology, CA, USA}
\address[lab21]{Department of Automation and Systems Engineering, Federal University of Santa Catarina, Florianopolis, SC, Brazil}
\address[lab23]{School of Geography, Archaeology and Environmental Studies, University of the Witwatersrand, Braamfontein, Johannesburg, South Africa}

\author[lab1,lab2]{Qian Yin}
\author[lab2]{Jiaxing Li}
\author[lab1]{Jiaqi Cheng}
\author[lab24]{Qizhang Luo}

\author[lab3]{Annalisa Riccardi}
\author[lab4]{Abhijit Chatterjee}

\author[lab6]{Rafael Vazquez}
\author[lab7]{Carlo Novara}
\author[lab80,lab8]{Michalis Mavrovouniotis}

\author[lab9]{Ponnuthurai Nagaratnam Suganthan}
\author[lab10]{Shengzhou Bai}
\author[lab11]{Xiaoxuan Hu}
\author[lab12]{Lining Xing}
\author[lab13]{Ming Xu}
\author[lab14]{Shuang Li}
\author[lab15]{Zixuan Zheng}
\author[lab16]{Xin Shen}
\author[lab17]{Xiaoyu Chen}
\author[lab1]{Yi Gu}
\author[lab18]{Yanjie Song}

\author[lab19]{Witold Pedrycz}
\author[lab20]{Evan L. Kramer}

\author[lab21]{Laio Oriel Seman}


\author[lab23]{Cletah Shoko}

\author[lab24]{Guohua Wu\corref{cor}}
\ead{guohuawu@csu.edu.cn}

\author[lab2]{Xinwei Wang\corref{cor}}
\ead{xinwei.wang@qmul.ac.uk}

\cortext[cor]{Corresponding authors}



		\begin{abstract}
      \noindent Earth observation satellite imaging scheduling is a challenging NP-hard combinatorial optimisation problem that is critical to space mission operations. 
      While next-generation agile Earth observation satellites (EOS) enhance flexibility, they also increase scheduling complexity. 
      Because the community currently lacks a unified, open-source benchmark, algorithm comparisons across private instance sets remain difficult to interpret consistently. 
      This paper introduces EOS-Bench, a comprehensive framework for the systematic and reproducible evaluation of scheduling algorithms.
      By integrating high-fidelity orbital dynamics and diverse platform constraints, EOS-Bench generates 1,390 scenarios and 13,900 benchmark instances, ranging from small-scale exact-validation cases to large coordination tasks featuring 1,000 satellites and 10,000 requests. 
      In addition, we introduce a scenario characterisation scheme that estimates scenario-level structural difficulty by aggregating instance-level descriptors describing feasible opportunity density, task flexibility, pairwise conflict, and satellite-side congestion. 
      We also propose a multidimensional protocol to evaluate solvers across five practical metrics: task profit, completion rate, workload balance, timeliness, and runtime. 
      We comprehensively evaluate the framework across both agile and non-agile scenarios using exact mixed-integer programming, constructive heuristics, meta-heuristics, and deep reinforcement learning. 
      The reported results indicate that EOS-Bench can differentiate solver behaviour across scales and operating conditions, highlight practical trade-offs between solution quality and runtime, and provide structural context for interpreting scenario difficulty. 
      Taken together, these features make EOS-Bench a useful open testbed for future research on Earth observation satellite scheduling. The code and data are available at \url{https://github.com/Ethan19YQ/EOS-Bench}.
		\end{abstract}


	\end{frontmatter}

\section{Introduction}

Earth observation satellites (EOSs) play an essential role in a wide range of civil and strategic applications, including climate monitoring, disaster response, precision agriculture, environmental surveillance, and security-related missions~\cite{wang2020agile}.
As Earth observation systems expand from single platforms to large constellations, planners must allocate limited observation opportunities and onboard resources under orbital, temporal, and platform constraints~\cite{gu2022large,GU2026131815}.
This leads to the Earth observation satellite scheduling problem (EOSSP), where, given a set of EOSs and a set of ground targets to be observed, imaging requests must be selected, allocated, and sequenced under constraints such as visibility, platform manoeuvrability, and onboard resources. The objective of EOSSP is typically to optimise performance metrics, for example by maximising the number of observed targets or minimising the total mission completion time.
The EOSSP is therefore inherently combinatorial, computationally challenging, and are widely regarded as NP-hard~\cite{lemaitre2002selecting,wang2020agile}.
Consequently, over the past three decades, extensive research has been devoted to the modelling and solution of EOSSPs in settings ranging from single-satellite operations to multi-satellite constellations and integrated observation planning~\cite{ferrari2025satellite,chen2020eosis}.

    \begin{figure}[htb]
        \centering
        \begin{tabular}{cc}
            \includegraphics[width=0.48\textwidth]{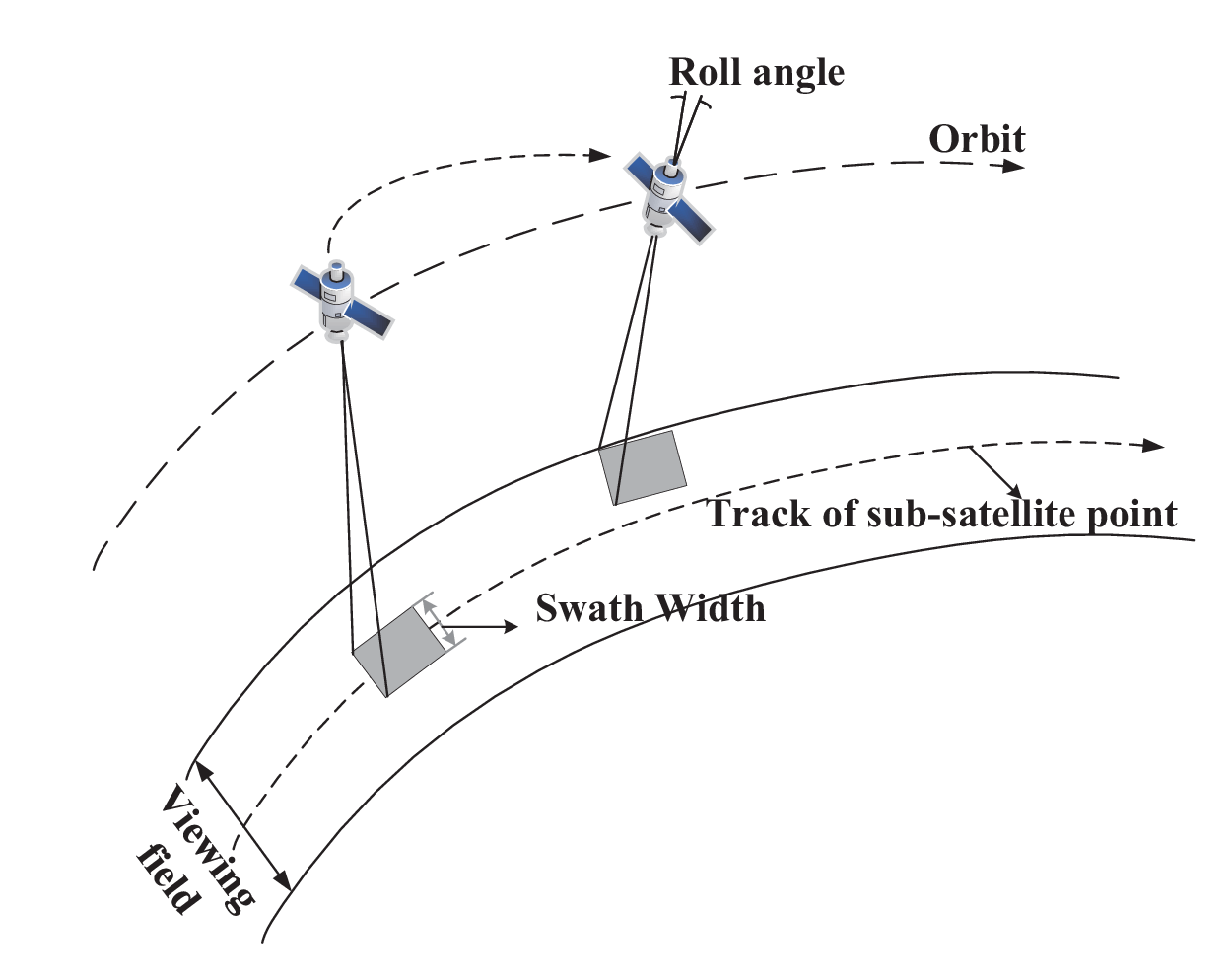} &
            \includegraphics[width=0.48\textwidth]{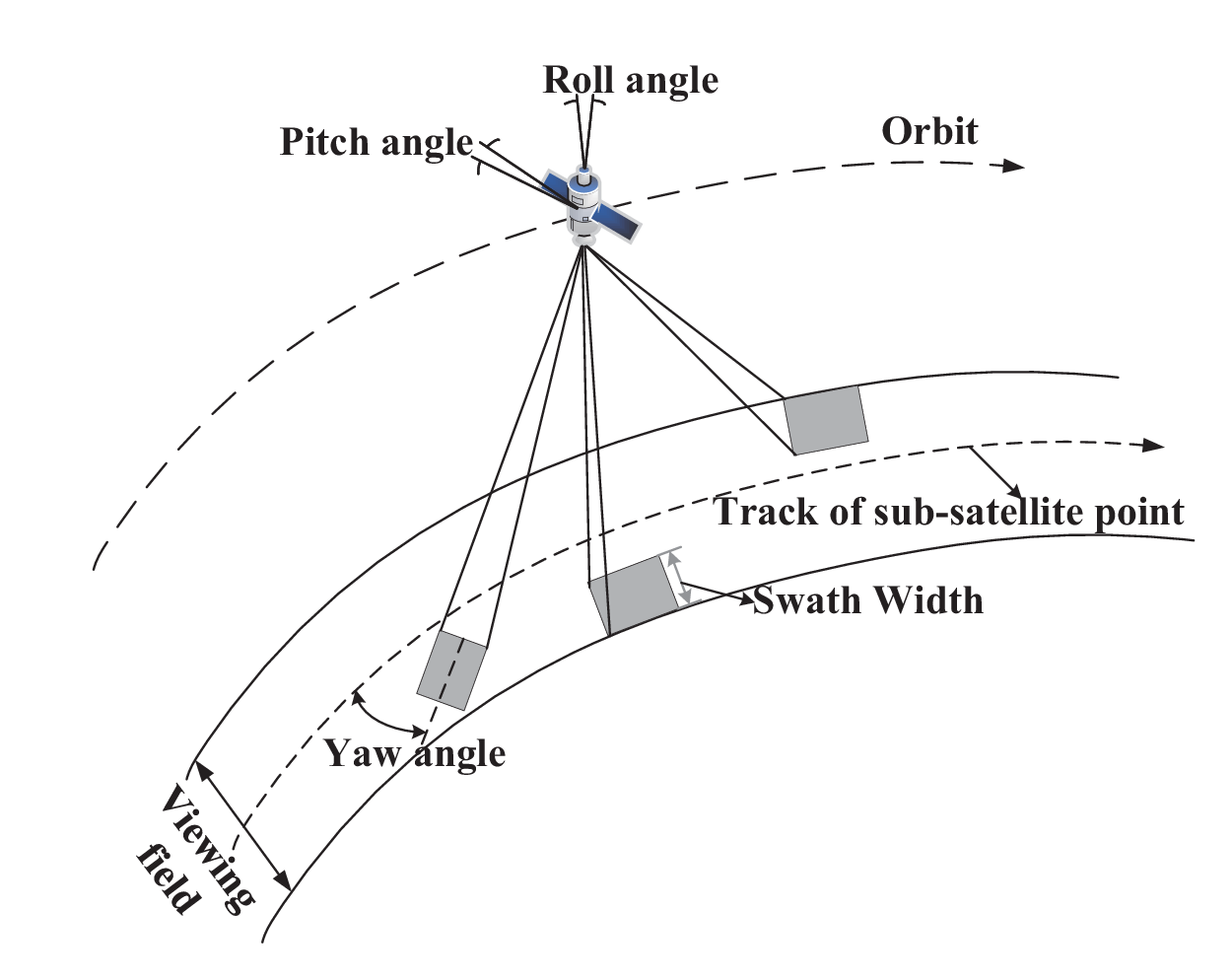} \\
            (a) Non-agile satellite & (b) Agile satellite \\
        \end{tabular}
        \caption{Differences between non-agile and agile satellites. Non-agile satellites mainly observe near nadir, whereas agile satellites can actively manoeuvre to image more targets within a pass. }
        \label{fig:satellites}
    \end{figure}

A central distinction in the EOSSP literature is between non-agile and agile observation platforms, as shown in Fig.~\ref{fig:satellites}.
Non-agile satellites typically observe near nadir, with limited off-nadir pointing capability and comparatively stable viewing geometry.
By contrast, agile satellites can actively manoeuvre about multiple attitude axes and revisit different ground locations within a short time interval, thereby substantially enlarging the feasible observation set within a pass.
This additional flexibility improves mission responsiveness, but also makes the scheduling problem considerably harder because feasibility depends not only on visibility, but also on attitude transitions, time-dependent observation opportunities, and coupled resource usage.
Non-agile and agile satellites therefore represent two related but distinct EOSSP settings, with the latter extending the former through greater pointing flexibility and more complex operational constraints.
Recent surveys~\cite{wang2020agile,ferrari2025satellite} show that EOSSP research spans this full spectrum, from simple single-satellite formulations to integrated constellation-level problems with profits and multiple resource couplings.

To solve these problems, researchers have developed a wide range of optimisation and learning-based methods.
Exact approaches, especially mixed-integer programming (MIP) and decomposition-based methods, can provide optimal or near-optimal solutions for small and medium-sized instances, but their computational cost increases rapidly with constellation size, task density, and planning horizon.
For larger operational settings, heuristic and meta-heuristic methods such as greedy construction, simulated annealing, tabu search, and large neighbourhood search are widely used~\cite{wang2020agile}.
He et al.~\cite{he2018alns} proposed an adaptive large neighbourhood search method for multi-satellite scheduling with time-dependent transition constraints.
Furthermore, learning-based approaches, particularly deep reinforcement learning (DRL) and related neural decision models, have attracted increasing attention because they offer the possibility of fast policy inference across diverse instances~\cite{wei2021drl,herrmann2023rl}.

However, despite the methodological diversity of the field, most studies still evaluate their methods on private or highly problem-specific instances\footnote{Throughout this paper, a Scenario refers to a specific benchmark setting defined by a unique combination of structural parameters, whereas an Instance refers to a concrete realisation of a Scenario. Formal definitions are provided in Table~\ref{tab:definitions}.}.
This lack of common evaluation standards makes it difficult to determine whether reported performance differences are due to genuinely stronger algorithms or simply to differences in modelling assumptions, instance construction, and evaluation criteria~\cite{ferrari2025satellite}. 
This gap has motivated several recent benchmark-oriented efforts.
Rocha et al.~\cite{rocha2025integrated} released benchmark instances for an integrated agile EOS scheduling model together with an MIP-based heuristic.
Vasegaard and Larsen~\cite{vasegaardEOSpython2024} introduced EOSpython, an open-source framework that connects instance generation with scheduling solvers.
However, while it provides the construction methodology, it lacks specific benchmark instances and remains mainly oriented towards agile EOS imaging.
Wang et al.~\cite{wang2025aeosbench} proposed AEOS-Bench, a high-fidelity benchmark environment built on Basilisk simulation~\cite{kenneally2020basilisk}.
However, AEOS-Bench focuses primarily on simulation-driven environments tailored for reinforcement learning, lacking a general combinatorial representation and multidimensional evaluation suitable for a broader range of exact and meta-heuristic solvers.

Another equally critical deficiency in current EOSSP benchmarking practice is the over-reliance on nominal scenario descriptors. 
Most existing literature relies solely on the number of satellites, tasks, or planning-horizon length to define problem scale. Yet, these superficial metrics fail to capture the intrinsic structural difficulty of an instance. Scenarios of identical nominal size may still differ substantially in feasible opportunity density, task flexibility, conflict intensity, and timeline congestion. Without explicitly characterising these underlying topological and temporal bottlenecks, algorithmic performance remains difficult to interpret consistently across different studies.

In a nutshell, while these studies improve reproducibility within particular EOSSP settings, they still lack the comprehensive features necessary for a robust benchmarking ecosystem  (see Table~\ref{tab:benchmark_comparison}). 
Specifically, existing frameworks generally omit explicit scenario characterisation to quantify structural difficulty, lack a standardised multi-dimensional evaluation protocol, and fail to provide integrated results visualisation tools for intuitive schedule analysis.
Therefore, the field still lacks a unified benchmark framework that simultaneously supports agile and non-agile platforms, spans substantially different constellation scales, and enables systematic comparison across fundamentally different solver classes under a shared evaluation protocol.

\begin{table}[htbp]
  \centering
  \footnotesize
  \caption{Comparison of EOS-Bench with existing EOSSP benchmarks.}
  \label{tab:benchmark_comparison}

  \begin{threeparttable}
  \begin{tabular}{
    >{\raggedright\arraybackslash}p{0.3\textwidth}
     >{\centering\arraybackslash}p{0.15\textwidth}
    >{\centering\arraybackslash}p{0.15\textwidth}
    >{\centering\arraybackslash}p{0.15\textwidth}
    >{\centering\arraybackslash}p{0.15\textwidth}
  }
    \toprule
    \textbf{Feature} & \textbf{EOSpython~\cite{vasegaardEOSpython2024}}& \textbf{Rocha et al.~\cite{rocha2025integrated}} & \textbf{AEOS-Bench~\cite{wang2025aeosbench}} & \textbf{EOS-Bench (Ours)} \\
    \midrule
    \textbf{Benchmark size} & / & 32 &192\tnote{a} & 13,900 \\
    \textbf{Satellite scale} & / & 1$\sim$4 &1$\sim$50 & 1$\sim$1000 \\
    \textbf{Task scale} & / & 200$\sim$800 & 50$\sim$300 & 10$\sim$10000 \\
    \textbf{Time horizon} & / & 24h$\sim$72h & 1h & 1h$\sim$168h \\
    \textbf{Agile EOS imaging} & $\checkmark$ & $\checkmark$ & $\checkmark$ & $\checkmark$ \\
    \textbf{Non-agile EOS imaging} & $\times$ & $\times$ & $\times$ & $\checkmark$ \\
    \textbf{Evaluation protocol} & $\checkmark$ & $\times$ & $\checkmark$ & $\checkmark$ \\
    \textbf{Results visualisation} & $\checkmark$ & $\times$ &$\times$ & $\checkmark$ \\
    \textbf{Scenario characterisation} & $\times$ & $\times$ & $\times$ & $\checkmark$ \\
    \bottomrule
  \end{tabular}

  \begin{tablenotes}[flushleft]
    \footnotesize
    \item[a] The AEOS-Bench work involves a total of 16,410 instances: 16,218 are randomly generated for reinforcement learning training (with no detailed information for reproduction), and the remaining 192 serve as a benchmark.
    \item \textit{Note.} ``/'' indicates that the corresponding information is not explicitly reported in the original reference.
  \end{tablenotes}
  \end{threeparttable}
\end{table}

To address this gap, this paper introduces EOS-Bench, a unified benchmark framework for EOSSP.
The aim of EOS-Bench is not to propose yet another scheduling algorithm, but to provide a common and reproducible experimental basis for evaluating optimisation-based, heuristic, meta-heuristic, and learning-based solvers.
The framework generates parametrised benchmark scenarios along four principal dimensions: satellite platform (agile or non-agile), planning horizon, constellation configuration, and mission load.
{Within this structure, it supports both agile and non-agile platforms under a common modelling architecture, while incorporating realistic orbital dynamics, visibility generation, attitude-related feasibility, and coupled energy/storage constraints.
Since agile platforms introduce complex time-dependent transition constraints that represent the most computationally demanding regime of EOSSP, our detailed analysis in this paper deliberately focuses on a representative subset of these agile scenarios. 
Comprehensive results for both agile and non-agile platforms are fully provided in~\ref{app:first}.}
Beyond solver-oriented evaluation, EOS-Bench also provides scenario characterisation descriptors by aggregating instance-level structural indicators under each scenario, thereby supporting more interpretable comparisons across scenario families.

Specifically, this paper makes four main contributions:
\begin{itemize}
\item We develop an open-source and extensible benchmark framework that standardises the modelling and evaluation of Earth observation satellite scheduling. The framework integrates orbital dynamics, visibility generation, platform constraints, and task modelling within a unified architecture that supports both agile and non-agile platforms.

\item We construct a comprehensive benchmark library comprising 1,390 scenarios and 13,900 instances covering a broad range of operational settings.
The scenarios are organised into: 1,104 Standard Scenarios associated with 11,040 instances, which systematically vary platform type, planning horizon, constellation size, and mission load; and 286 Specific Scenarios associated with 2,860 instances, which support controlled sensitivity analysis with respect to resource capacity, manoeuvrability, and constellation configuration.

\item We introduce a scenario characterisation scheme in which scenario-level descriptors are estimated from instance-level structural indicators computed before optimisation.
The proposed descriptors capture key properties such as feasible opportunity structure, task flexibility, conflict intensity, and satellite-side congestion, thereby providing a structural context for more interpretable comparison of scheduling methods across scenario families.

\item We establish a multi-dimensional evaluation protocol based on five complementary metrics: task profit, task completion rate, workload balance, timeliness, and runtime. 
By demonstrating the framework's discriminative power through extensive baselines, this study enables a structured analysis of the scalability, bottleneck vulnerabilities, and practical trade-offs of existing solution paradigms.
\end{itemize}

The remainder of this paper is organised as follows. 
Section~\ref{sec:literature} reviews existing work on EOSSP models, algorithms and relevant evaluation metrics, and benchmarking practices. 
Section~\ref{sec:framework} presents the benchmark framework, including the scenario space and the underlying models. 
Section~\ref{sec:generation} describes the generation process for both Standard and Specific Scenarios. 
Section~\ref{sec:scenario_characterisation} introduces structural descriptors used to characterise benchmark scenarios before optimisation. 
Section~\ref{sec:evaluation} establishes the unified evaluation protocol and details the specific performance metrics. 
Section~\ref{sec:experiments} reports comparative experiments on exact, heuristic, meta-heuristic, and learning-based schedulers. 
Section~\ref{sec:directions} outlines potential directions for future benchmark development. 
Finally, Section~\ref{sec:conclusion} concludes the paper.

\section{Literature Review}
\label{sec:literature}

This section reviews the literature on EOSSP from four interconnected perspectives.
First, it outlines the main problem characteristics and introduces representative modelling variants, with a particular focus on platform agility and resource couplings.
Second, it surveys the principal classes of solution approaches and algorithms, ranging from exact optimisation to meta-heuristics and learning-based methods.
Third, it examines experimental settings and benchmarking practices in the existing literature, highlighting the fragmentation of current instance designs.
Fourth, it discusses the evaluation protocols and metrics used to assess mission effectiveness and computational performance.
The section concludes by summarising the main observations that motivate the benchmark framework proposed in this paper.

\subsection{Problem Modelling and Variants}

Earth observation satellite scheduling has long been studied in both the operations research and aerospace communities, and remains one of the most active application areas in satellite operations.
At its core, the EOSSP concerns the selection, assignment, and sequencing of imaging tasks subject to visibility, temporal, platform, and resource constraints.
Early studies already showed that these problems are computationally difficult and widely considered as NP-hard~\cite{ferrari2025satellite}. Furthermore, practical formulations often involve multiple competing objectives, such as coverage, profit, fairness, and responsiveness~\cite{tangpattanakulMultiobjective2015}.

Recent survey papers have helped clarify the breadth of this research area, showing that EOS scheduling has evolved from relatively simple single-satellite formulations to integrated multi-satellite problems with complex dynamic elements~\cite{wang2020agile,ferrari2025satellite}.
A central distinction in this evolution is between non-agile and agile observation platforms, a feature that also serves as a primary structural dimension in the proposed EOS-Bench. 
Non-agile satellites typically observe near nadir with comparatively stable viewing geometries. By contrast, agile satellites can actively manoeuvre about multiple attitude axes within a short time interval. While this flexibility improves mission responsiveness, it makes the scheduling problem considerably harder. For agile platforms, the feasibility of consecutive observations depends strongly on attitude manoeuvres and the associated transition times. Liu et al.~\cite{liuAdaptive2017} and Peng et al.~\cite{pengSolving2022}, for example, model time-dependent slews between targets by using a piecewise linear function of the total angular change to estimate the transition time between observations. Peng et al.~\cite{pengAgile2019,pengExact2020} further incorporate time-dependent task profits, making the problem closely related to an orienteering model with time-dependent travel times and rewards. 

Beyond platform agility, other important modelling directions concern the definition of observation targets and environmental uncertainty. 
Many studies formulate EOS scheduling as an area-coverage problem over continuous geographic regions~\cite{perea2015swath}. 
In contrast, studies involving highly agile constellations often discretise demand into point-target requests to manage geometric and combinatorial complexity. 
Although this approach frequently creates redundant target formulations, such redundancy can be effectively managed using complex network theory~\cite{wang2016scheduling}.
Additionally, extensions have addressed uncertainty in the observation environment, such as cloud-cover stochasticity~\cite{valickaMixedinteger2019,wangRobust2021}, as well as explicit couplings between subsystems like attitude, energy, and onboard storage~\cite{heReasoningbased2023}. 
Overall, EOS scheduling models now span a wide spectrum, from compact point-target representations to highly integrated systems involving multiple coupled resources, directly motivating the need for a versatile benchmarking framework.

\subsection{Algorithmic Solution Methods}

The increasing complexity of EOS scheduling models has led to the development of a wide range of solution methods.
Exact approaches, particularly those based on mixed-integer programming and decomposition, remain important because they provide structural insight and high-quality baselines.
Peng et al.~\cite{pengExact2020,pengSolving2022} develop exact and branch-and-price algorithms for agile satellites with time-dependent profits and transition times, while Rigo et al.~\cite{rigoBranchandprice2022} apply branch-and-price to nano-satellite task scheduling.
Similarly, discrete-time and fixed-interval mathematical models have been formulated to address multiple observations across multi-satellite systems~\cite{wang2018fixed}.
For smaller missions, continuous-time and energy-aware models have also been proposed~\cite{camponogaraContinuoustime2022,semanImproving2023}.
Related work includes mixed open- and closed-loop planning models and stochastic models for constellation planning under cloud and weather uncertainty~\cite{tharmarasaMixed2019,valickaMixedinteger2019}.

These exact and related approaches can often produce optimal or near-optimal solutions for small- to medium-sized instances.
However, their computational burden rises sharply with constellation size, task density, and planning horizon.
In practical EOS settings, especially those involving large constellations or multi-day horizons, exact optimisation often becomes difficult to apply within reasonable computational budgets.
To address larger and more operationally demanding instances, many studies turn to heuristic and meta-heuristic methods.
Adaptive large neighbourhood search is one of the most successful frameworks for agile EOS scheduling with time-dependent transition constraints~\cite{he2018alns,wuImproved2024}.
Other representative approaches include multi-objective local search~\cite{tangpattanakulMultiobjective2015}, hybrid genetic and memetic algorithms for constellation and integrated scheduling~\cite{barkaouiNew2020,changIntegrated2021}, divide-and-conquer ensembles combining meta-heuristics with exact subroutines~\cite{wuEnsemble2022}, and decompose-and-learn strategies for large constellations~\cite{qiDecomposeandlearn2025}.
Further work has explored distributed game-negotiation mechanisms~\cite{liuStudy2021}, task-network aggregation strategies~\cite{fanNovel2022}, and approaches tailored to emergency observation tasks~\cite{sunAgile2021}.
More recently, quantum-inspired and quantum-enhanced meta-heuristics have also been investigated for agile EOS scheduling~\cite{stollenwerkAgile2021,rainjonneauQuantum2023,sunAdaptivestrategiesbased2025,sunMultiadaptive2024,sn_earth_obsv_2025}.

In parallel, learning-based methods, especially deep reinforcement learning (DRL), have become an increasingly active research direction.
Early studies formulated EOS scheduling as a Markov decision process and learned control policies for single satellites and small constellations~\cite{huangRevising2021,herrmannComparative2023}.
Subsequent work extended this idea to more complex settings, including multi-type observation scheduling~\cite{songGeneralized2024} and hybrid actor--critic schemes for joint imaging decisions~\cite{wenScheduling2023,wenScheduling2024}.
More recent studies have considered DRL for rolling-horizon control~\cite{liDynamic2025}, deep Q-learning combined with heuristic repair~\cite{liOptimizing2025}, two-stage scheduling architectures~\cite{liuTwostage2025}, and pre-trained MDP models for online decision-making~\cite{liPTMB2024}.
Learning-based methods have also been applied to time-sensitive constellation-level observation~\cite{liuLearningbased2024}, multi-point imaging with spiking neural networks and proximal policy optimisation~\cite{yaoSpiking2025}, and conflict-aware resource allocation in imaging missions~\cite{chenDeep2025,chenConflict2025}.

Overall, the EOSSP literature now contains a rich methodological spectrum, including exact optimisation, heuristics, meta-heuristics, learning-based, and quantum-based approaches.
This diversity makes standardised evaluation increasingly important, because fair comparison across such different solver classes is difficult without shared scenario families and consistent performance metrics.

\subsection{Performance Evaluation Metrics}

The evaluation of EOS scheduling methods typically relies on a combination of mission-oriented and computational-performance metrics. Mission-oriented metrics assess how effectively a schedule fulfils operational objectives, whereas computational metrics determine the practical viability of the underlying algorithm, particularly for large-scale deployments.

Within mission-oriented evaluation, the most frequently reported metric is the total mission reward, typically defined as the sum of task weights, priorities, or profits~\cite{pengAgile2019,pengExact2020,wuImproved2024}. Closely related measures include the absolute number of completed tasks and the overall completion ratio, which are especially prevalent in point-target scheduling~\cite{heImproved2018,barkaouiNew2020}. As mission models expand to encompass continuous areas or regions, coverage-specific metrics such as total covered area, mapping completeness, and regional workload balance become necessary~\cite{chenMultiobjective2020,chenLargescale2023,luMultiple2023,wuMultiregion2024,zengLongterm2026}. Furthermore, in time-sensitive or dynamic contexts, response time, tracking continuity, revisit intervals, and service timeliness are often emphasized~\cite{kimMission2015,hanContinuous2023,congAutonomous2025}. For environments characterised by uncertainty, evaluation criteria frequently shift towards expected mission value, success rates under cloud-cover stochasticity, or robustness against operational disturbances~\cite{wangRobust2020,wangRobust2021,hanSimulated2023}. Finally, in integrated scheduling studies, indicators such as downlink completion, data latency, and memory utilisation are increasingly reported to capture end-to-end service efficiency~\cite{heJoint2022,yuResearch2025}.

Alongside mission effectiveness, computational-performance metrics are essential for assessing algorithm scalability. Runtime remains the standard benchmark, usually reported as wall-clock time or average execution time across multiple instances~\cite{xuMultisatellite2020,chenLargescale2023,qiDecomposeandlearn2025}. For heuristic and meta-heuristic approaches, convergence speed and iteration efficiency provide additional algorithmic insights~\cite{heImproved2018,wuImproved2024}. In online or rolling-horizon applications, the ability to generate feasible decisions within strict decision cycles often supersedes the pursuit of global optimality~\cite{wangDynamic2015,liDynamic2025}. Moreover, because EOS scheduling is inherently multi-objective, relying on a single aggregated score is often inadequate. Consequently, there is a growing consensus that evaluation protocols should report multiple independent objective values, Pareto fronts, or dominance statistics, thereby making trade-offs among reward, fairness, timeliness, and computational cost more transparent~\cite{changIntegrated2021,changThree2025,songGeneralized2024}.

To illustrate the fragmentation of current evaluation standards, Table~\ref{tab:metric_usage} provides an illustrative summary of how representative EOS scheduling studies relate to the main evaluation dimensions considered in this paper. It shows that existing studies often emphasise different subsets of performance criteria, and that the corresponding definitions are not always directly aligned.

\begin{table}[htb]
  \centering
  \footnotesize
  \caption{Representative use of evaluation dimensions in the EOSSP literature.}
  \label{tab:metric_usage}
  \begin{threeparttable}
  \setlength{\tabcolsep}{6pt}
  \begin{tabular}{p{0.25\textwidth}p{0.1\textwidth}p{0.1\textwidth}p{0.1\textwidth}p{0.1\textwidth}p{0.1\textwidth}}
    \toprule
    \textbf{Reference} & \textbf{TP} & \textbf{TCR} & \textbf{BD} & \textbf{TM} & \textbf{RT} \\
    \midrule
    Wu et al.~\cite{wuEnsemble2022}                     & $\surd$ & $\surd$ & $\times$ & $\times$ & $\surd$ \\
    Qi et al.~\cite{qiDecomposeandlearn2025}            & $\surd$ & $\times$ & $\surd$ & $\times$ & $\times$ \\
    Barkaoui and Berger~\cite{barkaouiNew2020}          & $\times$ & $\surd$ & $\times$ & $\times$ & $\times$ \\
    Li et al.~\cite{liMultiobjective2024}               & $\times$ & $\surd$ & $\times$ & $\times$ & $\times$ \\
    Ren et al.~\cite{renCompetitive2022}                & $\times$ & $\times$ & $\surd$ & $\times$ & $\times$ \\
    Li et al.~\cite{liRemotesensing2024}                & $\surd$ & $\times$ & $\surd$ & $\times$ & $\surd$ \\
    Sun et al.~\cite{sunAgile2021}                      & $\times$ & $\surd$ & $\times$ & $\surd$ & $\times$ \\
    Li et al.~\cite{liDynamic2025}                      & $\times$ & $\times$ & $\times$ & $\surd$ & $\times$ \\
    Liu et al.~\cite{liuLearningbased2024}              & $\times$ & $\times$ & $\times$ & $\surd$ & $\times$ \\
    Herrmann and Schaub~\cite{herrmannComparative2023}  & $\surd$ & $\times$ & $\times$ & $\times$ & $\surd$ \\
    Stollenwerk et al.~\cite{stollenwerkAgile2021}      & $\surd$ & $\times$ & $\times$ & $\times$ & $\surd$ \\
    EOS-Bench (Ours)                                     & $\surd$ & $\surd$ & $\surd$& $\surd$ & $\surd$ \\
    \bottomrule
  \end{tabular}
  \begin{tablenotes}[flushleft]
    \footnotesize
    \item \textit{Note.} TP = Task Profit; TCR = Task Completion Rate; BD = Balance Degree; TM = Timeliness; RT = Runtime.
  \end{tablenotes}
  \end{threeparttable}
\end{table}

In summary, while the community has developed a rich set of evaluation criteria to capture diverse operational and computational goals, the selective reporting of these metrics across different studies heavily impedes direct algorithmic comparisons. A method that excels in total profit might struggle with computational scalability or workload fairness, yet such trade-offs remain hidden when studies report only a narrow subset of indicators. This highlights the critical need for a standardised, multi-dimensional evaluation protocol, such as the one proposed in this framework, that systematically and simultaneously assesses value, completion rate, fairness, timeliness, and runtime to provide a holistic and transparent view of solver performance.

\subsection{Benchmarking and Experiments}

Despite the richness of the modelling and algorithmic literature, benchmarking practice in EOS scheduling remains uneven.
Many studies evaluate their methods on bespoke instances designed for a particular problem formulation, platform type, or operational assumption.
While such case-specific evaluation is often reasonable for demonstrating a new method, it makes comparison across papers difficult.
For example, Cho et al.~\cite{choOptimizationBased2018} study optimisation-based methods for agile constellations over planning horizons of up to one week, whereas Barkaoui and Berger~\cite{barkaouiNew2020} consider a 12-hour horizon for a three-satellite constellation.
Valicka et al.~\cite{valickaMixedinteger2019} examine 24-hour schedules under cloud uncertainty for one or two satellites, while Tharmarasa et al.~\cite{tharmarasaMixed2019} analyse mixed open- and closed-loop planning over a single day.
Other studies focus on particular system architectures, such as multi-agent planning for multi-satellite systems~\cite{bonnetMultisatellite2015,povedaEvolutionary2019}, autonomous goal-based onboard planning~\cite{cividanesExtended2021}, or intelligent planning for distributed satellite systems~\cite{hiltonIntelligent2024}.
Although some of these papers describe their experimental settings carefully, reusable generators and public datasets remain relatively uncommon.
As a consequence, many published test sets cannot readily serve as common benchmarks for the wider community.

To better understand current experimental practice and to inform the design of EOS-Bench, we conduct a structured review of representative EOS scheduling studies.
For each study, we record the planning horizon, the number of satellites, the platform type (agile or non-agile), and the mission scale.
The resulting summary is presented in Table~\ref{tab:addlabel}.
A short horizon of 1--3 hours is often used in studies concerned with rolling or online decision-making for relatively small constellations and task sets.
Medium horizons of 12--24 hours are the most common in the literature, and appear in both exact and meta-heuristic studies for agile and non-agile platforms, with problem sizes ranging from a single satellite to several tens of satellites and from a few tens to several thousand tasks.
Longer horizons, such as 96 or 168 hours, are more often associated with higher-level constellation planning or large-scale mission-planning studies.

\begin{table}[htb]
  \centering
  \footnotesize
  \caption{Representative experimental settings in the EOSSP literature.}
  \label{tab:addlabel}
  \begin{threeparttable}
    \begin{tabular}{@{} ccccc @{\hspace{2em}} ccccc @{}}
    \toprule
    Period & Sats & Type & Tasks & Ref. & Period & Sats & Type & Tasks & Ref. \\
    \cmidrule(r){1-5} \cmidrule(l){6-10}
    1h    & 4$\sim$7   & non-agile & 50$\sim$400   & \cite{liDynamic2025,wuEnsemble2022} & 24h & 1          & agile     & 50$\sim$300   & \cite{liuTwostage2025} \\
    1h    & 7          & non-agile & 50$\sim$150   & \cite{renCompetitive2022,liuAdaptive2017} & 24h & 6          & agile     & 300$\sim$1000 & \cite{liuLargescale2023,liuAdaptive2017} \\
    1.5h  & 1          & agile     & 55$\sim$60    & \cite{heReasoningbased2023}         & 24h & 4          & agile     & 150$\sim$300  & \cite{wangRobust2021} \\
    1.5h  & 1          & non-agile & 25$\sim$50    & \cite{herrmannComparative2023}      & 24h & 1$\sim$2   & non-agile & 223$\sim$520  & \cite{valickaMixedinteger2019} \\
    3h    & 10         & agile     & 100$\sim$250  & \cite{huangReinforcement2025}       & 24h & 1          & agile     & 100           & \cite{eddyMarkov2020} \\
    3h    & 1          & agile     & 50$\sim$200   & \cite{wangStrategy2024}             & 24h & 1          & agile     & 50$\sim$200   & \cite{liuAdaptive2017} \\
    12h   & 3          & non-agile & 10$\sim$60    & \cite{barkaouiNew2020}              & 24h & 6          & agile     & 100$\sim$2000 & \cite{he2018alns} \\
    12h   & 3          & non-agile & 10$\sim$400   & \cite{tharmarasaMixed2019}          & 96h & 30         & non-agile & 140           & \cite{seckerAutomated2025} \\
    21h   & 10         & non-agile & 1500          & \cite{hiltonIntelligent2024}        & 168h & 1$\sim$3   & non-agile & 10$\sim$700   & \cite{choOptimizationBased2018} \\
    24h   & 3          & non-agile & 100$\sim$300  & \cite{chenDeep2025}                 & 168h & 12         & non-agile & 1000          & \cite{rocha2025integrated} \\
    24h   & 3$\sim$5   & non-agile & 100$\sim$1000 & \cite{songGeneralized2024}          & /   & 1          & non-agile & 9$\sim$24     & \cite{rigoBranchandprice2022} \\
    24h   & 10         & non-agile & 400$\sim$1100 & \cite{qiDecomposeandlearn2025,wuEnsemble2022} & /   & 1          & non-agile & 5$\sim$10     & \cite{camponogaraContinuoustime2022} \\
    24h   & 4          & agile     & 100$\sim$200  & \cite{hosseinabadiScheduling}       & /   & 1          & non-agile & 10$\sim$25    & \cite{cividanesExtended2021} \\
    24h   & 60         & non-agile & 1000$\sim$3000& \cite{liTwostage2025,wuEnsemble2022} & /   & 5$\sim$30  & non-agile & 500$\sim$3000 & \cite{herrmannReinforcement2024} \\
    24h   & 1          & agile     & 100$\sim$600  & \cite{pengIterated2018}             & /   & 1          & agile     & 100$\sim$600  & \cite{pengAgile2019} \\
    24h   & 10         & non-agile & 400$\sim$800  & \cite{liMultiobjective2024,wuEnsemble2022,liuAdaptive2017} & /   & 1          & agile     & 5$\sim$10     & \cite{mokHeuristicbased2019} \\
    24h   & 5          & non-agile & 500           & \cite{liPTMB2024}                   & /   & 2 const.   & non-agile & 2000          & \cite{povedaEvolutionary2019} \\
    24h   & 10         & non-agile & 200$\sim$1600 & \cite{wuEnsemble2022}               & /   & 2$\sim$10  & non-agile & 848           & \cite{bonnetMultisatellite2015} \\
    24h   & 10         & non-agile & 100$\sim$5000 & \cite{changIntegrated2021,liuAdaptive2017} & /   & 4$\sim$6   & agile     & 40$\sim$1000  & \cite{kimIntegrated2021} \\
    24h   & 2$\sim$10  & non-agile & 848           & \cite{bonnetMultisatellite2015}     &     &            &           &               & \\
    \bottomrule
    \end{tabular}
    \begin{tablenotes}[flushleft]
      \footnotesize
      \item \textit{Note.} ``/'' in the \textit{Period} column indicates that the scheduling horizon is not explicitly specified in the original reference. ``Sats'' denotes the number of satellites (Satellite Scale), and ``Tasks'' denotes the number of tasks (Mission Scale).
    \end{tablenotes}
  \end{threeparttable}
\end{table}

This review confirms that there is still no common standard for instance design in EOS scheduling.
Moreover, most studies report only nominal descriptors such as planning horizon, satellite count, and task volume, while providing little information about structural properties such as feasible opportunity sparsity, task flexibility, pairwise conflict, or satellite-side congestion.
As a result, two scenarios of similar nominal scale may still differ substantially in intrinsic scheduling difficulty, which makes cross-study interpretation of algorithmic performance more difficult.
Even when studies use similar planning horizons, such as 24 hours, they may differ substantially in constellation size, agility assumptions, resource models, and mission scale.
The diversity reflected in Table~\ref{tab:addlabel} illustrates both the breadth of the field and the difficulty of making direct comparisons across published results.
Many studies on large multi-satellite scenarios with thousands of tasks rely on advanced meta-heuristics or DRL, whereas exact and hybrid methods are typically evaluated only on small- and medium-scale instances.
Although some recent work~\cite{rocha2025integrated,wang2025aeosbench} has begun to provide benchmark-style instance sets for specific integrated formulations, the literature still lacks a unified, open, and extensible benchmark framework that spans multiple EOSSP settings under a common evaluation protocol.

\subsection{Summary}

The literature on EOSSP reveals five broad patterns. First, modelling has become progressively richer and closer to operational reality, with increasing attention to platform agility, time-dependent transition times, coupled resource constraints, and multi-satellite coordination~\cite{liuAdaptive2017,pengSolving2022,pengAgile2019,pengExact2020}.
Second, the range of solution approaches has expanded considerably, from exact mixed-integer optimisation and decomposition to meta-heuristics, DRL, imitation learning, and quantum-inspired methods~\cite{barkaouiNew2020,wuEnsemble2022,qiDecomposeandlearn2025}.
Third, evaluation metrics vary significantly depending on the specific study's focus. 
The community employs a rich diversity of evaluation metrics, including total profit, response time, and scalability. 
However, there remains a lack of standardised, multi-dimensional protocols capable of simultaneously assessing mission reward, completion rates, workload balance, timeliness, and computational efficiency across different solver classes.
Fourth, benchmarking practice remains fragmented: planning horizons, constellation scales, task volumes, and resource assumptions vary widely across studies, as illustrated in Table~\ref{tab:addlabel}.
Fifth, existing studies generally overlook the formal characterisation of intrinsic scenario difficulty. Most literature relies solely on nominal scale descriptors (such as the number of tasks or satellites), which fail to capture underlying structural bottlenecks like feasible opportunity sparsity, task flexibility, pairwise conflict intensity, or satellite-side congestion. Consequently, algorithms are often evaluated without a clear understanding of why nominally similar instances can induce vastly different computational challenges.

This literature review on EOSSP therefore indicates that the main difficulty is no longer a lack of modelling ideas or algorithmic techniques, but rather the absence of a common experimental basis on which such methods can be compared systematically.
This gap motivates the benchmark framework developed in this paper, which combines scenario-rich benchmark generation, structural scenario characterisation, and a multi-metric evaluation protocol for Earth observation satellite scheduling.

\section{Benchmark Framework Design}\label{sec:framework}

This section presents the benchmark framework underlying EOS-Bench, which is designed to support both agile and non-agile scenarios within a unified architecture. 
The section begins by defining the scenario space—the design domain encompassing planning horizons, constellation settings, mission loads, and related factors. 
It then describes the core modelling components: the satellite model, the task model, and the visibility module responsible for converting orbital motion and attitude constraints into feasible observation windows.

\begin{figure}[htbp]
    \centering
    \includegraphics[width=1.0\textwidth]{ 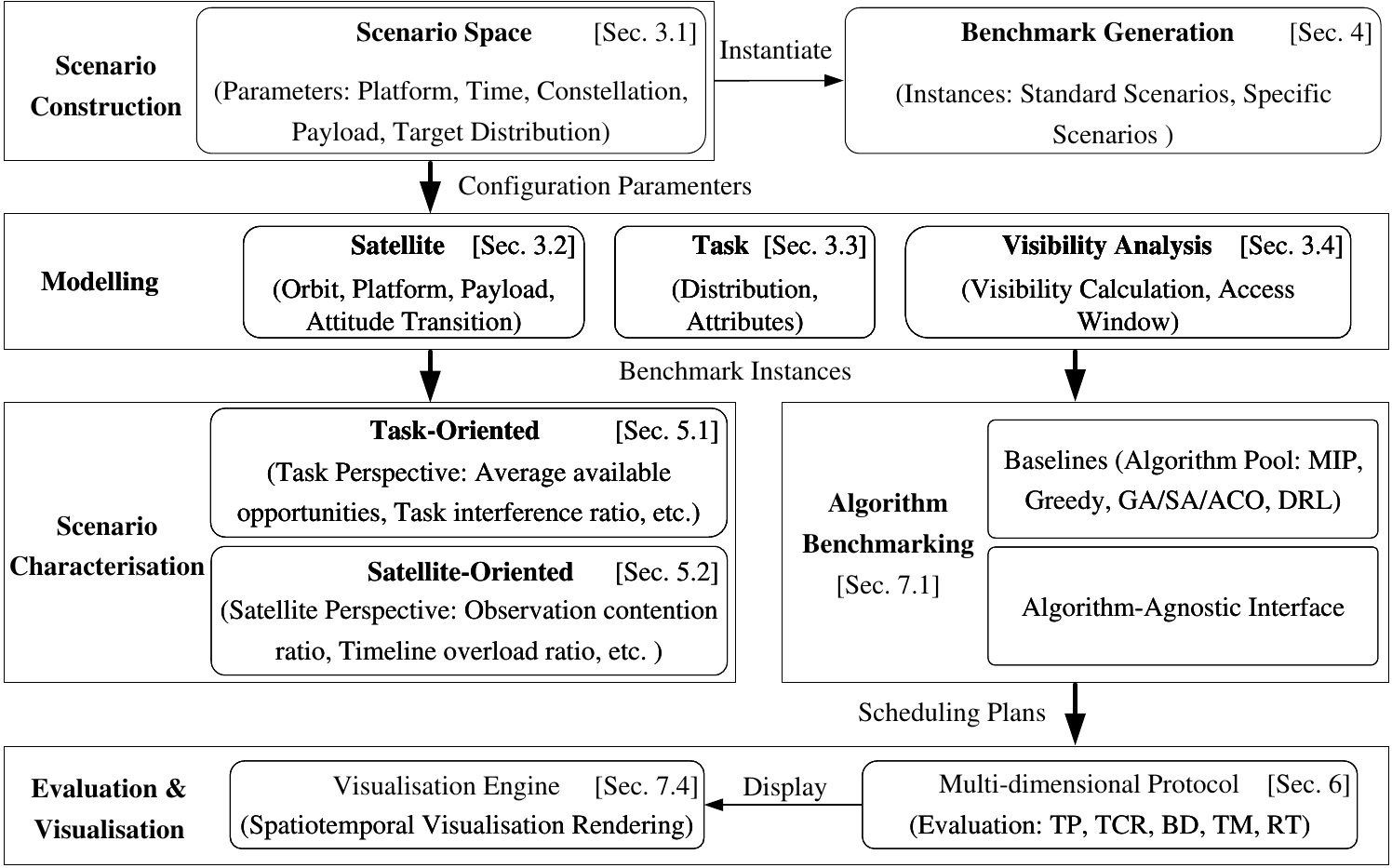 }
    \caption{Modular architecture and end-to-end workflow of the EOS-Bench framework, from initial modelling to multi-dimensional evaluation. A key component is the pre-optimisation characterisation layer used to quantify intrinsic scenario difficulty. Abbreviations: MIP = Mixed Integer Programming, GA = Genetic Algorithm, SA = Simulated Annealing, ACO = Ant Colony Optimisation, DRL = Deep Reinforcement Learning,  TP = Task Profit, TCR = Task Completion Rate, BD = Balance Degree, TM = Timeliness, RT = Runtime.}
    \label{fig:framework}
\end{figure}

To facilitate reading, Table \ref{tab:definitions} summarises the principal terms used throughout this section, and Figure \ref{fig:framework} provides a visual overview of the framework. 
Furthermore, the complete source code and examples of the main modules are available on GitHub (\url{https://github.com/Ethan19YQ/EOS-Bench}).

\begin{table}[htb]
\centering
\footnotesize
\caption{Key terms and definitions used in the EOS-Bench framework.}
\label{tab:definitions}
\begin{tabular}{p{0.23\textwidth} p{0.68\textwidth}}
\toprule
Term & Definition \\
\midrule
Scenario space
  & The overall design domain of the benchmark, that is, the range of satellite platform types (agile or non-agile), planning horizons, constellation settings, mission loads, target distributions, and other configurable factors from which benchmark scenarios are constructed. \\[3pt]
Scenario
  & A specific benchmark setting defined by one unique combination of structural parameters, such as platform type, planning horizon, constellation setting, mission load, target distribution, and the relevant platform or environment settings. \\[3pt]
Instance
  & A concrete realisation of a scenario, obtained by selecting a specific target subset, random seed, or other stochastic variation while keeping the structural scenario parameters fixed. \\[3pt]
Task/Target
  & A geographically referenced imaging request, defined by its target location, required observation duration, priority, and profit. \\[3pt]
Agile satellite
  & A satellite that can actively slew about roll, pitch, and yaw so as to point substantially away from nadir, thereby enabling the observation of different ground targets within a single pass~\cite{ferrari2025satellite,wang2020agile}. \\[3pt]
Non-agile satellite
  & A satellite whose attitude is fixed or only weakly manoeuvrable, so that the payload points at nadir or only slightly off nadir and observations are mainly made along the ground track~\cite{ferrari2025satellite,wang2020agile}. \\[3pt]
Standard Scenarios
  & A family of benchmark scenarios that systematically vary planning horizon, number of satellites, and number of tasks in order to support broad comparison of algorithms across problem scales. \\[3pt]
Specific Scenarios
  & A family of benchmark scenarios that vary one aspect of the system at a time (for example, resource capacity, manoeuvrability profile, or constellation geometry) in order to study algorithm sensitivity and robustness. \\
\bottomrule
\end{tabular}
\end{table}

\subsection{Benchmark Scenario Space}

To provide a systematic testbed for algorithmic evaluation, EOS-Bench constructs a multi-dimensional scenario space. 
Rather than relying on isolated test cases, the framework systematically varies fundamental operational parameters to create a broad spectrum of scheduling environments. 
This parameterised design allows for the precise isolation of algorithmic performance drivers under increasing levels of structural complexity.

For mathematical clarity, we write the benchmark scenario space as
{
\begin{equation}
\label{eq:scenario_space}
\Sigma = \mathcal{P}\times\mathcal{H}\times\mathcal{C}\times\mathcal{L}\times\mathcal{D}\times\mathcal{Z},
\end{equation}
}
where $\mathcal{P}$ is the set of satellite platform types, $\mathcal{H}$ is the set of planning horizons, $\mathcal{C}$ is the set of constellation configurations, $\mathcal{L}$ is the set of mission-load settings, $\mathcal{D}$ is the set of target-distribution patterns, and $\mathcal{Z}$ collects any additional scenario-specific parameters used in the Specific Scenarios.
A scenario is therefore one element
{
\begin{equation}
\label{eq:scenario_tuple}
\sigma=(p,h,c,\ell,d,z)\in\Sigma.
\end{equation}
}
Given a scenario $\sigma$ and a random seed $\xi$, the instance generator produces one concrete benchmark instance
{
\begin{equation}
\label{eq:instance_generator}
\mathcal{I}(\sigma,\xi)=\bigl(\mathcal{S},\mathcal{T},\mathcal{W},h,\Theta^{\mathrm{sat}},\Theta^{\mathrm{task}}\bigr),
\end{equation}
}
\noindent where $\mathcal{S}$ is the satellite set, $\mathcal{T}$ is the task/target set (note that the terms task and target are used interchangeably throughout the paper.
), $\mathcal{W}$ is the family of feasible access-window sets, and $\Theta^{\mathrm{sat}}$ and $\Theta^{\mathrm{task}}$ collect the associated satellite-side and task-side parameters.
This notation makes explicit that EOS-Bench is a distribution over structured scheduling instances rather than a single fixed dataset.

Specifically, along the temporal dimension ($\mathcal{H}$), the benchmark includes short planning horizons of 1, 12, and 24 hours, as well as extended horizons of 72 hours and 7 days. Short horizons approximate a rolling or online decision-making setting, restricting attention to local choices within a limited window, whereas longer horizons enable the examination of cumulative effects, such as repeated resource usage and long-term schedule stability.

Along the resource and mission dimensions ($\mathcal{C}$ and $\mathcal{L}$), scenarios are grouped into three size classes. Small-scale scenarios involve 1--20 satellites and up to 1,000 tasks, serving as testbeds for exact or near-exact methods. Medium-scale scenarios feature 50--200 satellites and up to 2,000 tasks, representing moderately complex operational settings. Finally, large-scale scenarios encompass 500--1,000 satellites and 500--10,000 tasks, designed to evaluate scalability and push algorithms toward their computational limits. By systematically varying these elements, EOS-Bench forms a structured grid that supports comparisons across problem scales.

\subsection{Satellite Modelling}

The satellite model provides a unified representation of orbital dynamics, platform kinematics, and payload behaviour across all benchmark scenarios. 
Its primary purpose is to define when a satellite can observe a ground target, which attitude manoeuvres are admissible, and how imaging activities consume onboard resources. 
Formally, to encapsulate these operational characteristics, EOS-Bench associates each satellite $s\in\mathcal{S}$ with a parameter tuple:
{
\begin{equation}
\label{eq:sat_tuple}
\theta_s=\bigl(o_s,a_s,r_s,q_s\bigr),
\end{equation}
}
where $o_s$ denotes the orbital description, $a_s$ the admissible attitude envelope, $r_s$ the onboard resource capacities, and $q_s$ the payload-consumption parameters.
The benchmark is solver-neutral in the sense that these quantities are defined independently of any optimisation or learning architecture and are supplied directly as instance data.

\subsubsection{Orbital Information and Constellation Design}

The orbital component of EOS-Bench is generated using a hybrid strategy that combines real satellites for small-scale scenarios with synthetic Walker--Delta constellations for medium- and large-scale scenarios.
This design balances physical realism with controlled coverage properties and repeatable large-scale experimentation.

\begin{figure} [htb]
    \centering
    \includegraphics[width=0.85\textwidth]{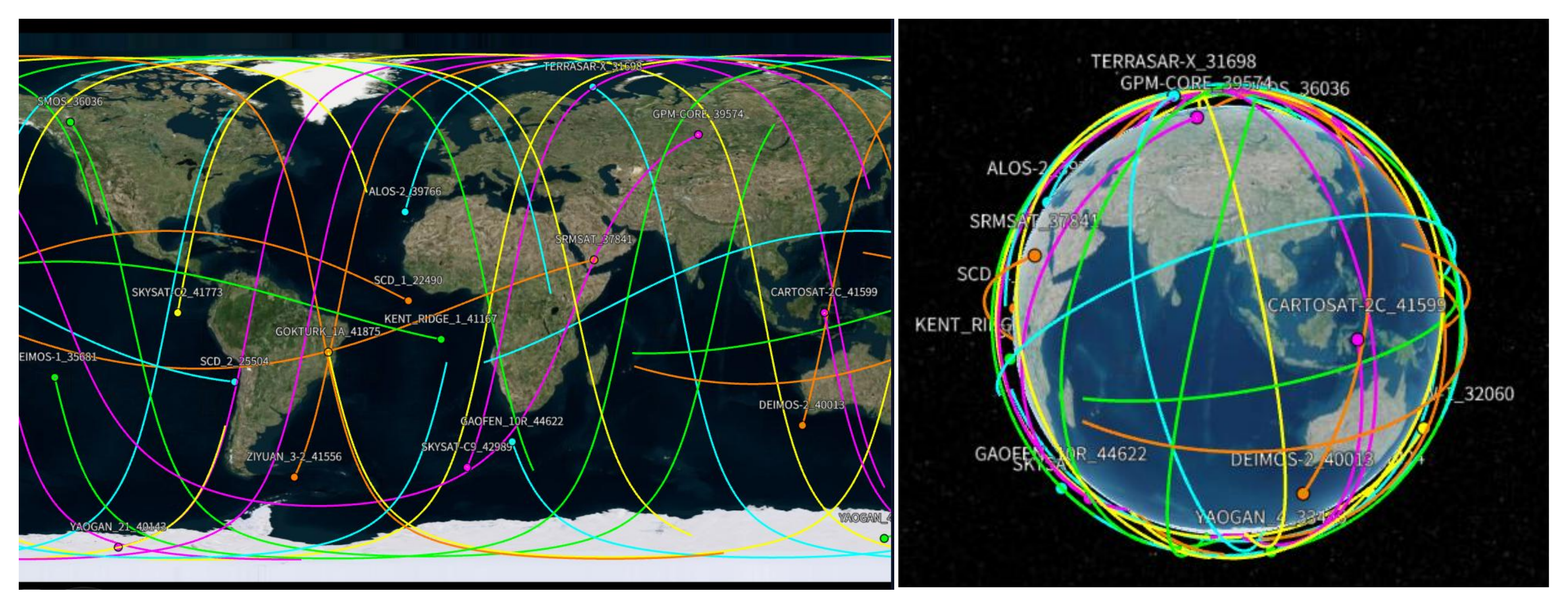}
    \caption{Orbital ground tracks of the 20 real satellites. These ground tracks illustrate the orbital diversity of the real satellite set adopted for the small-scale benchmark scenarios.}\label{fig:flow}
\end{figure}

To instantiate the orbital component $o_s$ defined in Equation~\eqref{eq:sat_tuple}, EOS-Bench employs real-world satellite data for small-scale scenarios (i.e., configurations involving 1, 3, 5, 10, and 20 satellites). 
The raw data are sourced from the ``Earth Resources'' category on CelesTrak\footnote{\url{https://celestrak.org/NORAD/elements/}}, which encompasses a diverse range of operational Earth observation platforms. For standardisation, these data are extracted as the six classical Keplerian elements at a unified reference epoch of 18 Nov 2025 12:00:00 UTC. 
The selected satellites and their specific Keplerian elements are detailed in Table~\ref{tab1}, and their corresponding ground tracks are visualised in Figure~\ref{fig:flow}.

\begin{table}[htbp]
  \centering
  \footnotesize
  \caption{Orbital elements of the real satellites.}
  \begin{tabular}{@{}l l r r r r r r@{}}
    \toprule
    \textbf{Satellite} & \textbf{Country} & \textbf{\makecell{Semi-major\\Axis}} & \textbf{\makecell{Eccen-\\tricity}} & \textbf{\makecell{Incli-\\nation}} & \textbf{\makecell{RAAN}} & \textbf{\makecell{Arg of\\Perigee}} & \textbf{\makecell{True\\Anomaly}} \\
    \midrule
    ALOS-2 & Japan & 7013.62362 & 0.000898 & 98.04 & 57.345 & 101.516 & 96.459 \\
    AQUA & USA   & 7054.608686 & 0.002665 & 98.264 & 282.658 & 100.906 & 181.914 \\
    CARTOSAT-2C & India & 6885.388452 & 0.002331 & 97.526 & 19.64 & 129.493 & 269.449 \\
    DEIMOS-1 & Spain & 7026.003591 & 0.001598 & 97.832 & 123.909 & 99.999 & 287.41 \\
    DEIMOS-2 & Spain & 6955.196768 & 0.002894 & 97.569 & 198.785 & 83.505 & 180.934 \\
    GAOFEN\_10R & China & 7004.506326 & 0.002046 & 97.874 & 269.586 & 82.424 & 282.074 \\
    GOKTURK\_1A & Turkey & 7061.790586 & 0.001387 & 98.061 & 214.868 & 112.033 & 286.074 \\
    GPM-CORE & Japan & 6801.857205 & 0.001294 & 64.827 & 293.591 & 284.823 & 176.995 \\
    KENT\_RIDGE\_1 & Singapore & 6884.956782 & 0.000964 & 15.043 & 31.097 & 207.516 & 57.259 \\
    SCD\_1 & Brazil & 7121.14563 & 0.003755 & 25.1  & 68.138 & 73.8  & 157.166 \\
    SCD\_2 & Brazil & 7123.258992 & 0.001882 & 24.859 & 291.105 & 37.112 & 275.948 \\
    SKYSAT-C2 & USA   & 6833.433187 & 0.00167 & 97.006 & 347.712 & 69.009 & 149.715 \\
    SKYSAT-C9 & USA   & 6825.950288 & 0.00325 & 97.567 & 92.251 & 106.838 & 162.537 \\
    SMOS & ESA   & 7129.70257 & 0.000962 & 98.521 & 147.596 & 18.842 & 100.232 \\
    SRMSAT & India & 7236.742741 & 0.002142 & 19.826 & 254.192 & 78.797 & 37.063 \\
    TERRASAR-X & Germany & 6877.192296 & 0.000926 & 97.37 & 327.892 & 6.52  & 107.452 \\
    WORLDVIEW-1 & USA   & 6874.558678 & 0.001374 & 97.524 & 79.993 & 89.387 & 281.492 \\
    YAOGAN\_21 & China & 6866.60274 & 0.003193 & 97.197 & 6.525 & 115.547 & 184.675 \\
    YAOGAN\_4 & China & 6993.513131 & 0.00264 & 97.779 & 252.523 & 84.41 & 221.626 \\
    ZIYUAN\_3-2 & China & 6866.915774 & 0.002603 & 97.497 & 36.721 & 72.615 & 202.817 \\
    \bottomrule
    \end{tabular}%
  \label{tab1}%
\end{table}%

For medium- and large-scale scenarios, EOS-Bench constructs synthetic Walker--Delta constellations in order to provide controlled and repeatable global coverage.
The real satellite ALOS-2 in Table~\ref{tab1} is used as the seed orbit.
ALOS-2 flies in a sun-synchronous orbit that is representative of many modern Earth observation missions.
To ensure realistic orbital topologies, the Walker pattern parameters, specifically the number of orbital planes, the number of satellites per plane, and the relative phasing, are purposefully structured to maintain a balanced spatial-temporal distribution of assets. Rather than arbitrarily populating orbits, these configurations scale the number of planes and satellites symmetrically. This design choice ensures that as the constellation grows, both the global revisit frequency (driven by orbital plane diversity) and the instantaneous observation capacity (driven by intra-plane density) improve proportionally, thereby accurately reflecting the architectural principles of contemporary and next-generation commercial EO mega-constellations. 
The specific Walker--Delta configurations, selected to generate constellations with progressively higher operational densities, are detailed in Table~\ref{tab:walker_configs}.

\begin{table}[htb]
  \centering
  \small
  \caption{Synthetic Walker--Delta constellation configurations across medium and large scales.}
  \label{tab:walker_configs}
  \begin{tabular}{lccc}
    \toprule
    \textbf{Scale Category} & \textbf{Total Satellites} & \textbf{Orbital Planes} & \textbf{Satellites per Plane} \\
    \midrule
    \multirow{3}{*}{Medium-scale} & 50   & 10 & 5 \\
                                  & 100  & 10 & 10 \\
                                  & 200  & 20 & 10 \\
    \midrule
    \multirow{2}{*}{Large-scale}  & 500  & 25 & 20 \\
                                  & 1000 & 50 & 20 \\
    \bottomrule
  \end{tabular}
\end{table}

All satellites in the benchmark constellations carry an optical imaging payload.
The current release adopts this single payload type in order to maintain consistency across benchmark scales.
At the same time, the framework retains a dedicated payload interface, allowing additional payload models, such as SAR, to be incorporated in future work without changing the rest of the benchmark structure.

\subsubsection{Satellite Platform Model}\label{subsec:sat_platform}

The satellite platform model details the admissible attitude envelope $a_s$ defined in Equation~\eqref{eq:sat_tuple}, accommodating both agile and non-agile configurations. 
In EOS-Bench, this envelope is parametrised by bounding limits on the roll, pitch, and yaw axes, which collectively determine the instantaneous field of regard. 

For agile satellites, the model assumes a highly flexible three-axis envelope. 
Based on representative values from the literature~\cite{liuAdaptive2017,wang2020agile,pengAgile2019}, EOS-Bench adopts the symmetric bounds summarised in Table~\ref{tab:platform}. 
For non-agile satellites, rather than employing an ambiguous nadir-dominant characterisation, the model explicitly parametrises them as roll-only manoeuvrable platforms. 
This means pitch and yaw capabilities are disabled (fixed at $0^\circ$), while the roll axis retains the bounds specified in Table~\ref{tab:platform} to enable standard cross-track targeting.

\begin{table}[htb]
  \centering
  \small
  \caption{Satellite platform parametrisation. Default attitude-envelope parameters for representing satellite pointing capability in EOS-Bench.}
  \label{tab:platform}
  \begin{tabular}{p{0.2\textwidth}p{0.15\textwidth}p{0.7\textwidth}}
    \toprule
     Parameter & Default Value & Rationale \\
    \midrule
       Max roll angle & $\pm 45^\circ$ & Typical field-of-regard for agile Earth observation satellites~\cite{liuAdaptive2017,wang2020agile} \\
       Max pitch angle & $\pm 45^\circ$ & Symmetric along-track pointing envelope~\cite{pengAgile2019} \\
       Max yaw angle & $\pm 90^\circ$ & Additional cross-track flexibility for agile platforms~\cite{pengAgile2019} \\
    \bottomrule
  \end{tabular}%
\end{table}%

\subsubsection{Satellite Payload Model}

The payload model describes how imaging activities consume onboard resources.
Following common practice in EOSSP~\cite{liuAdaptive2017,pengAgile2019,liuTwostage2025,liDynamic2025}, EOS-Bench expresses energy and data budgets in normalised units rather than in physical joules or bits.
For scheduling purposes, what matters is the relative scale between observation time, slewing effort, and per-orbit resource limits, rather than the absolute physical values.
The parameters adopted in the benchmark are listed in Table~\ref{tab:payload}.

\begin{table}[htb]
  \centering
  \small
  \caption{Payload and imaging parametrisation. Default resource-consumption and capacity parameters used to model imaging activities in EOS-Bench.}
  \label{tab:payload}
  \begin{tabular}{p{0.35\textwidth}p{0.15\textwidth}p{0.5\textwidth}}
    \toprule
     Parameter & Default Value & Rationale \\
    \midrule
       Observation energy rate & 1 unit/s & Normalised per-second energy draw in imaging mode~\cite{liuAdaptive2017,liuTwostage2025} \\
       Observation memory rate & 1 unit/s & Normalised data-acquisition rate~\cite{liDynamic2025} \\
       Energy capacity per orbit & 200 units & Normalised energy budget per orbit, consistent with per-orbit modelling in~\cite{liuAdaptive2017,liuTwostage2025} \\
       Storage capacity per orbit & 2400 units & Normalised data buffer capacity, sufficient for multiple long observations~\cite{liDynamic2025} \\
       Attitude manoeuvre energy rate & 1 unit/deg & Linearised energy cost for slews, as in~\cite{liuAdaptive2017,pengAgile2019} \\
    \bottomrule
  \end{tabular}%
\end{table}%

\subsubsection{Attitude Transition-Time Model}\label{subsec:transition}

For agile satellites, the time required to change the line-of-sight between two observations has a strong effect on schedule feasibility.
EOS-Bench adopts a piecewise linear transition-time function linking transition time to the total change in look angle, following the time-dependent models proposed by~\cite{liuAdaptive2017} and~\cite{pengSolving2022,pengAgile2019,pengExact2020}.

Let $\Delta g = |\Delta\gamma| + |\Delta\pi| + |\Delta\psi|$ denote the sum of the absolute changes in roll, pitch, and yaw angles between two imaging attitudes.
The transition time is then defined as
\begin{equation}
\label{eq_1}
Trans(\Delta g) =
\begin{cases}
11.66, & \Delta g \le 10,   \\
a_1 + \dfrac{\Delta g}{v_1}, & 10 < \Delta g \le 30, \\
a_2 + \dfrac{\Delta g}{v_2}, & 30 < \Delta g \le 60, \\
a_3 + \dfrac{\Delta g}{v_3}, & 60 < \Delta g \le 90, \\
a_4 + \dfrac{\Delta g}{v_4}, & \Delta g > 90,
\end{cases}
\end{equation}
Here $v_1,\dots,v_4$ are angular velocities in degrees per second, and $a_1,\dots,a_4$ are offsets chosen so that $Trans(\Delta g)$ is continuous and non-decreasing at the breakpoint angles.
EOS-Bench uses the parameter values reported in~\cite{pengSolving2022,pengAgile2019,pengExact2020}, with $v_1 = 1.5^\circ/\mathrm{s}$, $v_2 = 2.0^\circ/\mathrm{s}$, $v_3 = 2.5^\circ/\mathrm{s}$ and $v_4 = 3.0^\circ/\mathrm{s}$.
The offsets $a_1$, $a_2$, $a_3$, and $a_4$ are set so that $Trans(\Delta g)$ remains continuous at all breakpoints.
For two consecutive observations $i$ and $j$ with start times $t_i$ and $t_j$, the required slew time is written as $Trans_{i,j}(t_i,t_j)$, and is enforced in the scheduling model as a minimum time-gap constraint between the two tasks.

This transition-time model is widely used in the agile EOSSP literature~\cite{ferrari2025satellite,pengSolving2022,pengAgile2019,pengExact2020}, so adopting it here keeps the kinematic assumptions of EOS-Bench consistent with recent work.
The model is intended as a benchmark-level approximation rather than as a full attitude-control simulation.
Specifically, $\Delta g=|\Delta\gamma|+|\Delta\pi|+|\Delta\psi|$ provides a conservative axis-wise estimate of manoeuvre cost, and the transition is evaluated from a representative viewing geometry rather than from a continuously updated line of sight throughout the slew.
These simplifications improve tractability for large-scale benchmark generation while retaining the main time-dependent features relevant to EOSSP.

In contrast to the dynamic reorientation of agile platforms, non-agile satellites operate under much stricter kinematic constraints. 
As widely adopted in non-agile EOSSP literature (e.g., the operational constraints of platforms like SPOT-5)~\cite{lemaitre2002selecting}, continuous angle-dependent transition calculation is unnecessary. 
Consequently, EOS-Bench applies a fixed transition time of 10 seconds between any two consecutive non-agile observations. 
This constant duration represents the standard hardware stabilisation delay required for instrument reset and minor cross-track mirror adjustments.

\subsection{Task Modelling}

The task model is designed to reflect the diversity and time-dependent character of real Earth observation demand while retaining a problem structure compatible with most existing scheduling methods.
In EOS-Bench, each task is represented as a point target together with an associated priority, profit weight, and required observation duration.
In the current release, EOS-Bench assumes an offline setting in which all task information is available at the initial decision time $t_0$.
Specifically, the full task set for a scenario is known in advance, including target location, profit, priority, and required observation duration.

Formally, for each task $i\in\mathcal{T}$, we write
{
\begin{equation}
\label{eq:task_tuple}
\tau_i=(z_i,\rho_i,p_i,d_i),
\end{equation}
}
where $z_i\in\mathbb{R}^2$ is the geodetic target location, $\rho_i$ is the priority level, $p_i$ is the task profit, and $d_i$ is the required continuous observation duration.
At the benchmark level, EOS-Bench therefore treats task generation as the construction of a finite set $\mathcal{T}=\{\tau_1,\dots,\tau_{N_T}\}$ together with the access-window family induced by the orbital and visibility models.

In the EOSSP literature, a number of studies consider area-type requests such as regions or strips~\cite{gu2022large,chenLargescale2023}.
These requests are often discretised into sets of point targets or small tiles for scheduling.
For example, several agile-satellite studies sample representative points within polygons or along strips in order to approximate area requests~\cite{nagScheduling2018,mokHeuristicbased2019,gu2022large}.
Recent large-scale constellation studies follow the same principle and convert region demands into collections of point-like missions~\cite{fengMultisatellite2024,wenScheduling2024}.
However, to follow the point-target convention and for simplicity, the first release of EOS-Bench focuses on point targets.
As detailed  in Section~\ref{sec:directions}, future releases could accommodate area and strip targets by generating suitable sets of point tasks, grouping related points into higher-level missions, or adopting set-covering formulations, without changing the core benchmark interfaces.

\subsubsection{Task Distribution}

EOS-Bench uses three simple synthetic spatial distributions of targets in order to test algorithms under different geographical demand patterns:
\begin{itemize}
    \item \textbf{Global-Random}: targets are sampled approximately uniformly over the Earth's land surface.
    \item \textbf{Region-Clustered}: targets are grouped within five broad continental regions (Asia, Europe, Africa, the Americas, and Oceania).
    \item \textbf{Hybrid}: each task set combines some Global-Random targets with others drawn from the Region-Clustered pattern.
\end{itemize}

For each distribution pattern, a pool of 10{,}000 candidate target locations is first generated.
This pool is then divided into 10 disjoint subsets, each containing 1{,}000 targets.
For any given benchmark instance, the required number of tasks is drawn from one chosen subset.
As a result, when the task count increases, the additional tasks come from the same subset, so the mission load changes while the spatial pattern remains comparable.

In addition to these synthetic pools, EOS-Bench also includes a set of 1{,}000 real-world city targets.
These cities comprise national capitals and first-level administrative capitals.
Their coordinates are obtained from the GeoNames geographical database\footnote{\url{https://download.geonames.org/export/}}.

\subsubsection{Task Attributes}

Each observation task is described by a small set of attributes specifying its spatial location and relative importance.
These attributes are summarised in Table~\ref{tab3}.

\begin{table}[htp]
  \centering
  \small
  \caption{Task attributes. Main attributes and their mathematical symbols used to define observation tasks $\tau_i$ in EOS-Bench.}
    \begin{tabular}{p{0.25\textwidth} p{0.1\textwidth} p{0.55\textwidth}}
    \toprule
    \textbf{Attribute Name} & \textbf{Symbol} & \textbf{Description and Values} \\
    \midrule
    Task ID & $i$ & Unique identifier for the task. \\
    Target Location & $z_i$ & Geodetic latitude and longitude coordinates. \\
    Priority & $\rho_i$ & Integer from 1 to 10, indicating task urgency. \\
    Profit & $p_i$ & Integer from 1 to 10, indicating the value gained upon task completion. \\
    Observation Duration & $d_i$ & Random integer between 5 and 15 seconds, indicating the required continuous imaging time. \\
    \bottomrule
    \end{tabular}%
  \label{tab3}%
\end{table}%

Priority and profit jointly define a multi-objective optimisation landscape, encouraging algorithms to trade off the service of urgent requests against the maximisation of total reward.
Variation in observation duration introduces heterogeneity in resource consumption, since longer tasks consume more energy and storage and may reduce the number of tasks that can be accommodated within a given orbit.
In a more detailed physical model, observation duration may depend on the satellite--task pair, because it is influenced by relative motion, viewing geometry, and the spatial extent of the target region.
In the current release of EOS-Bench, however, observation duration is treated as a task-level attribute and generated randomly within a predefined range.
More detailed region-based and geometry-dependent duration models are left to future extensions.

\subsection{Visibility Analysis Model}

The visibility module links constellation orbital dynamics to the spatio-temporal availability of ground targets, discretising continuous trajectories into valid access windows based on line-of-sight and FOV conditions. 
Following established models~\cite{hanVisibility2019,wangOnboard2019,hanNovel2021}, EOS-Bench applies scalable geometric tests to generate candidate contacts, which are then filtered against the attitude constraints of both agile and non-agile platforms.

Currently, environmental uncertainty is not explicitly modelled. By excluding time-varying cloud cover and post-scheduling execution failures, access windows are generated under strictly deterministic assumptions. 
This ensures a standardised and reproducible baseline for algorithm evaluation, preserving the framework's flexibility to support stochastic extensions in the future.

\subsubsection{Visibility Calculation Method}

The visibility module links the orbital trajectories of the constellation to the spatio-temporal availability of ground targets. Because orbit propagation is a mature technology, EOS-Bench relies on the standard SGP4 model to efficiently generate baseline satellite positions. For any given satellite--target pair, instantaneous visibility is determined by a binary predicate:
\begin{equation}
\label{eq:visibility_predicate}
\chi_{si}(t)=\mathbf{1}\!\left(\mathrm{LOS}_{si}(t)\wedge\mathrm{FOV}_{si}(t)\wedge\mathrm{ATT}_{si}(t)\right),
\end{equation}
where $\mathrm{LOS}_{si}(t)$ denotes line-of-sight feasibility, $\mathrm{FOV}_{si}(t)$ denotes payload field-of-view feasibility, and $\mathrm{ATT}_{si}(t)$ denotes platform attitude feasibility. Crucially, $\mathrm{ATT}_{si}(t)$ explicitly reflects the specific kinematic constraints of either the agile or non-agile platforms defined in Section~\ref{subsec:sat_platform}.

Consecutive visible epochs satisfying Equation~\eqref{eq:visibility_predicate} are merged into discrete access windows, which serve as the direct input for all downstream scheduling algorithms. While the standard SGP4 geometric engine provides a robust and reproducible default, the modular architecture of EOS-Bench allows users to seamlessly integrate advanced visibility accelerators~\cite{hanNovel2021,wangOnboard2019,gu2019kriging} without altering the scenario structure or solver interfaces.

\subsubsection{Access Window Representation}

For each satellite--target pair, the visibility module returns a set of access windows that serves as the standard input for all downstream scheduling algorithms in EOS-Bench.
To accurately capture the distinct kinematic capabilities of different platforms, the data structure of these windows is tailored according to satellite agility, as detailed in Table~\ref{tab:access_window}.

While core temporal and identification attributes are shared across all platforms, the representation of pointing geometry diverges. 
For agile satellites, the window incorporates a discrete-time dynamic attitude sequence. This high-resolution profile is essential for scheduling algorithms to precisely evaluate the time-dependent reorientation costs between consecutive tasks.
Conversely, the representation for non-agile platforms is strictly simplified to a single, fixed roll angle, implicitly assuming that pitch and yaw remain locked at $0^\circ$ throughout the entire observation. 

By relying on these transparent and standardised data structures, EOS-Bench ensures that the resulting benchmark instances remain solver-agnostic and computationally manageable for large-scale constellation studies.

\begin{table}[htp]
  \centering
  \small
  \caption{Access-window attributes. Standardised data fields defining observation opportunities for agile and non-agile platforms in EOS-Bench.}
    \begin{tabular}{p{0.22\textwidth} p{0.18\textwidth} p{0.5\textwidth}}
    \toprule
    \textbf{Attribute Name} & \textbf{Platform Type} & \textbf{Description and Values} \\
    \midrule
    Task ID & Both & Unique identifier for the observation task. \\
    Satellite ID & Both & Identifier of the satellite executing the task. \\
    Time Window & Both & $[t^{\text{start}}, t^{\text{end}}]$, the continuous feasible observation period. \\
    Attitude Sequence & Agile Only & A discrete-time sequence of $(\gamma, \pi, \psi)$ triplets defining the dynamic pointing profile required to track the target. \\
    Fixed Roll Angle & Non-agile Only & A single, constant roll value ($\gamma^*$) for the entire observation, with pitch and yaw fixed at $0^\circ$. \\
    \bottomrule
    \end{tabular}%
  \label{tab:access_window}%
\end{table}

\subsection{Formal Benchmark Specification}

The benchmark can now be formalised at the level of feasible assignments, schedules, and evaluation maps. This formalisation is intentionally solver-agnostic: it does not prescribe a specific optimisation model, but instead defines the rigorous mathematical structure that any solver interacting with EOS-Bench must process.

For a fixed instance $\mathcal{I}=(\mathcal{S},\mathcal{T},\mathcal{W},h,\Theta^{\mathrm{sat}},\Theta^{\mathrm{task}})$, define the set of candidate observation assignments as
\begin{equation}
\label{eq:candidate_assignments}
\mathcal{A}(\mathcal{I})=
\left\{(s,i,\ell,t): s\in\mathcal{S},\ i\in\mathcal{T},\ w_{si\ell}\in W_{si},\ t\in\mathcal{F}_{si\ell}\right\}.
\end{equation}
Each element $a=(s,i,\ell,t)\in\mathcal{A}(\mathcal{I})$ represents the decision to execute task $i$ on satellite $s$ by starting within access window $w_{si\ell}$ at time $t$.

EOS-Bench does not assume that all such assignments are jointly executable. Instead, it induces a compatibility relation $\mathcal{C}(a,a')\in\{0,1\}$, where $\mathcal{C}(a,a')=1$ if assignments $a$ and $a'$ can coexist without violating window overlap, minimum transition-time separation, or single-satellite exclusivity. Furthermore, EOS-Bench induces instance-dependent resource-feasibility maps, written abstractly as
\begin{equation}
\label{eq:resource_feasibility_maps}
R_E(X)\le B_E, \qquad R_M(X)\le B_M,
\end{equation}
where $R_E$ and $R_M$ aggregate energy and memory consumption over the selected assignments $X\subseteq\mathcal{A}(\mathcal{I})$, and $B_E,B_M$ denote the corresponding satellite capacity limits.

A feasible schedule is therefore any subset $X\subseteq\mathcal{A}(\mathcal{I})$ that simultaneously satisfies pairwise compatibility, task uniqueness (each task is executed at most once), and satellite-side resource feasibility:
\begin{equation}
\label{eq:feasible_schedule_conditions}
\forall a,a'\in X,\ a\neq a' \implies \mathcal{C}(a,a')=1,
\end{equation}
\begin{equation}
\label{eq:task_uniqueness_condition}
\forall i\in\mathcal{T},\qquad \left|\{a=(s,j,\ell,t)\in X: j=i\}\right|\le 1,
\end{equation}
\begin{equation}
\label{eq:resource_schedule_condition}
R_E(X)\le B_E,\qquad R_M(X)\le B_M.
\end{equation}

This definition yields the feasible-set view
\begin{equation}
\label{eq:feasible_set}
\mathfrak{F}(\mathcal{I})=
\left\{X\subseteq\mathcal{A}(\mathcal{I}) : X \text{ satisfies \eqref{eq:feasible_schedule_conditions}--\eqref{eq:resource_schedule_condition}}\right\},
\end{equation}
which is the central mathematical space presented by the benchmark to any downstream solver.

Finally, EOS-Bench operates as an evaluation pipeline over a distribution of instances. Let $P_\Sigma$ denote a scenario distribution over $\Sigma$, let $G$ be the instance generator, and let $f$ be any solver mapping instances to feasible schedules. The benchmark evaluates $f$ through
\begin{equation}
\label{eq:evaluation_map}
\mathcal{E}(f;\sigma)=
\mathbb{E}_{\xi\sim P(\cdot\mid \sigma)} \bigl[m\bigl(f(G(\sigma,\xi))\bigr)\bigr],
\end{equation}
where $m(\cdot)$ returns the vector of reported benchmark metrics. In practice, EOS-Bench approximates this expectation by averaging over a finite number of generated instances per scenario. This formulation clarifies that the benchmark is not merely a collection of handcrafted test cases, but an empirical protocol for estimating solver behaviour under controlled distributions of scheduling challenges.

\subsection{Scope and Limitations of EOS-Bench}

While EOS-Bench is designed to provide a comprehensive and extensible benchmark for EOSSP, the current release reflects a set of deliberate modelling choices that define its scope.

\begin{itemize}
\item {Presentation focus on agile scheduling:}
Although EOS-Bench supports both agile and non-agile satellite configurations and extensive experiments were conducted for both, the detailed analysis presented in the main text focuses on agile scenarios due to space limitations. The complete non-agile experimental results are provided in the benchmark's public repository.

\item {Offline and deterministic setting:}
The benchmark adopts an offline scheduling formulation with deterministic access windows and known task parameters. It does not currently model execution uncertainty, such as cloud cover, task failure, or onboard replanning, which are relevant in operational settings.

\item {Simplified task representation:}
The current version focuses on point targets, enabling controlled evaluation across large scenario sets. More complex task types, such as area or strip imaging, are not explicitly modelled in this release.

\item {Abstracted resource modelling:}
Energy and storage capacities are represented using normalized units rather than hardware-calibrated values. While this supports scalability and comparability across scenarios, it abstracts away mission-specific resource constraints.

\item {Benchmark-level operational modelling:}
Attitude transition and resource consumption are modelled using simplified but representative approximations. The benchmark is not intended to serve as a high-fidelity spacecraft simulator, but rather as a controlled environment for algorithm evaluation.

\item {Limited environmental and sensing factors:}
The current release does not explicitly account for environmental factors such as weather conditions, lighting variability beyond geometric constraints, or sensor-specific characteristics. These aspects can influence real-world scheduling outcomes.

\item {Selective reporting of empirical results:}
Although the released library covers both platform types and a comprehensive cross-platform experimental campaign was conducted, the main text discusses only a selected agile subset to ensure a focused narrative. Users interested in the full comparative assessment across both agile and non-agile platforms should refer to the supplementary data repository. Future extensions could further incorporate environmental uncertainty and heterogeneous constellations to strengthen external validity.
\end{itemize}
These design choices are intentional to ensure tractability, reproducibility, and broad applicability of the benchmark. Future extensions of EOS-Bench can progressively incorporate richer task models, uncertainty, and additional operational constraints, which are discussed in Section~\ref{sec:directions}.

\color{black}

\section{Benchmark Generation}\label{sec:generation}


EOS-Bench can be viewed as a two-stage map for benchmark generation. First, a structural scenario template $\sigma\in\Sigma$ is fixed. Second, independent seeds $\xi_1,\dots,\xi_M$ are used to instantiate
{
\begin{equation}
\label{eq:instance_family}
\mathcal{I}^{(m)} = G(\sigma,\xi_m), \qquad m=1,\dots,M,
\end{equation}
}
where $G$ denotes the benchmark generator.
The resulting instance family shares the same structural parameters and differs only through the stochastic elements admitted by the template, such as target-subset selection or seed-dependent task sampling.

\begin{figure}[htb]
    \centering
    \includegraphics[width=0.8\textwidth]{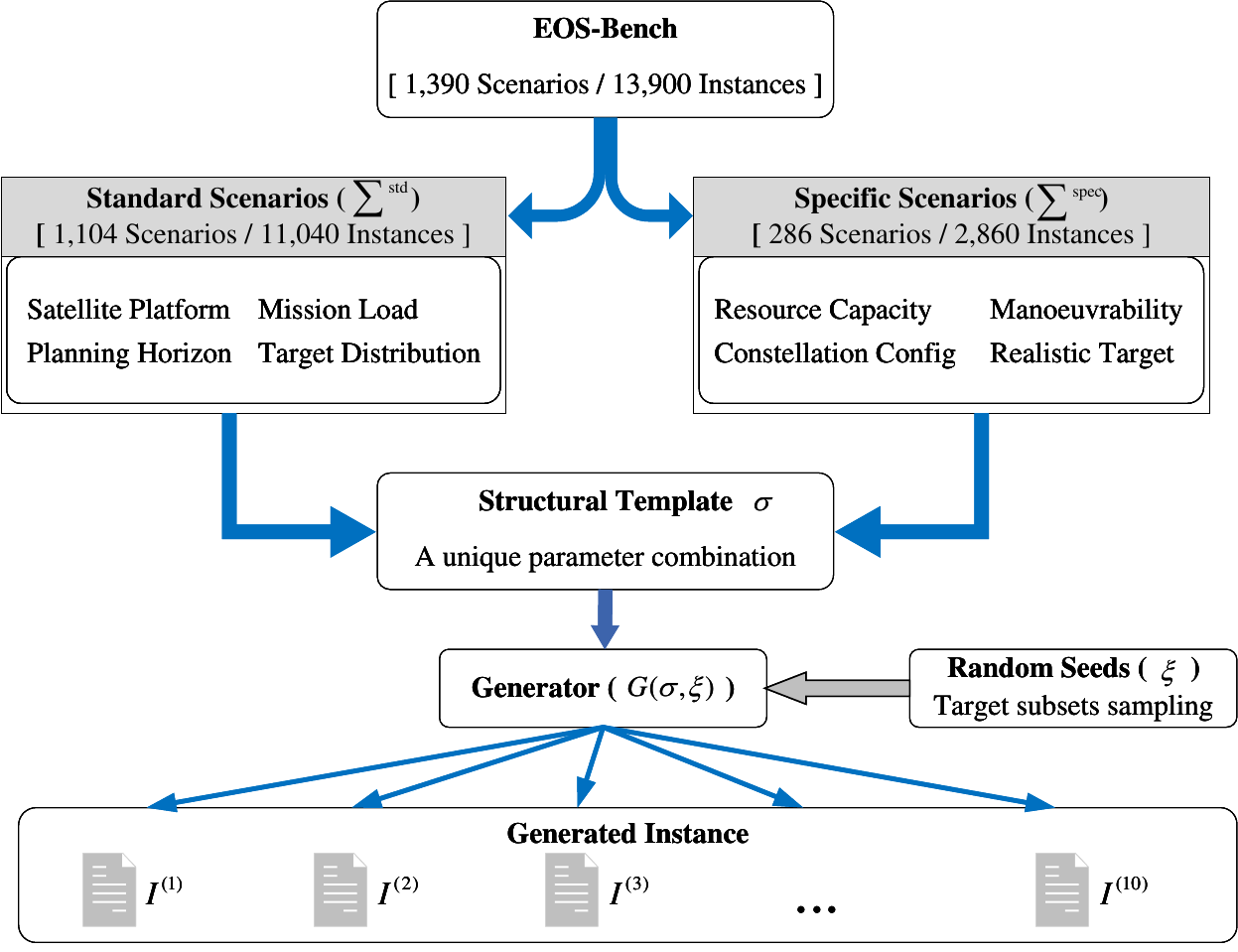}
    \caption{Hierarchical organisation and generation mechanism of the EOS-Bench. It illustrates the mapping of combinatorial parameters to structural templates ($\sigma$), which are instantiated via a stochastic generator ($G(\sigma, \xi)$) to form the Standard and Specific scenario libraries.}
    \label{fig:organize}
\end{figure}

Benchmark scenarios are constructed by combining the main modelling dimensions introduced in Section~\ref{sec:framework}:
satellite platform (agile or non-agile), the planning horizon, the constellation and platform configuration, and
the task set.
Each unique combination of these elements defines one benchmark scenario.
From a given scenario, multiple instances can then be generated by varying the sampled target subset or the random seed used in task generation.
To keep the benchmark library structured and interpretable, EOS-Bench divides scenarios into two groups and each scenario has generated 10 instances, as illustrated in Figure~\ref{fig:organize}.
Formally, we write
{
\begin{equation}
\label{eq:scenario_partition}
\Sigma = \Sigma^{\mathrm{std}} \cup \Sigma^{\mathrm{spec}},
\qquad
\Sigma^{\mathrm{std}} \cap \Sigma^{\mathrm{spec}} = \varnothing,
\end{equation}
}
where $\Sigma^{\mathrm{std}}$ denotes the Standard Scenario family and $\Sigma^{\mathrm{spec}}$ denotes the Specific Scenario family.
Standard Scenarios provide the main benchmark grid for broad comparison across solver classes, whereas Specific Scenarios are designed to isolate the effects of particular factors, including resource capacity, manoeuvrability, constellation geometry, and realistic target geography.
In total, EOS-Bench contains 1,390 scenarios and 13,900 instances, comprising 1,104 Standard Scenarios (11,040 instances) and 286 Specific Scenarios (2,860 instances).

\subsection{Standard Scenarios}\label{subsec:standard_scenarios}

The Standard Scenarios form the core of the benchmark library.
They systematically vary four principal dimensions: platform type, planning horizon, constellation size, and mission load.

Four planning horizons are considered, including $0.5$, $1$, $3$, and $7$ days.
Ten constellation sizes are included, specifically $1$, $3$, $5$, $10$, $20$, $50$, $100$, $200$, $500$, and $1000$ satellites.
For small constellations (1--20 satellites), the real satellites defined in Section~\ref{sec:framework} are used.
For medium- and large-scale constellations (50--1000 satellites), the default Walker--Delta architectures introduced in Section~\ref{sec:framework} are adopted.

For each satellite count, several task-load levels are defined in order to produce scenarios of increasing density.
The admissible task counts are listed in Table~\ref{tab5}.
Each combination of planning horizon, satellite count, and task count defines one structural Standard Scenario.
For each structural template, EOS-Bench distinguishes between agile and non-agile platform instantiations under the common framework introduced in Section~\ref{sec:framework}. In the agile branch, the benchmark uses the standard agile attitude and transition model. In the non-agile branch, it uses the nadir-dominant non-agile observation model described in Section~\ref{sec:framework}.

\begin{table}[htbp]
  \centering
  \caption{Mission-load levels used in Standard Scenario templates.}
    \begin{tabular}{cc}
    \toprule
    Satellite Count & Task Count \\
    \midrule
     1 & 10, 20, 30, 40, 50, 100 \\
     3 & 10, 50, 100, 200 \\
     5 & 50, 100, 200, 500 \\
     10 & 100, 200, 500, 1000 \\
     20 & 100, 200, 500, 1000 \\
     50 & 50, 100, 150, 200, 500 \\
     100 & 100, 300, 500, 1000, 2000 \\
     200 & 200, 500, 1000, 1500, 2000 \\
     500 & 500, 1000, 1500, 2000, 5000 \\
     1000 & 1000, 2000, 5000, 10000 \\
    \bottomrule
    \end{tabular}%
  \label{tab5}%
\end{table}%

Table~\ref{tab5} contains 46 distinct combinations of satellite count and task count.
Combined with the four planning horizons, this yields 184 structural scenario templates.
When the three target-distribution patterns and the two platform types (agile and non-agile) are also taken into account, the benchmark comprises 1,104 Standard Scenarios in total.
With 10 instances generated for each scenario, the Standard suite therefore contains 11,040 instances.
Because each target pool is pre-generated and split into 10 disjoint subsets for each distribution pattern, multiple instances can be produced for the same Standard Scenario without altering its structural parameters.

\subsection{Specific Scenarios}\label{subsec:specific_scenarios}

Whereas the Standard Scenarios span a broad range of benchmark settings, they keep the main platform parameters at their nominal values.
To vary these parameters in a controlled manner,  286 Specific Scenarios are further designed to examine individual settings through targeted sensitivity analysis.
EOS-Bench includes four families of Specific Scenarios: resource-capacity variants, manoeuvrability profiles, constellation-configuration variants, and realistic-target scenarios.
Among them, the manoeuvrability-profile family is specific to agile satellites.

\subsubsection{Resource-Capacity Variants}

The resource-capacity scenarios examine how algorithmic performance changes under different energy and storage budgets.
This scenario family is defined at the benchmark-library level for both agile and non-agile platforms under the same resource parametrisation.
The baseline per-orbit energy and storage capacities are those given in Table~\ref{tab:payload}.
To keep the number of scenarios manageable while still providing a comprehensive evaluation across different problem scales, four representative Standard Scenarios are selected as baseline configurations. As detailed in Table~\ref{tab:capacity_baselines}, these configurations capture a progressive scale from small-satellite clusters to massive constellations, all operating under a standard 24-hour planning horizon.

\begin{table}[htb]
\centering
\small
\caption{Baseline configurations for resource-capacity scenarios.}
\begin{tabular}{lccc}
\toprule
\textbf{Scale Category} & \textbf{Satellite Count} & \textbf{Task Count} & \textbf{Planning Horizon} \\
\midrule
Small-Scale  & 3   & 200  & 24 hours \\
Small-Scale  & 10  & 500  & 24 hours \\
Medium-Scale & 100 & 500  & 24 hours \\
Large-Scale  & 500 & 2000 & 24 hours \\
\bottomrule
\end{tabular}%
\label{tab:capacity_baselines}%
\end{table}

For each of these three baseline scenarios, all six capacity patterns are systematically evaluated using the scaling profiles listed in Table~\ref{tab6}.
These include three homogeneous variants (Low-Capacity, Standard-Capacity, and High-Capacity) and three heterogeneous variants (Mixed-Capacity A--C).
The Standard-Capacity setting coincides with the underlying Standard Scenario and is therefore not counted as an additional Specific Scenario.
This yields 120 distinct resource-capacity scenarios and 1,200 instances in total (4 baseline settings $\times$ 5 capacity patterns $\times$ 3 target distributions $\times$ 2 platform types, with 10 instances per scenario).

\begin{table}[htb]
\centering
\small
\caption{Resource-capacity scenario configurations.}
\begin{tabular}{p{0.2\textwidth}p{0.2\textwidth}p{0.6\textwidth}}
\toprule
\textbf{Pattern} & \textbf{Configuration Name} & \textbf{Capacity Distribution} \\
\midrule
\multirow{3}{*}{\textbf{Homogeneous}}
& Low-Capacity & All satellites: 50\% energy and storage capacity \\
& Standard-Capacity & All satellites: 100\% energy and storage capacity (baseline) \\
& High-Capacity & All satellites: 150\% energy and storage capacity \\
\midrule
\multirow{3}{*}{\textbf{Heterogeneous}}
& Mixed-Capacity A & 20\% high-capacity, 60\% standard-capacity, 20\% low-capacity \\
& Mixed-Capacity B & 25\% high-capacity, 50\% standard-capacity, 25\% low-capacity \\
& Mixed-Capacity C & 30\% high-capacity, 40\% standard-capacity, 30\% low-capacity \\
\bottomrule
\end{tabular}%
\label{tab6}%
\end{table}

These scenarios require algorithms to balance task selection, energy expenditure, and data-buffer usage under both uniformly constrained fleets and fleets with mixed resource levels.

\subsubsection{Manoeuvrability Profiles}

Satellite manoeuvrability determines how rapidly a platform can adjust its attitude between successive observations.
In EOS-Bench, attitude changes are modelled by the piecewise linear transition-time function $Trans(\Delta g)$ in Equation~\eqref{eq_1}, which is widely used in agile Earth observation satellite scheduling studies with time-dependent transition times and profits~\cite{liuAdaptive2017,pengAgile2019,pengExact2020,pengSolving2022}.
In this model, the transition time between two observations depends on the total change in look angle $\Delta g$ and on four angular velocities $v_1,\dots,v_4$, which describe the slew rate over different angular ranges.

Following~\cite{pengAgile2019,pengExact2020,pengSolving2022}, EOS-Bench fixes the breakpoint angles at $\Delta g = 10^\circ, 30^\circ, 60^\circ, 90^\circ$ and uses a minimum transition time of $11.66$~s for very small attitude changes.
The constant terms in Equation~\eqref{eq_1} are fixed at $a_1=5$, $a_2=10$, $a_3=16$, and $a_4=22$.
These values correspond to the reference agile optical satellite adopted in~\cite{pengAgile2019,pengExact2020,pengSolving2022} and give a continuous, monotonically non-decreasing transition-time curve under the standard parameter setting.

To examine the effect of agility on scheduling performance, EOS-Bench defines four manoeuvrability profiles by scaling the angular velocities $v_1,\dots,v_4$ within the ranges reported for agile satellites in~\cite{liuAdaptive2017,pengAgile2019,pengExact2020,pengSolving2022}.
The breakpoint angles and constants $a_k$ in Equation~\eqref{eq_1} are kept unchanged across all profiles, so that the profiles share the same basic functional form and minimum transition time, but differ in their effective slew rates.
The resulting velocity values are listed in Table~\ref{tab7}.

\begin{table}[htb]
\centering
\small
\caption{Manoeuvrability profiles and their angular-velocity settings.}
\begin{tabular}{p{0.25\textwidth}p{0.15\textwidth}p{0.15\textwidth}p{0.15\textwidth}p{0.15\textwidth}}
\toprule
\textbf{Profile} & \textbf{$v_1$ (°/s)} & \textbf{$v_2$ (°/s)} & \textbf{$v_3$ (°/s)} & \textbf{$v_4$ (°/s)} \\
\midrule
High-Agility     & 3.00 & 4.00 & 5.00 & 6.00 \\
Standard-Agility & 1.50 & 2.00 & 2.50 & 3.00 \\
Low-Agility      & 0.75 & 1.00 & 1.25 & 1.50 \\
Limited-Agility  & 0.50 & 0.67 & 0.83 & 1.00 \\
\bottomrule
\end{tabular}%
\label{tab7}%
\end{table}

For the agile Standard Scenarios, EOS-Bench uses the Standard-Agility profile. The non-agile Standard Scenarios use the corresponding non-agile platform model defined in Section~\ref{sec:framework}.
For the manoeuvrability-specific analysis, EOS-Bench again utilises the four representative baseline configurations established in the resource-capacity study (as detailed previously in Table~\ref{tab:capacity_baselines}).

For each baseline configuration, all four manoeuvrability profiles (High-Agility, Standard-Agility, Low-Agility, and Limited-Agility) are systematically evaluated.
This yields 36 distinct manoeuvrability-specific scenarios and 360 instances in total (4 baseline settings $\times$ 3 new manoeuvrability profiles $\times$ 3 target distributions, with 10 instances per scenario).
Because the constants $a_k$ are fixed and only the angular velocities $v_k$ are varied, the relative ordering of transition times across profiles is preserved: more agile profiles always produce shorter transition times for a given $\Delta g$, whereas less agile profiles produce longer ones.
This design provides a controlled setting in which to test the sensitivity of scheduling methods to satellite agility.

\subsubsection{Constellation-Configuration Variants}

This scenario family can be instantiated under either satellite platform type, since it changes orbital layout rather than the underlying observation model.
The constellation-configuration scenarios examine how different orbital layouts affect scheduling performance when the total number of satellites is held fixed.
All constellations follow Walker--Delta patterns and retain the seed orbit of ALOS-2, including altitude and inclination, so that the benchmark varies only the number of planes and the number of satellites per plane.

Five medium- and large-scale satellite counts are considered in this study: 50, 100, 200, 500, and 1000 satellites.
For each scale, three constellation architectures are defined: the Default pattern (identical to the baseline used in the Standard Scenarios), the Few-Planes pattern (fewer planes with more satellites per plane), and the Many-Planes pattern (more planes with fewer satellites per plane).
The corresponding Walker--Delta configurations and their assigned task loads are listed in Table~\ref{tab8}.
For each scale, the Default pattern coincides with the configuration already used in the Standard Scenarios, so the constellation-configuration study focuses on comparing it against the Few-Planes and Many-Planes alternatives.

\begin{table}[htb]
\centering
\small
\caption{Walker--Delta constellation configurations and task loads for constellation-configuration scenarios.}
\begin{tabular}{c c c c}
\toprule
\textbf{Satellite Count} & {\textbf{Task Count}} & \textbf{Configuration} & \textbf{Walker--Delta Pattern (planes $\times$ sats/plane)} \\
\midrule
\multirow{3}{*}{50} & \multirow{3}{*}{{100, 200}}
& Default & 10 planes $\times$ 5 sats \\
& & Few-Planes & 5 planes $\times$ 10 sats \\
& & Many-Planes & 25 planes $\times$ 2 sats \\
\midrule
\multirow{3}{*}{100} & \multirow{3}{*}{{100, 300}}
& Default & 10 planes $\times$ 10 sats \\
& & Few-Planes & 4 planes $\times$ 25 sats \\
& & Many-Planes & 20 planes $\times$ 5 sats \\
\midrule
\multirow{3}{*}{200} & \multirow{3}{*}{{100, 500}}
& Default & 20 planes $\times$ 10 sats \\
& & Few-Planes & 10 planes $\times$ 20 sats \\
& & Many-Planes & 40 planes $\times$ 5 sats \\
\midrule
\multirow{3}{*}{500} & \multirow{3}{*}{{100, 1000}}
& Default & 25 planes $\times$ 20 sats \\
& & Few-Planes & 10 planes $\times$ 50 sats \\
& & Many-Planes & 50 planes $\times$ 10 sats \\
\midrule
\multirow{3}{*}{1000} & \multirow{3}{*}{{100, 1000}}
& Default & 50 planes $\times$ 20 sats \\
& & Few-Planes & 10 planes $\times$ 100 sats \\
& & Many-Planes & 100 planes $\times$ 10 sats \\
\bottomrule
\end{tabular}%
\label{tab8}%
\end{table}

For the constellation-configuration study, the benchmark uses the five satellite scales above and selects two representative task loads (as specified in Table~\ref{tab8}) for each scale to reflect medium- and high-workload regimes, while fixing the planning horizon at 1 day.
For each scale and task load, the three Walker--Delta configurations in Table~\ref{tab8} are evaluated across all three target distributions.
This yields 120 constellation-configuration scenarios in total (5 satellite scales $\times$ 2 task-load levels $\times$ 2 new constellation configurations $\times$ 3 target distributions $\times$ 2 platform types).
For each scenario, 10 independent instances are generated by sampling from the task pools described in Section~\ref{sec:framework}, resulting in 1,200 instances overall.
To ensure fair comparison, the same task set is used across the three constellation configurations under the same satellite scale and task-load setting.

\subsubsection{Realistic-Target Scenarios}

EOS-Bench also includes scenarios based on real geographic targets.
As described in Section~\ref{sec:framework}, the benchmark constructs a set of 1{,}000 point targets from the GeoNames database\footnote{\url{https://download.geonames.org/export/}} by selecting national capitals and first-level administrative capitals worldwide.
These 1{,}000 targets are divided into 10 disjoint subsets, each containing 100 targets.

In the realistic-target scenarios, these targets are paired with the same real-satellite orbits used in the small-constellation settings.
These scenarios can be instantiated under either the agile or the non-agile platform model within the benchmark library.
Ten realistic-target scenarios are defined, corresponding to constellations of 1, 3, 5, 10, and 20 satellites under both agile and non-agile platform models, each with a 1-day horizon and 100 real targets.

For each scenario, the 10 target subsets generate 10 instances.
These instances share the same satellite set and planning horizon, and differ only in the selected subset of real targets.
Accordingly, the realistic-target family contains 10 scenarios and 100 instances in total.
These scenarios complement the synthetic target sets used elsewhere in the benchmark and allow algorithms to be evaluated under real satellite ground tracks and more realistic geographic demand patterns.

\section{Scenario Characterisation} \label{sec:scenario_characterisation}

As one of the major contributions of this work, EOS-Bench characterises the intrinsic structural difficulty of benchmark scenarios.
This is necessary because nominal scale descriptors, such as the number of tasks, the number of satellites, or the planning-horizon length, do not by themselves provide a reliable indication of scheduling difficulty.
Scenarios of similar nominal size may still differ substantially in effective difficulty because of differences in feasible opportunity structure, task flexibility, pairwise competition, and congestion on satellite timelines.
In a benchmark setting, these structural differences can strongly influence solver behaviour even when the nominal scenario scale is comparable.

For this reason, EOS-Bench supplements solver-oriented evaluation with a set of scenario characterisation descriptors.
These descriptors are intended to describe the intrinsic structure of a scenario rather than solution quality.
Because a scenario in EOS-Bench is a structural configuration, whereas an instance is a concrete realisation generated under that configuration, each descriptor is first computed at the instance level and then averaged over all instances belonging to the same scenario.
Accordingly, the reported scenario-level descriptor should be interpreted as an empirical estimate of the structural characteristics induced by that scenario template under the adopted generation process.
As the number of instances increases, this estimate becomes more stable and approaches the corresponding scenario-level expectation. Note that in the proposed EOS-Bench framework, each scenario comprises 10 independent instances.

The adopted descriptors describe scenario structure from two complementary perspectives:
a task-oriented perspective, which focuses on feasible opportunities, flexibility, and pairwise competition, and a satellite-oriented perspective, which focuses on congestion and overload on satellite timelines.
The full set of adopted descriptors is summarised in Table~\ref{tab:scenario_characterisation_metrics}.
Related ideas have appeared in EOSSP studies that analyse visible-window conflicts and feasible-interval interactions to understand instance difficulty and reduce problem complexity. 
For instance, complex network theory has been applied to model constraint graphs in multi-satellite settings, particularly to resolve intense conflicts in scenarios with oversubscribed targets~\cite{wang2016scheduling,wang2019scheduling,chen2020eosis}.
Such considerations are also consistent with broader discussions of EOS scheduling complexity in recent surveys~\cite{wang2020agile,ferrari2025satellite}.

\begin{table}[htb]
\centering
\footnotesize
\caption{Scenario characterisation descriptors used in EOS-Bench.}
\label{tab:scenario_characterisation_metrics}
\begin{tabular}{@{}p{0.2\textwidth} p{0.3\textwidth} c p{0.45\textwidth}@{}}
\toprule
\textbf{Perspective} & \textbf{Descriptor} & \textbf{Symbol} & \textbf{Description} \\
\midrule
\multirow{5}{*}{\textbf{Task-oriented}}
& Average Available Opportunities & $\Gamma_{\mathrm{ao}}$ & Average number of feasible execution opportunities per task. \\
& Opportunity Constrained Task Ratio & $\Gamma_{\mathrm{oc}}$ & Proportion of highly constrained tasks with at most two feasible windows. \\
& Task Interference Ratio & $\Gamma_{\mathrm{ti}}$ & Proportion of task pairs that possess at least one mutually conflicting opportunity. \\
& Average Task Pair Conflict Ratio & $\Gamma_{\mathrm{at}}$ & Average ratio of conflicting available window pairs to comparable available-window pairs among competing tasks. \\
& Task Elasticity Ratio & $\Gamma_{\mathrm{te}}$ & Ratio of the total number of feasible satellite window combinations to the total number of tasks. \\
\midrule
\multirow{5}{*}{\textbf{Satellite-oriented}}
& Observation Contention Ratio & $\Lambda_{\mathrm{oc}}$ & Degree of continuous-time overlap among tasks competing for satellite capacity. \\
& Conflict Satellite Ratio & $\Lambda_{\mathrm{cs}}$ & Proportion of satellites experiencing at least one scheduling conflict. \\
& Timeline Overload Ratio & $\Lambda_{\mathrm{to}}$ & Fraction of the aggregate constellation timeline occupied by conflicting episodes. \\
& Average Conflict Tasks & $\Lambda_{\mathrm{ac}}$ & Average number of tasks simultaneously competing during a local conflict. \\
& Excess Demand Ratio & $\Lambda_{\mathrm{ed}}$ & Ratio of the actual duration-weighted excess demand to the theoretical maximum excess demand. \\
\bottomrule
\end{tabular}
\end{table}

Let a scenario contain $M$ instances. 
For any instance-level descriptor $\phi^{(m)}$ computed on instance $m$, the corresponding scenario-level value is reported as
\begin{equation}
\bar{\phi}
=
\frac{1}{M}\sum_{m=1}^{M}\phi^{(m)}.
\end{equation}
In the following, all formulas are defined at the instance level; scenario-level values are then obtained by averaging these values over the $M$ instances generated under the same scenario.

\subsection{Task-Oriented Descriptors}

Task-oriented descriptors describe instance difficulty from the perspective of individual observation requests and their feasible execution opportunities.

Let $N_T$ denote the number of tasks in an instance.
For each task $i$, let $a_i$ denote the number of feasible available opportunities generated after applying the available-window construction rules of EOS-Bench.
Here, an \emph{available window} is defined as one executable assignment option for a task, i.e., one concrete feasible placement on a satellite timeline together with the associated execution information required by the benchmark model.
EOS-Bench first reports the average available opportunities per task:
\begin{equation}
\Gamma_{\mathrm{ao}}
=
\frac{1}{N_T}\sum_{i=1}^{N_T} a_i.
\end{equation}
This quantity measures the average size of the executable assignment space.
Smaller values indicate a sparser feasible opportunity structure and therefore a less flexible instance.

To also quantify the prevalence of highly constrained tasks, EOS-Bench defines the opportunity constrained task ratio at threshold $k\leq 2$ as
\begin{equation}
\Gamma_{\mathrm{oc}}
=
\frac{1}{N_T}\sum_{i=1}^{N_T}\mathbf{1}(a_i\leq 2),
\end{equation}
to measure the proportion of tasks with at most two feasible available opportunities.

To characterise pairwise competition, consider a task pair $(i,j)$.
Let $C_{ij}$ denote the number of comparable available-window pairs, defined as those available-window pairs that share at least one exclusive execution timeline and may therefore compete directly.
In the current benchmark, this is primarily the satellite observation timeline.
Let $F_{ij}$ denote the number of conflicting available-window pairs among them.
An available-window pair is regarded as conflicting if the two available opportunities overlap on the shared timeline, or, on the same satellite timeline, if the separation is shorter than the required transition time.

Based on these quantities, EOS-Bench defines the task interference ratio as
\begin{equation}
\Gamma_{\mathrm{ti}}
=
\frac{\sum_{i<j}\mathbf{1}(F_{ij}>0)}{\binom{N_T}{2}},
\end{equation}
which is the proportion of all task pairs for which at least one conflicting available-window pair exists.

For the subset of comparable task pairs,
\begin{equation}
\mathcal{P}_{\mathrm{comp}}
=
\{(i,j): i<j,\; C_{ij}>0\},
\end{equation}
the average task pair conflict ratio is defined as
\begin{equation}
\Gamma_{\mathrm{at}}
=
\frac{1}{|\mathcal{P}_{\mathrm{comp}}|}
\sum_{(i,j)\in\mathcal{P}_{\mathrm{comp}}}
\frac{F_{ij}}{C_{ij}},
\end{equation}
with the convention that it is set to zero when $|\mathcal{P}_{\mathrm{comp}}|=0$.
Unlike $\Gamma_{\mathrm{ti}}$, which only records whether a task pair conflicts at all, $\Gamma_{\mathrm{at}}$ measures the average severity of interference among comparable task pairs.

Task flexibility is described using both platform choice and feasible visible-window availability.
For each task $i$, let $s_i$ denote the number of optional satellites that can serve it, and let $w_i$ denote the number of feasible visible windows associated with that task.
EOS-Bench defines the task elasticity ratio as
\begin{equation}
\Gamma_{\mathrm{te}}
=
\frac{1}{N_T}\sum_{i=1}^{N_T} s_i w_i,
\end{equation}
where larger values indicate greater structural flexibility.

\subsection{Satellite-Oriented Descriptors}

Satellite-oriented descriptors describe instance difficulty from the perspective of congestion on satellite timelines.

The first descriptor is the observation contention ratio.
For each satellite $s$, let $n_s(t)$ denote the number of simultaneously active available opportunities on its timeline at time $t$.
EOS-Bench defines
\begin{equation}
\Lambda_{\mathrm{oc}}
=
\frac{
\sum_{s}\int \binom{n_s(t)}{2}\,dt
}{
\sum_{s}\int \max\{n_s(t),1\}\,dt
}.
\end{equation}
The numerator accumulates continuous-time pairwise overlap on satellite timelines, while the denominator normalises by the corresponding timeline exposure.
A larger value, therefore, indicates denser competition for observation capacity.

To characterise congestion more directly, EOS-Bench projects feasible opportunities onto a fixed analysis grid on each satellite timeline.
Let $\tau$ index the analysis steps.
A satellite--step pair $(s,\tau)$ is marked as conflicting if at least two tasks simultaneously compete for the same satellite timeline, either because their available opportunities overlap or because the available separation between consecutive observations is shorter than the required transition time.
Let $q_{s,\tau}$ denote the number of conflicting tasks at step $(s,\tau)$.

Using this definition, EOS-Bench reports the conflict satellite ratio
\begin{equation}
\Lambda_{\mathrm{cs}}
=
\frac{1}{|\mathcal{S}|}
\sum_{s\in\mathcal{S}}
\mathbf{1}\!\left(
\sum_{\tau}\mathbf{1}(q_{s,\tau}\ge 2)>0
\right),
\end{equation}
to measure the proportion of satellites whose timelines contain at least one conflict step.

Because adjacent conflict steps may belong to the same congested episode, EOS-Bench merges neighbouring conflict steps into conflict segments.
Let $\mathcal{L}$ denote the set of merged conflict segments, let $d_{\ell}$ be the duration of segment $\ell\in\mathcal{L}$, and let $T_{\mathrm{hor}}$ denote the planning-horizon length.
The timeline overload ratio is then defined as
\begin{equation}
\Lambda_{\mathrm{to}}
=
\frac{\sum_{\ell\in\mathcal{L}} d_{\ell}}{|\mathcal{S}|\,T_{\mathrm{hor}}},
\end{equation}
to measure the fraction of the aggregate satellite timeline covered by conflict segments.

EOS-Bench also reports the average conflict tasks:
\begin{equation}
\Lambda_{\mathrm{ac}}
=
\frac{
\sum_{s\in\mathcal{S}}\sum_{\tau}
q_{s,\tau}\,\mathbf{1}(q_{s,\tau}\ge 2)
}{
\sum_{s\in\mathcal{S}}\sum_{\tau}\mathbf{1}(q_{s,\tau}\ge 2)
},
\end{equation}
with the convention that $\Lambda_{\mathrm{ac}}=0$ when no conflict step is present.
This descriptor reflects the typical local overload depth when a conflict occurs.

Finally, let $q_{\ell}$ denote the number of simultaneously conflicting tasks associated with segment $\ell$, and let
\begin{equation}
q_{\max}=\max_{\ell\in\mathcal{L}} q_{\ell}
\end{equation}
denote the peak number of conflicting tasks across all conflict segments.
EOS-Bench then defines the excess demand ratio as
\begin{equation}
\Lambda_{\mathrm{ed}}
=
\frac{
\sum_{\ell\in\mathcal{L}} \max(q_{\ell}-1,0)\,d_{\ell}
}{
(q_{\max}-1)\sum_{\ell\in\mathcal{L}} d_{\ell}
},
\end{equation}
with the convention that $\Lambda_{\mathrm{ed}}=0$ when no conflict segment is present.
This descriptor measures the average Excess demand depth during conflict periods, normalised by the peak conflict depth of the instance.
It therefore distinguishes scenarios in which conflicts are typically close to their worst observed severity from those in which peak overload occurs only occasionally.


Overall, these task- and satellite-oriented descriptors offer a concise representation of scenario/instance difficulty prior to solver comparison. They enable a clear distinction between scenarios that are large yet flexible and those that, despite a smaller nominal size, are inherently opportunity-sparse, conflict-dense, or temporally congested. To help better understand the defined descriptors, a toy example illustrating the calculation of all descriptors is provided in~\ref{app:scenario_characterisation_example}. The defined descriptors are also employed to characterise the experimental scenarios in Section~\ref{sec:experiments}.



\section{Evaluation Protocol and Metrics}\label{sec:evaluation}

To compare Earth observation satellite scheduling methods consistently and transparently, EOS-Bench adopts a unified evaluation protocol. 
This protocol is completely platform-agnostic, applying equally to both agile and non-agile benchmark instances, and accommodates exact optimisation methods, heuristics, meta-heuristics, and learning-based schedulers.

Because EOS scheduling is inherently multi-objective, different application settings may prioritise different aspects of solution quality, such as total mission value, task completion, workload distribution, response timeliness, or computational cost. 
As highlighted in Section~\ref{sec:literature}, relying on a single aggregated score or selectively reporting metrics heavily obscures the practical trade-offs between fundamentally different solvers. 
For this reason, and in keeping with the purpose of a general benchmark, EOS-Bench reports five complementary scalar metrics rather than a single combined score. 
This multi-dimensional design preserves the structure of algorithmic performance, makes trade-offs more visible, and improves the interpretability of cross-method comparisons.

Accordingly, EOS-Bench systematically assesses scheduling performance across the following five dimensions: Task Profit (TP), which measures the total value of completed tasks; Task Completion Rate (TCR), which measures the fraction of tasks completed; Balance Degree (BD), which measures how evenly the workload is distributed across satellites; Timeliness (TM), which measures how early tasks are executed within the planning horizon; and Runtime (RT), which measures the computational time required to produce a schedule.

Taken together, these five metrics provide a framework that is both benchmark-oriented and interpretable. By keeping solution value, completion behaviour, balance, timeliness, and runtime as separate but complementary dimensions, the protocol supports reproducible comparison across solver families while allowing users to interpret performance in accordance with their own operational priorities.

\subsection{Task Profit (TP)}

Task Profit is one of the principal effectiveness measures in the protocol.
It sums the values of the tasks actually executed by a schedule, which is consistent with objective formulations widely used in EOS scheduling studies~\cite{liuAdaptive2017,pengAgile2019,pengExact2020,qiDecomposeandlearn2025}.
It is defined as
\begin{equation}
\label{eq_2}
TP = \sum_i p_i x_i
\end{equation}
where $p_i$ is the profit associated with task $i$, and $x_i$ equals 1 if task $i$ is executed and 0 otherwise.

Maximising $TP$ encourages a scheduling method to allocate limited observation opportunities to high-value tasks rather than simply to maximise the number of accepted tasks.
This is consistent with practical mission planning, where some requests are substantially more important than others, for example, emergency monitoring as opposed to routine mapping.
It also matches the weighted-profit formulations commonly used in exact and meta-heuristic EOS models~\cite{pengExact2020,wuEnsemble2022}.

\subsection{Task Completion Rate (TCR)}

Task Completion Rate measures the fraction of tasks completed, irrespective of their individual profit values.
It is defined as
\begin{equation}
\label{eq_3}
TCR = \frac{1}{N_T} \sum_i x_i
\end{equation}
where $N_T$ is the total number of tasks and $x_i$ is 1 if task $i$ is completed and 0 otherwise.

Whereas $TP$ captures the total value of completed tasks, $TCR$ captures the extent to which the schedule delivers service coverage and makes use of the available observation capacity.
Both measures are reported because different methods may exhibit very different behaviours.
One method may select only a small number of high-profit tasks, thereby achieving a high $TP$ but a relatively low $TCR$.
Another may favour broader coverage, thereby achieving a high $TCR$ at the expense of total value.
Both perspectives are common in EOS and constellation-planning studies~\cite{barkaouiNew2020,liMultiobjective2024,liuLargescale2023}, and together they help users identify methods that best match their operational objectives.

\subsection{Balance Degree (BD)}

In multi-satellite settings, a good schedule should not only achieve high value, but should also avoid concentrating most observations on a small subset of satellites.
This is particularly relevant when the platforms are broadly similar and balanced usage is desirable for operational or fairness-related reasons~\cite{renCompetitive2022,liRemotesensing2024}.
EOS-Bench measures workload balance using the Balance Degree:
\begin{equation}
\label{eq_4}
BD = 1 - \frac{\sigma_s}{\mu_s}
\end{equation}
where $\sigma_s$ is the standard deviation of the number of executed tasks per satellite and $\mu_s$ is the corresponding mean.

If all active satellites execute the same number of tasks, then $\sigma_s=0$ and $BD=1$, indicating perfect balance.
As the workload becomes more uneven, $\sigma_s$ increases and $BD$ decreases.
In highly uneven cases, $BD$ may become negative; the metric should therefore be interpreted comparatively, with larger values indicating more balanced schedules.
In our experiments, $BD$ is computed over satellites that execute at least one task.
If no task is executed, then $BD$ is not meaningful and is reported as undefined.

This metric is equivalent to one minus the coefficient of variation, and therefore provides a compact summary of relative workload dispersion.
It is particularly useful in studies concerned with fairness across satellites, users, or service regions~\cite{renCompetitive2022}.

\subsection{Timeliness (TM)}

In many EOS missions, especially emergency and time-critical applications, the execution time of a task matters almost as much as whether the task is completed at all~\cite{sunAgile2021,liDynamic2025,liuLearningbased2024}.
EOS-Bench captures this aspect using the Timeliness score:
\begin{equation}
\label{eq_5}
TM = \frac{\sum_i (t_i - t_s) x_i + \sum_i T_{\mathrm{hor}} (1 - x_i)}{T_{\mathrm{hor}} \times N_T}
\end{equation}
Here, $t_i$ is the start time of task $i$, $t_s$ is the scenario start time, $T_{\mathrm{hor}}$ is the planning-horizon length, and $N_T$ is the total number of tasks.

If a task is completed ($x_i=1$), it contributes $(t_i-t_s)/T_{\mathrm{hor}}$, that is, its normalised execution time within the horizon.
Earlier execution, therefore, leads to a smaller contribution.
If a task is not completed ($x_i=0$), a penalty of 1 is assigned, which is equivalent to treating the task as if it were executed at the end of the horizon.
It follows that $TM \in [0,1]$, with smaller values indicating earlier execution on average and/or fewer missed tasks.

This metric is consistent with latency- and waiting-time-style measures used in dynamic and rolling-horizon EOS scheduling~\cite{liDynamic2025,wenScheduling2023}.
Its main advantage is that it penalises both delayed execution and non-execution within a single scalar quantity.
It is therefore particularly useful in scenarios with urgent requests or rapidly decaying task value, where a method may perform well on $TP$ but still be operationally unsatisfactory if it systematically postpones important tasks.

\subsection{Runtime (RT)}

Many EOS scheduling studies report computational runtime for exact, meta-heuristic, and learning-based methods~\cite{wuEnsemble2022,herrmannComparative2023,stollenwerkAgile2021}.
Runtime is an important indicator of practical usability, especially in operational settings where schedules must be produced or updated within tight time limits.
For learning-based methods, it is often useful to distinguish between offline training time and per-instance inference time.
In the experiments reported in this paper, however, $RT$ refers to the per-instance wall-clock time required to generate a schedule, that is, the online solution time.
For classical optimisation and heuristic methods, this corresponds directly to solver runtime; for trained learning-based methods, it corresponds to inference time per benchmark instance.
This definition facilitates comparison across solver families in a common operational setting.
Including $RT$ explicitly is particularly important in benchmarking, because methods with similar solution quality may differ substantially in their computational cost and hence in their practical deployability.

\section{Algorithm Experimentation and Analysis}\label{sec:experiments}
The purpose of this section is not merely to compare a small set of baseline solvers, but to validate that EOS-Bench produces informative distinctions across solver classes, scenario families, and performance metrics. Accordingly, the experimental results should be interpreted primarily as evidence that the benchmark is discriminative, scalable, and structurally interpretable, rather than as a definitive ranking of all available EOS scheduling algorithms.
To this end, Section~\ref{subsec:algorithms} introduces the selected baseline algorithms, and Section~\ref{subsec:experimental_scenarios} characterises the structural difficulty of the experimental scenarios. Following this, Section~\ref{subsec:experimental_results} presents the comparative benchmarking results across standard, specific, and realistic settings. Finally, Section~\ref{subsec:visualisation} showcases the framework's interactive visualisation capabilities.

\subsection{Solution Algorithms}\label{subsec:algorithms}

EOS-Bench is designed to be algorithm-agnostic. To support a wide variety of solution approaches, all solvers interact with the framework through a standardised interface that explicitly defines the required inputs and expected outputs. For any given benchmark instance, the input provided to the scheduler includes the planning horizon, the satellite set (along with their platform and resource parameters), the task set (with corresponding attributes), and the precomputed access windows generated by the visibility module. In return, the algorithm must output a concrete scheduling plan. This plan specifies exactly which tasks are selected, the satellite assigned to execute each task, the specific access window utilised, and the exact execution time or observation interval. 

Once generated, the returned plan is assessed using the multi-dimensional evaluation protocol defined in Section~\ref{sec:evaluation}. By decoupling the solution methods from the scenario generation and evaluation modules, this standardised design makes it straightforward to incorporate new algorithms and ensures that all methods are compared fairly under identical conditions. 
{Note that while the framework interface seamlessly handles both agile and non-agile platforms and all algorithms were tested on both, the detailed analysis in the remainder of this paper focuses exclusively on a selected agile subset due to space constraints.}

Table~\ref{tab:algorithms} summarises the solver classes and baseline algorithms considered in this study. The experiments comprise four groups of solution algorithms: exact optimisation (MIP variants), constructive heuristics (greedy rules), meta-heuristics including Genetic Algorithm (GA), Simulated Annealing (SA) and Ant Colony Optimisation (ACO), and a learning-based algorithm. Specifically, the exact optimisation group includes three MIP variants with different objective functions.
The constructive heuristic group includes four greedy baselines.
The meta-heuristic group includes three methods, each instantiated with three objective-specific variants.
Finally, the learning-based group contains one proximal policy optimisation (PPO)-based DRL algorithm. Details of these 17 algorithms are provided below in this subsection. 

These solution algorithms provide a representative cross-section of solver paradigms commonly used in EOSSP research. Although only these specific algorithms are evaluated here, the proposed framework's modular interface ensures that additional algorithms can be seamlessly incorporated in future work.

\begin{table}[htb]
\centering
\footnotesize
\caption{17 specific algorithms evaluated in EOS-Bench.}
\label{tab:algorithms}
\setlength{\tabcolsep}{5pt}
\begin{tabular}{p{0.25\textwidth}p{0.15\textwidth}p{0.5\textwidth}}
\toprule
\textbf{Algorithm class} & \textbf{Method} & \textbf{Specific algorithms}  \\
\midrule
Exact optimisation & MIP & \textit{MIP-TP}, \textit{MIP-TCR}, \textit{MIP-ALL}  \\
Constructive heuristics & Greedy rules & \textit{Greedy-TP}, \textit{Greedy-TCR}, \textit{Greedy-TM}, \textit{Greedy-BD}  \\
Meta-heuristics & GA, SA, ACO & \textit{GA-TP}, \textit{GA-TCR}, \textit{GA-ALL}; \textit{SA-TP}, \textit{SA-TCR}, \textit{SA-ALL}; \textit{ACO-TP}, \textit{ACO-TCR}, \textit{ACO-ALL} \\
Learning-based baseline & PPO-based DRL & \textit{RL-TP}  \\
\bottomrule
\end{tabular}
\end{table}


\subsubsection{Class~1: Exact MIP}

For small benchmark scenarios and for selected moderate cases involving up to 20 satellites, we implement exact MIP models as reference baselines.
These formulations follow the standard discrete-window MIP framework used in EOSSP research~\cite{CHEN2019Amixed,pengExact2020,rigoBranchandprice2022,ferrari2025satellite}.
Binary decision variables indicate whether a satellite executes a given task within a specific access window.
The models enforce that each task is executed at most once and that each satellite performs at most one activity at any time.
Energy and onboard storage constraints are imposed over the planning horizon, while precomputed visibility windows define feasible assignments and transition times are enforced through compatibility and precedence constraints between activities.

To reflect the multi-metric evaluation protocol, three exact MIP variants are considered.
MIP-TP maximises total task profit; MIP-TCR maximises the number of completed tasks, and hence task completion rate; and MIP-ALL uses a composite objective that jointly considers TP, TCR, BD, and TM.
{In MIP-ALL, these four metrics are first normalised and then combined through a weighted-sum objective. Specifically, Total Profit (TP) is divided by the sum of all task profits, while the Completion Rate (CR) and Balance Degree (BD) are retained in their original forms. The Time Margin (TM) is transformed as $1 - TM$ to align the optimisation direction.
In our experiments, all four normalised metrics are assigned equal weights.}
The models are implemented in Python using the PuLP~\cite{mitchell2011pulp} optimisation modelling interface and solved with the CBC~\cite{kaliszewski2022idol} MIP solver.
When the CBC terminates with an optimality certificate, the corresponding MIP solution is globally optimal with respect to the stated objective.
For harder instances solved under a time limit, the reported results should instead be interpreted as best feasible solutions returned by the solver.

\subsubsection{Class~2: Constructive Heuristics}

We include four simple constructive greedy heuristics as fast baseline methods.
All four start from an empty plan and insert tasks sequentially, but they differ in the objective emphasised during construction.
Specifically,
Greedy-TP prioritises total profit,
Greedy-TCR prioritises task completion,
Greedy-TM prioritises early execution,
and Greedy-BD promotes workload balance across satellites.
All four follow a sequential construction pattern with lightweight priority rules, which is common in EOSSP literature~\cite{wang2025multipleobs}.

\paragraph{Profit-first greedy (Greedy-TP)}
Greedy-TP processes tasks in a random order and attempts to place each task in its best access window.
Candidate windows are scored primarily by task profit, with ties broken using simple time slack-based or early-start preferences.
A candidate is accepted only if it remains feasible with respect to the current partial plan and all resource constraints.
The construction is repeated under several random task orders, and the plan with the highest $TP$ is retained.

\paragraph{Completion-first greedy (Greedy-TCR)}
Greedy-TCR aims to schedule as many tasks as possible, even when some tasks carry relatively low profit.
It prioritises tasks with fewer feasible windows, or tighter temporal flexibility, and then assigns them as early as possible.
As with Greedy-TP, several randomised passes are performed and the plan with the highest $TCR$ is retained.

\paragraph{Timeliness-first greedy (Greedy-TM)}
Greedy-TM prioritises timeliness by favouring tasks that are urgent or whose feasible opportunities occur early in the horizon.
For each task, candidate windows are ranked using an urgency-aware score.
The heuristic then processes tasks in descending urgency, for example, according to shorter slack or fewer late opportunities, and assigns each task to its earliest feasible window that preserves the feasibility of the current partial plan.
Several randomised tie-breaking passes are carried out, and the plan with the best timeliness-oriented objective is retained.

\paragraph{Balance-first greedy (Greedy-BD)}
Greedy-BD seeks to balance workload and resource usage across satellites during construction.
At each insertion step, candidate windows are evaluated not only by task value but also by a balance penalty reflecting the current unevenness of assigned workload and remaining resource budgets across satellites.
The heuristic chooses the feasible assignment that produces the smallest increase in imbalance, thereby reducing the risk of early over-commitment of a small subset of platforms.
As with the other greedy algorithms, multiple randomised passes are performed and the plan with the best balance-aware score is retained.

\subsubsection{Class~3: Meta-heuristics}

The third group consists of stochastic search methods that improve scheduling plans when exact optimisation becomes too expensive at larger scales.
These methods can explore many alternative task--window assignments without changing the underlying benchmark model.
In EOS-Bench, we include three representative meta-heuristic families, i.e., GA, SA and ACO.
To remain consistent with the multi-metric evaluation protocol, each family is instantiated in three objective-specific variants: a profit-oriented variant (-TP), a completion-oriented variant (-TCR), and a composite variant (-ALL).
Accordingly, the benchmark includes GA-TP, GA-TCR, GA-ALL, SA-TP, SA-TCR, SA-ALL, ACO-TP, ACO-TCR, and ACO-ALL.
The -ALL variants optimise a normalised weighted combination of TP, TCR, BD, and TM, in the same spirit as MIP-ALL.
These methods are widely used in satellite scheduling research and provide useful baselines for larger instances where exact solvers may not terminate within practical time limits~\cite{ferrari2025satellite,wang2020agile}.

\paragraph{Genetic algorithm (GA)} Genetic and memetic designs have been widely used in EOS and constellation scheduling as effective population-based baselines~\cite{barkaouiNew2020,changIntegrated2021}.
The GA maintains a population of candidate scheduling plans and improves them over generations through selection, crossover, and mutation.
In our implementation, each encodes a feasible set of task assignments, including the chosen satellite and access window for each scheduled task.
The population is initialised using constructive plans from Class~2 together with random perturbations, so that the search begins from valid and diverse solutions.
At each generation, the GA selects high-performing individuals, applies crossover to mix assignment patterns, and applies mutation to modify a small number of task placements.
After each operator, a repair step enforces feasibility with respect to time conflicts, transition-time separations, and resource limits.

\paragraph{Simulated annealing (SA)}
SA is a trajectory-based method that improves a single current scheduling plan by local moves, while allowing occasional worsening moves to escape local optima~\cite{kirkpatrick1983optimization}. It has been repeatedly applied to satellite observation scheduling, including agile settings with clustering and time-dependent feasibility checks~\cite{wuAdaptiveSA2017,ferrari2025satellite}.
In EOS-Bench, the state is a complete scheduling plan.
Neighbourhood moves are chosen to match the EOS setting directly, such as moving a task to another feasible access window, swapping two tasks on the same satellite timeline, or removing and reinserting a small set of tasks.
Improving moves are accepted deterministically, whereas worsening moves are accepted with a probability controlled by the temperature.
The temperature is reduced according to a geometric cooling schedule, so that the search becomes increasingly selective over time.

\paragraph{Ant colony optimisation (ACO)}
ACO is a constructive population-based method in which multiple ants build scheduling plans step by step under a shared pheromone model~\cite{dorigo2002ant}. ACO-style ideas have been used in satellite scheduling, including multi-satellite observation settings and divide-and-conquer frameworks combining constructive search with improvement procedures~\cite{wuMultisatellite2012,wuTwophase2013,wuEnsemble2022}.
In our implementation, each ant repeatedly selects an unscheduled task and then chooses one feasible satellite--window option for that task.
The choice probability combines (i) pheromone values that record which assignments have appeared in high-quality plans and (ii) a simple heuristic score reflecting profit and local feasibility slack.
After all ants have completed their plans, pheromones associated with the best solutions are reinforced, and evaporation is applied to limit premature stagnation.

\subsubsection{Class~4: Learning-based (RL)}

In this paper, we include one DRL baseline based on PPO, denoted RL-TP~\cite{schulman2017ppo}.
PPO provides a simple and stable policy-gradient method for large discrete decision spaces, and it has been used in recent EOSSP studies to learn task-selection and sequencing policies from simulated instances~\cite{herrmann2023rltaes,huang2021revising}.

Within EOS-Bench, the RL-TP agent constructs a scheduling plan sequentially.
At each decision step, the agent observes the current planning time, the remaining task set, and the remaining satellite resource budgets.
It then selects one feasible action from a masked action set, where each action corresponds to assigning a task to a specific satellite access window, or skipping when no promising move exists.
The environment applies the same feasibility checks as the optimisation and heuristic solvers, so the agent cannot bypass attitude, energy, memory, or window-compatibility constraints.
The reward is defined in terms of task profit, and the learned policy is therefore trained specifically to maximise TP rather than the other evaluation metrics.
The policy is trained on a designated training split of benchmark scenarios, and its test-time performance is reported using the same evaluation protocol as for the other solver classes.

The benchmark currently contains 13,900 instances (comprising 7,130 agile and 6,770 non-agile instances) whose sizes span 1--1,000 satellites, 10--10,000 tasks, and horizons from 1 to 168 hours.
To provide sufficient instance diversity for policy learning, we construct an on-the-fly training generator by recombining satellite and task assets sampled from the benchmark library, as shown in Algorithm~\ref{alg:rl_train_scenario_gen}.
{Because policy learning is conducted separately for agile and non-agile platforms, the generation process operates exclusively within the corresponding platform-specific subset. 
For each training episode, the generation proceeds in four explicit steps. 
First, we randomly sample a target episode scale ($N_S, N_T, T_{\mathrm{hor}}$) within the defined benchmark ranges. 
Second, we filter the platform-specific sub-library to form a candidate pool, retaining only those instances whose satellite count, task count, and planning horizon all meet or exceed the generated $N_S$, $N_T$, and $T_{\mathrm{hor}}$, respectively. 
Third, we randomly select one valid instance from this candidate pool to serve as the ``parent''. 
Fourth, we construct the new instance by uniformly sampling exactly $N_S$ satellites and $N_T$ tasks from this chosen parent without replacement. 
To ensure compatibility with the sampled horizon length $T_{\mathrm{hor}}$, the time windows of the selected tasks are stored in a normalised form and re-mapped to the new horizon via a random temporal shift, preserving their relative release--deadline structures. 
Finally, the resulting episode is rigorously screened: it is accepted only if it passes a basic feasibility check (ensuring a minimum fraction of valid access windows) and is strictly non-identical to any existing instance in the library; otherwise, the episode is resampled.} 
In our empirical implementation, the maximum resampling limit $K$ in Algorithm~\ref{alg:rl_train_scenario_gen} is set to $2,000$.

\begin{algorithm}[htbp]
\caption{Training instance generation via sub-sampling}
\label{alg:rl_train_scenario_gen}
\DontPrintSemicolon
\KwIn{
Instance library $\mathcal{D}$;
ranges $N_S\!\in[1,1000]$, $N_T\!\in[10,10000]$, $T_{\mathrm{hor}}\!\in[1,168]$ hours;
maximum resampling attempts $K$;
feasibility thresholds $\Theta$ (e.g., min.\ feasible-task ratio $\alpha$).
}
\KwOut{A training instance $\mathcal{I}$ for one RL episode.}

\BlankLine
\For{$k \leftarrow 1$ \KwTo $K$}{
    Sample horizon $T_{\mathrm{hor}} \sim \mathcal{U}(1,168)$ hours\;
    Sample satellite count $N_S \sim \mathcal{U}\{1,1000\}$\;
    Sample task count $N_T \sim \mathcal{U}\{10,10000\}$\;

    \tcp{(1) Filter library and select a sufficiently large parent instance}
    $\mathcal{D}_{\text{valid}} \leftarrow \{ I=(P_S, P_T, \dots) \in \mathcal{D} \mid |P_S| \ge N_S \text{ and } |P_T| \ge N_T \}$\;
    \If{$\mathcal{D}_{\text{valid}} = \emptyset$}{
        \textbf{continue}\;
    }
    Sample a parent instance $I_{\text{parent}}=(P_S, P_T, \dots) \sim \mathcal{U}(\mathcal{D}_{\text{valid}})$\;

    \tcp{(2) Sub-sample satellites from the parent instance}
    $\mathcal{S}_S \leftarrow$ \texttt{SubSample}$(P_S, N_S)$\;
    \tcp{Optional: randomise initial phase/time origin while keeping physical parameters}
    $\mathcal{S}_S \leftarrow$ \texttt{RandomiseInitialStates}$(\mathcal{S}_S)$\;

    \tcp{(3) Sub-sample tasks and re-map their temporal windows to the sampled horizon}
    $\mathcal{S}_T^{\text{raw}} \leftarrow$ \texttt{SubSample}$(P_T, N_T)$\;
    $\mathcal{S}_T \leftarrow \emptyset$\;
    Sample a temporal shift $\Delta \sim \mathcal{U}(0,T_{\mathrm{hor}})$\;
    \For{$t \in \mathcal{S}_T^{\text{raw}}$}{
        \tcp{Each sampled task stores a normalised window $(\hat r(t),\hat d(t))$}
        $r \leftarrow \texttt{clip}(\hat r(t)\cdot T_{\mathrm{hor}} + \Delta,\;0,\;T_{\mathrm{hor}})$\;
        $d \leftarrow \texttt{clip}(\hat d(t)\cdot T_{\mathrm{hor}} + \Delta,\;0,\;T_{\mathrm{hor}})$\;
        \If{$d \ge r + \texttt{dur}(t)$}{
            Add task $t'=(\texttt{loc}(t), \texttt{dur}(t), r, d, \texttt{attr}(t))$ to $\mathcal{S}_T$\;
        }
    }
    \If{$|\mathcal{S}_T| < N_T$}{
        \tcp{Not enough valid tasks after re-mapping; resample}
        \textbf{continue}\;
    }

    \tcp{(4) Build access windows and screen the scenario for basic feasibility}
    $\mathcal{W} \leftarrow$ \texttt{BuildAccessWindows}$(\mathcal{S}_S,\mathcal{S}_T,T_{\mathrm{hor}})$\;
    $q \leftarrow$ \texttt{EvaluateFeasibility}$(\mathcal{S}_S,\mathcal{S}_T,\mathcal{W})$\;
    \If{$q \ge \alpha$}{
        \Return{$\mathcal{I}=(\mathcal{S}_S,\mathcal{S}_T,\mathcal{W},T_{\mathrm{hor}})$}\;
    }
}
\Return{\texttt{FallbackScenario}$(\mathcal{D})$}\;
\end{algorithm}

    \subsection{Experimental Scenarios and Characterisation}\label{subsec:experimental_scenarios}
    The benchmark defines a large number of scenarios, making it unnecessary and impractical to present results on every scenario in this work.
    In this subsection, we therefore employ a selected subset of the scenarios, that preserves coverage of the main modelling dimensions while keeping the total computational cost manageable.
    The selection follows three principles.
    First, we cover multiple planning horizons and constellation scales.
    Second, we include the target distributions used in the Standard Scenarios.
    Third, we include a limited number of scenarios from the Specific Scenarios to isolate the effects of key parameters.
    All experiments use the same scenario description format, characterisation descriptors, evaluation metrics, and visualisation pipeline.

 Specifically, the selected subset consists of 96 Standard Scenarios (960 instances) and 154 Specific Scenarios (1,540 instances), yielding a total of 250 scenarios and 2,500 independent instances. 
   All numerical experiments were carried out on a server with an Intel Xeon Gold 5218R processor (40 logical cores, 2.10~GHz), 256~GB of memory, and CentOS Linux~7.


    \subsubsection{Standard Scenarios}
    The core experimental set is drawn from the Standard Scenarios and serves as the main comparison testbed for all solver classes.
    Scenarios are selected to jointly cover variations in planning horizon, constellation scale, and mission load.
    We consider three planning horizons ($0.5$, $1$, and $3$~days) and four constellation sizes ($3$, $10$, $100$, and $500$ satellites).
    For each constellation size, two task loads are selected from Table~\ref{tab5} to represent moderate and high workload regimes.
    Specifically, we consider $100$ and $200$ tasks for $3$ satellites, $200$ and $500$ tasks for $10$ satellites,
    $500$ and $2000$ tasks for $100$ satellites,
    and $2000$ and $5000$ tasks for $500$ satellites.
    All selected configurations are evaluated under the two synthetic target distributions defined in the benchmark, namely Global-Random and Region-Clustered.

    Accounting for both satellite platforms, the core experimental set contains 96 Standard Scenarios in total (48 agile and 48 non-agile).
    For each scenario, 10 independent instances are generated by sampling different target subsets, resulting in 960 instances overall.

    \subsubsection{Specific Scenarios}
    To analyse algorithm sensitivity beyond the nominal settings of the Standard Scenarios, we include four focused scenario subsets, in total 154 scenarios, drawn from the Specific Scenarios.
    Each subset modifies one modelling factor at a time while keeping the remaining parameters unchanged.
    These scenarios are evaluated with the heuristic, meta-heuristic, and learning-based solvers.

    (i) Capacity set:
    We select four baseline configurations with $3$ satellites with $200$ tasks, $10$ satellites and $500$ tasks, $100$ satellites and $500$ tasks, and $500$ satellites and $2000$ tasks, all under a $1$-day planning horizon, and apply the five non-standard capacity patterns defined in Table~\ref{tab6}, evaluating them across two target distributions under both agile and non-agile platform models.
    This combinatorial design yields $80$ distinct capacity-specific scenarios (4 baselines $\times$ 5 patterns $\times$ 2 distributions $\times$ 2 platform types).
    With $10$ instances generated per scenario, this set contains $800$ instances in total.

    (ii) Agility set:
    Using the same satellite--task configurations as in the capacity set, we apply the three non-standard manoeuvrability profiles (High-, Low-, and Limited-Agility) defined in Table~\ref{tab7}, under a $1$-day planning horizon.
    This combinatorial design yields $24$ distinct agility-specific scenarios (4 baselines $\times$ 3 profiles $\times$ 2 distributions).
    With $10$ instances generated per scenario, the agility set contains $240$ instances in total.

    (iii) Orbit-architecture set:
    We consider five constellation scales, each paired with two specific task loads: $50$ satellites ($100$ and $200$ tasks), $100$ satellites ($100$ and $300$ tasks), $200$ satellites ($100$ and $500$ tasks), $500$ satellites ($100$ and $1000$ tasks), and $1000$ satellites ($100$ and $1000$ tasks). 
    For each of these $10$ base combinations, we test the two non-default Walker--Delta configurations (Few-Planes and Many-Planes) defined in Table~\ref{tab8} under a fixed $1$-day planning horizon. 
    This analysis is conducted for both agile and non-agile platform types using a single baseline target distribution. 
    Consequently, this yields $40$ distinct orbital architecture scenarios (10 combinations $\times$ 2 configurations $\times$ 2 platform types).
    With $10$ instances generated per scenario, this set contains $400$ instances in total.

    (iv) Realistic target set:
    In addition to synthetic targets, we include a realistic target set based on city-level locations and real satellite ground tracks.
    Each scenario is defined by a fixed constellation size, planning horizon, and task count, combined with real geographic targets.
    We consider five satellite scales with $1$, $3$, $5$, $10$, and $20$ satellites, associated with task loads of $10$, $10$, $50$, $100$, and $100$ tasks, respectively.
    By evaluating these configurations across both agile and non-agile platform models, we generate $10$ distinct realistic-target scenarios (5 baselines $\times$ 2 platform types).
    For each scenario, $10$ instances are generated using different non-overlapping subsets of city targets, resulting in $100$ instances in total.

    \subsubsection{Characterisation of the Experimental Scenarios} \label{subsec:exp_scenario_characterisation}

    Before comparing solver performance, we first examine the intrinsic structure of the experimental scenarios using the scenario characterisation descriptors introduced in Section~\ref{sec:scenario_characterisation}. To systematically evaluate how structural difficulty evolves, we first isolate a representative configuration (with 100 satellites, 500 tasks and the agile satellite platform) and analyse its properties across ten key dimensions.
    
    \begin{table}[htb]
      \centering
      \footnotesize
      \caption{Scenario characterisation descriptors for a representative configuration (100 satellites, 500 tasks) across varying horizons and distributions.}
      \label{tab:scenario_characterisation}
      \setlength{\tabcolsep}{4pt}
      \begin{tabular}{p{0.35\textwidth} c c c c c c}
        \toprule
        \multirow{2}{*}{{\textbf{Descriptor}}} & \multicolumn{3}{c}{\textbf{Global-Random (Horizon)}} & \multicolumn{3}{c}{\textbf{Region-Clustered (Horizon)}} \\
        \cmidrule(lr){2-4} \cmidrule(lr){5-7}
        & \textbf{12h} & \textbf{24h} & \textbf{72h} & \textbf{12h} & \textbf{24h} & \textbf{72h} \\
        \midrule
        \textbf{Task-Oriented} & & & & & & \\
        {Average Available Opportunities} ($\Gamma_{\mathrm{ao}}$) & 4.21 & 8.54 & 26.12 & 3.85 & 7.92 & 24.33 \\
        {Opportunity Constrained Task Ratio} ($\Gamma_{\mathrm{oc}}$) & 0.352 & 0.124 & 0.015 & 0.410 & 0.165 & 0.022 \\
        {Task Interference Ratio} ($\Gamma_{\mathrm{ti}}$) & 0.012 & 0.025 & 0.068 & 0.045 & 0.088 & 0.215 \\
        {Average Task Pair Conflict Ratio} ($\Gamma_{\mathrm{at}}$) & 0.003 & 0.003 & 0.002 & 0.012 & 0.011 & 0.009 \\
        {Task Elasticity Ratio} ($\Gamma_{\mathrm{te}}$) & 18.50 & 38.20 & 120.40 & 15.20 & 32.60 & 105.80 \\
        \midrule
        \textbf{Satellite-Oriented} & & & & & & \\
        {Observation Contention Ratio} ($\Lambda_{\mathrm{oc}}$) & 0.154 & 0.142 & 0.130 & 0.685 & 0.650 & 0.612 \\
        {Conflict Satellite Ratio} ($\Lambda_{\mathrm{cs}}$) & 0.420 & 0.650 & 0.880 & 0.750 & 0.920 & 1.000 \\
        {Timeline Overload Ratio} ($\Lambda_{\mathrm{to}}$) & 0.015 & 0.014 & 0.012 & 0.082 & 0.078 & 0.070 \\
        {Average Conflict Tasks} ($\Lambda_{\mathrm{ac}}$) & 2.15 & 2.12 & 2.08 & 3.45 & 3.38 & 3.25 \\
        {Excess Demand Ratio} ($\Lambda_{\mathrm{ed}}$) & 0.250 & 0.235 & 0.210 & 0.580 & 0.545 & 0.490 \\
        \bottomrule
      \end{tabular}
    \end{table}

    \noindent \textbf{(1) Overall structural differences.}

    Table~\ref{tab:scenario_characterisation} shows that the selected scenarios differ along two largely independent structural axes. The first axis is temporal opportunity richness: longer horizons substantially increase the number of feasible windows and reduce the proportion of hard tasks. The second axis is spatial congestion: clustered demand sharply increases the fraction of timeline segments in conflict and the number of tasks competing at those instances. As a result, scenarios with similar nominal sizes can induce qualitatively different optimisation challenges. Some are dominated by task-side scarcity, in which the main difficulty is finding any feasible execution opportunity, whereas others are dominated by satellite-side congestion, in which many individually feasible tasks compete for the same short temporal segments. This distinction is precisely the kind of structural information that nominal descriptors such as satellite count, task count, and horizon length cannot convey on their own.

    \noindent \textbf{(2) Effect of planning horizon.}

    Extending the planning horizon from 12h to 72h fundamentally alters the combinatorial nature of the tasks. From a task-oriented perspective, temporal expansion significantly enhances individual flexibility. The average available opportunities ($\Gamma_{\mathrm{ao}}$) and task elasticity ratio ($\Gamma_{\mathrm{te}}$) scale almost linearly with the horizon. Consequently, the opportunity constrained task ratio ($\Gamma_{\mathrm{oc}}$) declines sharply; for instance, in the Global-Random scenario, it drops from 0.352 at 12h to 0.015 at 72h. This indicates that over 3 days, almost all tasks possess abundant feasible insertion opportunities.

    However, this increased flexibility introduces a paradoxical challenge: conflict accumulation. As the feasible windows for individual tasks multiply, the probability of intersection between any two tasks increases, driving up the task interference ratio ($\Gamma_{\mathrm{ti}}$) from 0.012 (12h) to 0.068 (72h) in the Global setting, and up to 0.215 in the Clustered setting. Interestingly, from a satellite-oriented perspective, extending the horizon slightly alleviates the instantaneous intensity of conflicts. The observation contention ratio ($\Lambda_{\mathrm{oc}}$) and peak-normalized excess demand ($\Lambda_{\mathrm{ed}}$) show a slight downward trend as the horizon extends (e.g., $\Lambda_{\mathrm{oc}}$ drops from 0.685 to 0.612 in clustered scenarios), because the total workload is spread over a longer temporal timeline, effectively diluting the immediate resource bottlenecks.

    \noindent \textbf{(3) Effect of target distribution.}

    While the horizon length governs temporal flexibility, the spatial distribution of targets dictates timeline congestion. Comparing the Global-Random and Region-Clustered columns in Table~\ref{tab:scenario_characterisation} reveals that spatial clustering acts as a massive multiplier for satellite-oriented conflicts.

    Under Region-Clustered demand, the observation contention ratio ($\Lambda_{\mathrm{oc}}$) surges to 0.650 (at 24h), compared to only 0.142 under Global-Random demand. This indicates that over 60\% of the active observation timeline is subject to pairwise competition. Furthermore, the conflict depth intensifies significantly: the average conflict tasks ($\Lambda_{\mathrm{ac}}$) rise from approximately 2.1 to 3.4, and the peak-normalised excess demand ($\Lambda_{\mathrm{ed}}$) more than doubles. Physically, this means that when satellites fly over dense target clusters (e.g., a specific continent), they are consistently bombarded by multiple mutually exclusive requests within the same narrow time window.

    \noindent \textbf{(4) Temporal structure of conflicts.}

    Figure~\ref{fig:merged_conflict_windows} complements the aggregate descriptors by showing that conflicts are not distributed uniformly over time. Instead, they appear in bursts separated by relatively quiet intervals. This temporal clustering matters for solver behaviour. Myopic constructive methods are especially vulnerable when an early decision inside a dense conflict episode blocks several downstream alternatives, whereas iterative search methods can benefit from the existence of quieter intervals that make local repairs easier to absorb. The figure, therefore, provides an intuitive bridge between the scalar descriptors and the algorithmic behaviour observed later in the experiments: not only the amount of conflict, but also its temporal concentration, shapes practical scheduling difficulty.

    \begin{figure}[htb]
        \centering
        \includegraphics[width=1\textwidth]{merged_conflict_GAOFEN_10R.png}
        \caption{Merged continuous conflict windows in a representative scenario. Specifically, data for a single satellite, GAOFEN\_10R, is selected for display. To enhance visual clarity, time periods with absolutely no conflicts have been omitted, displaying only the compressed timeline of active conflicts. The conflict types are categorized as follows: ``Overlap'' indicates time overlap conflicts; ``Transition'' refers to transition time conflicts; and ``Mixed'' represents the simultaneous occurrence of both overlap and transition conflicts.}
        \label{fig:merged_conflict_windows}
    \end{figure}

    This visual pattern also reinforces why scenario characterisation should accompany nominal scale descriptors in benchmark reporting: two scenarios with comparable counts of satellites and tasks may still produce very different conflict episode lengths, densities, and repair opportunities.

\noindent \textbf{(5) Effect of constellation scale under fixed demand.}

\begin{figure}[htb]
\centering
\includegraphics[width=1\textwidth]{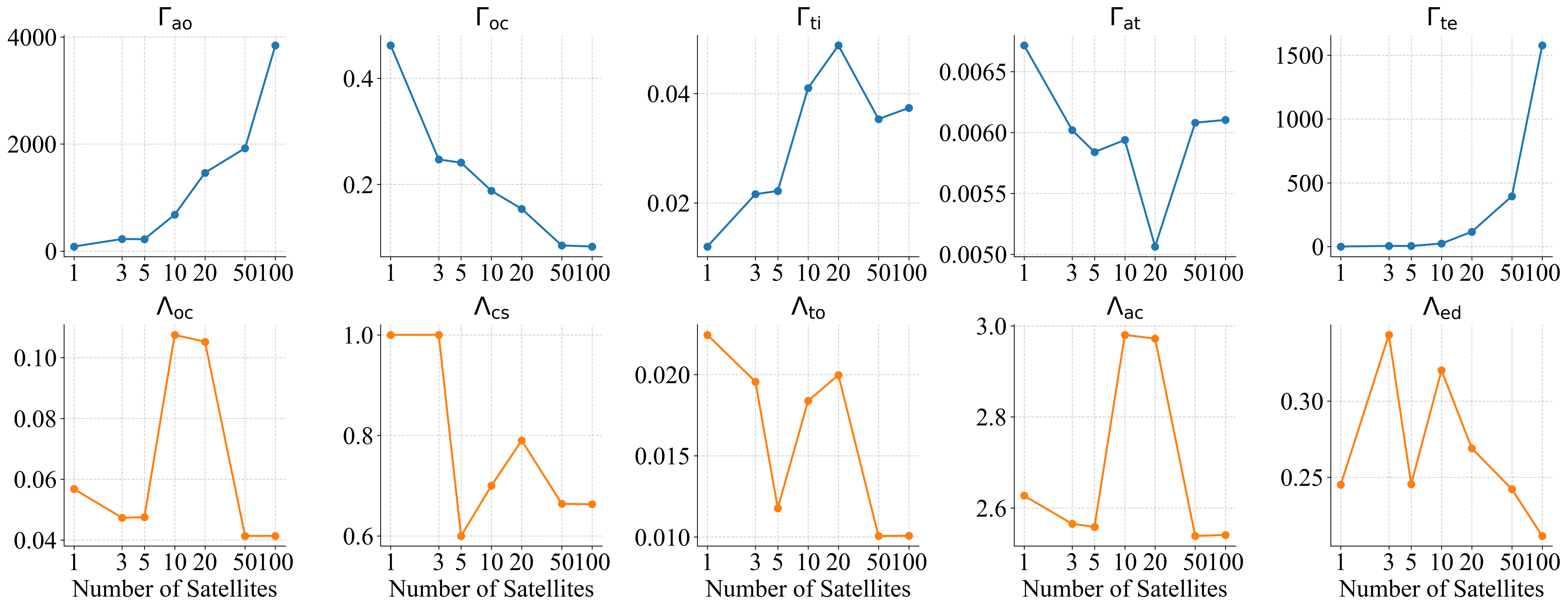}
\caption{Evolution of the ten scenario characterisation descriptors as constellation size scales from 1 to 100 satellites under a fixed load of 100 tasks and a 24h horizon. The vertical axis represents the value of each corresponding descriptor. The evaluated metrics are task-oriented ($\Gamma_{\mathrm{ao}}$: Average Available Opportunities, $\Gamma_{\mathrm{oc}}$: Opportunity Constrained Task Ratio, $\Gamma_{\mathrm{ti}}$: Task Interference Ratio, $\Gamma_{\mathrm{at}}$: Average Task Pair Conflict Ratio, $\Gamma_{\mathrm{te}}$: Task Elasticity Ratio) and satellite-oriented ($\Lambda_{\mathrm{oc}}$: Observation Contention Ratio, $\Lambda_{\mathrm{cs}}$: Conflict Satellite Ratio, $\Lambda_{\mathrm{to}}$: Timeline Overload Ratio, $\Lambda_{\mathrm{ac}}$: Average Conflict Tasks, $\Lambda_{\mathrm{ed}}$: Excess Demand Ratio).}
\label{fig:scale_effect}
\end{figure}

To understand the structural transition from a resource-scarce regime to a resource-abundant regime, we then isolate a baseline configuration with a fixed demand of 100 tasks and a 24-hour horizon, while scaling the constellation size from 1 to 100 satellites. Because this experiment tracks continuous evolution across seven discrete constellation sizes, the structural descriptors are visualised as line trajectories in Figure~\ref{fig:scale_effect} rather than tabulated.

From a task-oriented perspective (top row of Figure~\ref{fig:scale_effect}), expanding the constellation monotonically increases individual assignment flexibility. As the number of satellites grows, the Average Available Opportunities ($\Gamma_{\mathrm{ao}}$) and the Task Elasticity Ratio ($\Gamma_{\mathrm{te}}$) scale almost linearly. Consequently, the Opportunity Constrained Task Ratio ($\Gamma_{\mathrm{oc}}$) drops rapidly to zero, indicating that tasks are no longer limited by a lack of visibility. However, this abundance of opportunities triggers a secondary effect: the Task Interference Ratio ($\Gamma_{\mathrm{ti}}$) rises significantly. Because each task now possesses multiple feasible windows across the constellation, the probability that any two tasks overlap on at least one shared satellite timeline increases, weaving a denser web of potential assignment conflicts.

From a satellite-oriented perspective (bottom row of Figure~\ref{fig:scale_effect}), scaling the constellation effectively dilutes temporal congestion. By distributing the same 100 tasks across a progressively larger aggregate timeline, the Observation Contention Ratio ($\Lambda_{\mathrm{oc}}$), Timeline Overload Ratio ($\Lambda_{\mathrm{to}}$), and Average Conflict Tasks ($\Lambda_{\mathrm{ac}}$) all exhibit sharp concave decays. Notably, the Conflict Satellite Ratio ($\Lambda_{\mathrm{cs}}$) initially peaks before declining, because in very small constellations nearly all platforms are congested, whereas in a 100-satellite constellation, many platforms experience zero conflict.

Taken together, Figure~\ref{fig:scale_effect} illustrates a fundamental phase transition in instance difficulty. At $N_S \le 5$, the solver faces a \textit{congestion-dominated} challenge: timelines are heavily overloaded ($\Lambda_{\mathrm{ed}}$ is high), and the primary difficulty is deciding which tasks to discard. At $N_S \ge 50$, the solver faces a \textit{routing-dominated} challenge: individual satellite overload vanishes, but the combinatorial space of feasible task-satellite assignments ($\Gamma_{\mathrm{te}}$) explodes. These descriptors thereby provide a quantitative explanation for why scalability algorithms (such as meta-heuristics or decentralised learning) become necessary at larger scales, even when the underlying task demand remains constant.

\noindent \textbf{(6) Effect of orbital architecture under fixed constellation scale.}

\begin{figure}[htb]
\centering
\includegraphics[width=1\textwidth]{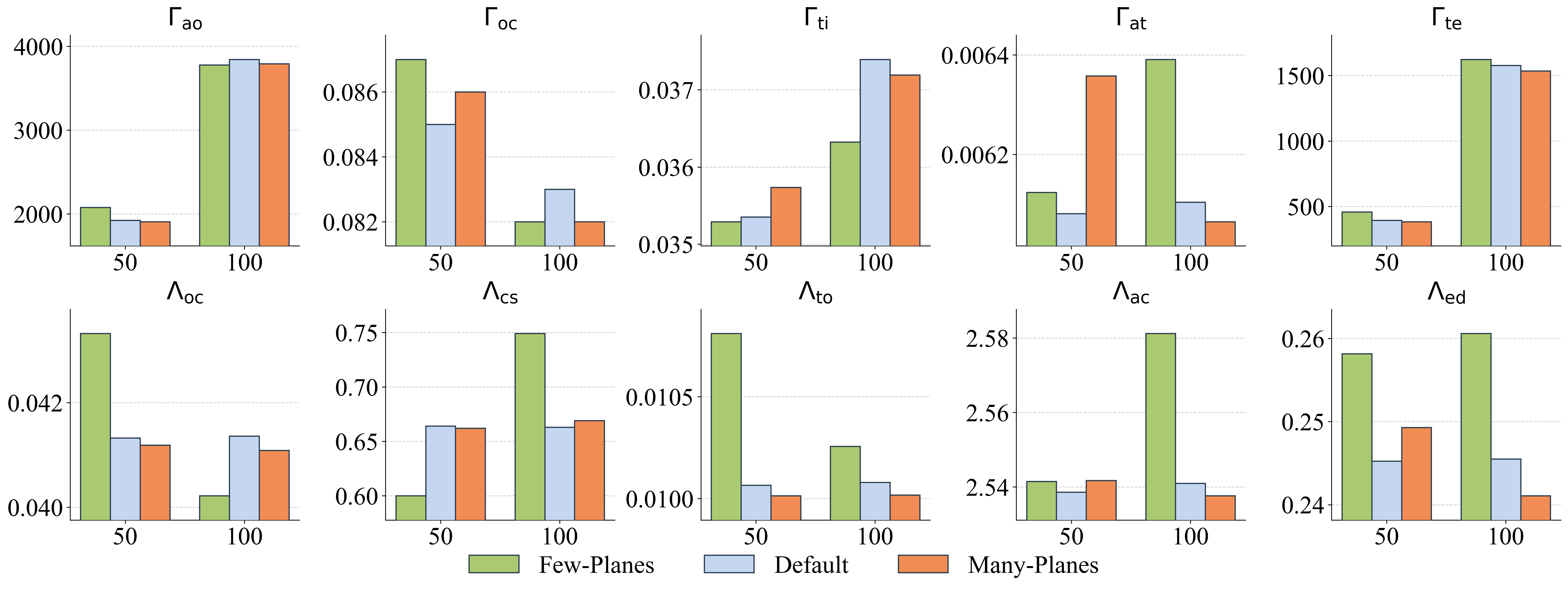}
\caption{Effect of constellation orbital architecture on structural descriptors at 50- and 100-satellite scales. Bars represent different Walker--Delta configurations: Few-Planes (dense orbital corridors), Default (balanced), and Many-Planes (sparse distribution). The horizontal axis denotes the constellation scale (50 and 100 satellites), while the vertical axis represents the value of each corresponding descriptor. The evaluated metrics are task-oriented ($\Gamma_{\mathrm{ao}}$: Average Available Opportunities, $\Gamma_{\mathrm{oc}}$: Opportunity Constrained Task Ratio, $\Gamma_{\mathrm{ti}}$: Task Interference Ratio, $\Gamma_{\mathrm{at}}$: Average Task Pair Conflict Ratio, $\Gamma_{\mathrm{te}}$: Task Elasticity Ratio) and satellite-oriented ($\Lambda_{\mathrm{oc}}$: Observation Contention Ratio, $\Lambda_{\mathrm{cs}}$: Conflict Satellite Ratio, $\Lambda_{\mathrm{to}}$: Timeline Overload Ratio, $\Lambda_{\mathrm{ac}}$: Average Conflict Tasks, $\Lambda_{\mathrm{ed}}$: Excess Demand Ratio).}
\label{fig:orbit_effect}
\end{figure}

Beyond the sheer number of satellites, the geometric layout of the constellation fundamentally alters the temporal distribution of observation opportunities. To isolate this effect, we further examine two fixed satellite scales ($N_S=50$ and $N_S=100$) under a constant demand of 100 tasks and a 24-hour horizon. For each scale, we compare three Walker--Delta architectures: a Default (Balanced) pattern, a Few-Planes (Dense) pattern containing more satellites per plane, and a Many-Planes (Sparse) pattern distributing satellites across more orbital planes. The resulting structural descriptors are visualised in Figure~\ref{fig:orbit_effect}.

From a task-oriented perspective (top row of Figure~\ref{fig:orbit_effect}), the geometric layout subtly but consistently shifts assignment flexibility. At both the 50- and 100-satellite scales, the Few-Planes configurations (e.g., 5 planes $\times$ 10 sats) yield noticeably higher Task Elasticity Ratios ($\Gamma_{\mathrm{te}}$) than the Many-Planes configurations (e.g., 25 planes $\times$ 2 sats). This occurs because packing satellites densely into fewer planes creates continuous ``platoons'' of sensors that pass over targets in quick succession, yielding highly redundant and overlapping access windows for individual tasks.

However, this spatial grouping comes at a cost from a satellite-oriented perspective (bottom row of Figure~\ref{fig:orbit_effect}). The redundant overlapping accesses generated by the Few-Planes topology force more simultaneous tasks onto the grouped satellite timelines, thereby exacerbating local congestion. Consequently, the Timeline Overload Ratio ($\Lambda_{\mathrm{to}}$) and Excess Demand Ratio ($\Lambda_{\mathrm{ed}}$) are elevated in the Few-Planes configurations. Conversely, the Many-Planes topology spreads the satellite assets more evenly across the globe. While this slightly reduces peak redundancy for individual tasks, it smooths out temporal contention and helps alleviate localised timeline bottlenecks.

This analysis confirms that the proposed scenario characterisation descriptors are sensitive not only to nominal problem size, but also to the underlying orbital mechanics. It mathematically reveals why constellations of identical size can still exhibit distinct scheduling behaviours: topologies that clump satellites together (Few-Planes) enhance local opportunity redundancy but intensify timeline congestion, whereas topologies that distribute them broadly (Many-Planes) favour workload balance over peak flexibility.

    \subsection{Experimental Results}\label{subsec:experimental_results}

    Extensive experiments were conducted to comprehensively evaluate the performance of the various scheduling algorithms/solvers over the selected 250 scenarios/2,500 instances. However, due to space constraints, 
all results reported in this subsection are based on a smaller, representative subset of agile satellite scenarios drawn from the selected 250 scenarios. The complete experimental results are publicly available in the data directory of our GitHub repository\footnote{\url{https://github.com/Ethan19YQ/EOS-Bench}}, with access instructions provided in~\ref{app:first}.

    \subsubsection{Standard Scenarios}
    \label{subsec:exp_standard}

    We selected a small-scale scenario (3 satellites, 100 tasks) to establish exact-solver baselines using exact MIP solvers, and a large-scale scenario (100 satellites, 2,000 tasks) to represent a high-dimensional operational setting where exact methods are intractable, focusing instead on the scalability of heuristic and learning-based approaches.
    The analysis is organised along four dimensions: solution quality assessment, computational scalability, distributional robustness, and multi-objective trade-off analysis.

\noindent \textbf{(1) Solution quality and optimality assessment}

\begin{figure}[htb]
        \centering
        \begin{subfigure}[b]{0.8\textwidth}
            \centering
            \includegraphics[width=\textwidth]{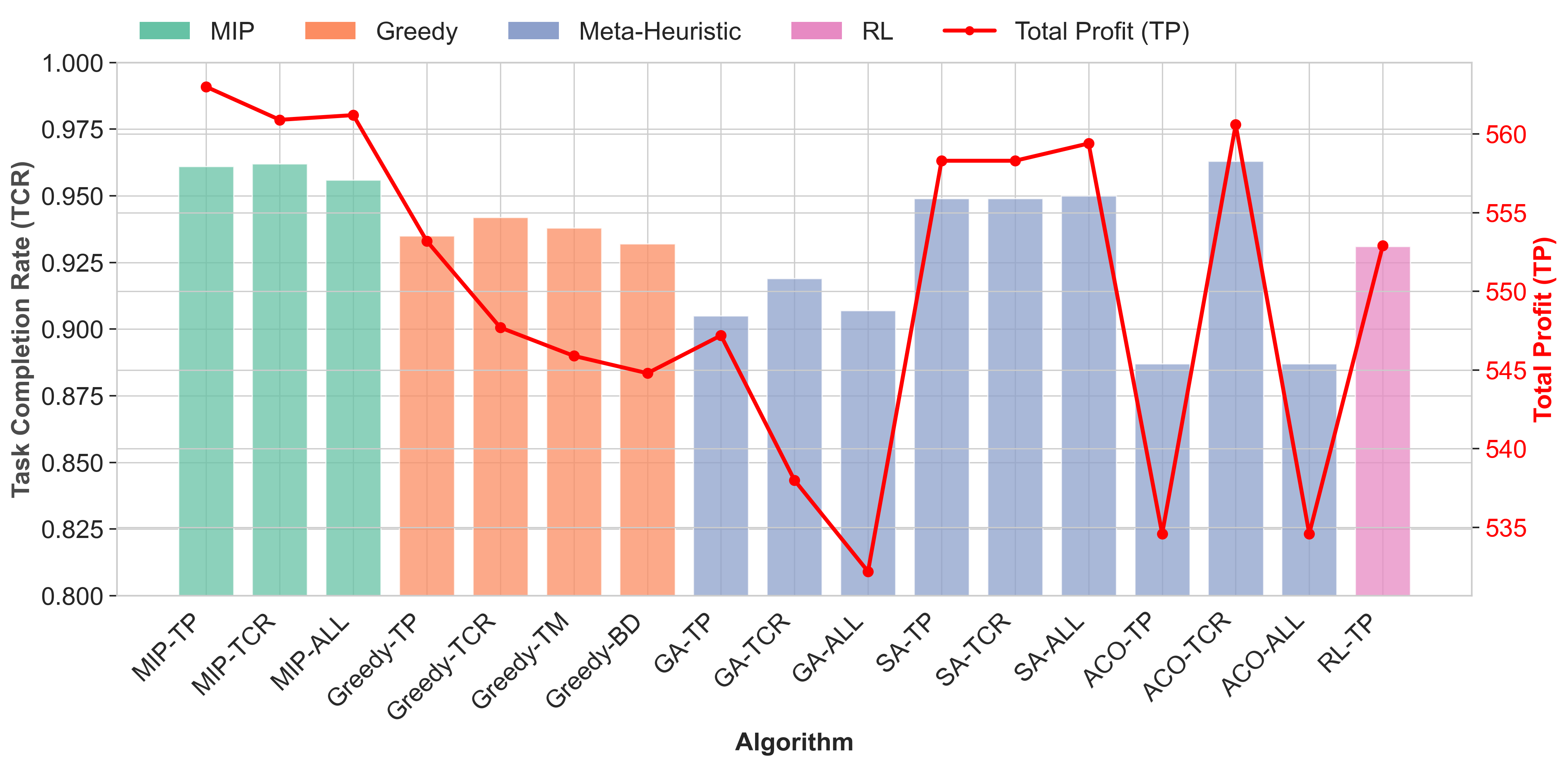}
            \caption{Small-Scale Scenario ($N_S=3, N_T=100, T_{\mathrm{hor}}=24\mathrm{h}$)}
            \label{fig:overview_small}
        \end{subfigure}

        \vspace{0.25cm}

        \begin{subfigure}[b]{0.8\textwidth}
            \centering
            \includegraphics[width=\textwidth]{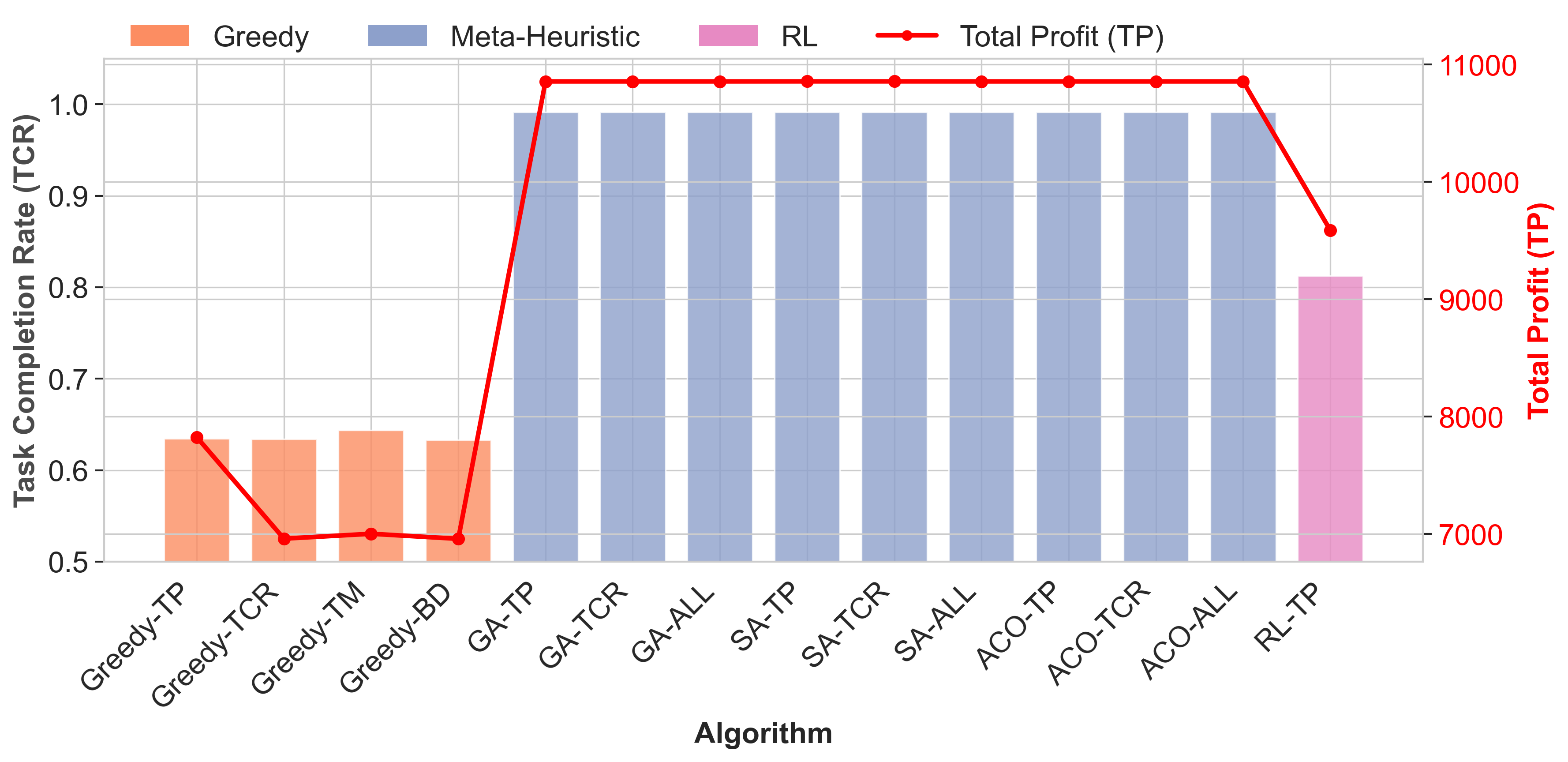}
            \caption{Large-Scale Scenario ($N_S=100, N_T=2000, T_{\mathrm{hor}}=24\mathrm{h}$))}
            \label{fig:overview_large}
        \end{subfigure}

        \caption{Solution quality and optimality assessment.}
        \label{fig:overview}
    \end{figure}

    The first dimension evaluates the benchmark's ability to measure the optimality gap and discriminate between solver efficacies.
    Figure~\ref{fig:overview} presents the comparative performance in terms of Task Completion Rate (TCR) and Task Profit (TP).

   In the Small-Scale case (Figure~\ref{fig:overview}a), the benchmark provides exact MIP references, which serve as useful optimality anchors for interpreting the heuristic results. Within this setting, SA-based methods appear closest to the MIP frontier, while {ACO-TCR} also performs competitively on the reported metrics. By contrast, the Greedy methods tend to attain lower TP and TCR values, which is consistent with the limitations of myopic constructive decisions even in relatively moderate problem instances. RL performs better than the weaker heuristic baselines in these experiments, but it does not match the strongest SA/ACO variants on the reported scenarios.

    In the Large-Scale case (Figure~\ref{fig:overview}b), where exact MIP baselines are no longer available, the benchmark primarily reflects the relative scalability of different solver classes. The Greedy methods show noticeable performance degradation as the problem size increases, whereas GA, SA, and ACO maintain higher TP and TCR values in the reported experiments, forming the leading group in this regime. RL again occupies an intermediate position: it improves on the Greedy baselines, but remains below the best meta-heuristic results on these instances. Overall, the large-scale scenarios appear sufficiently discriminative to separate fast constructive methods from more computationally intensive iterative search procedures.

    \noindent \textbf{(2) Computational effort and scalability analysis}

\begin{figure}[htb]
        \centering
        \begin{subfigure}[b]{0.49\textwidth}
            \centering
            \includegraphics[width=\textwidth]{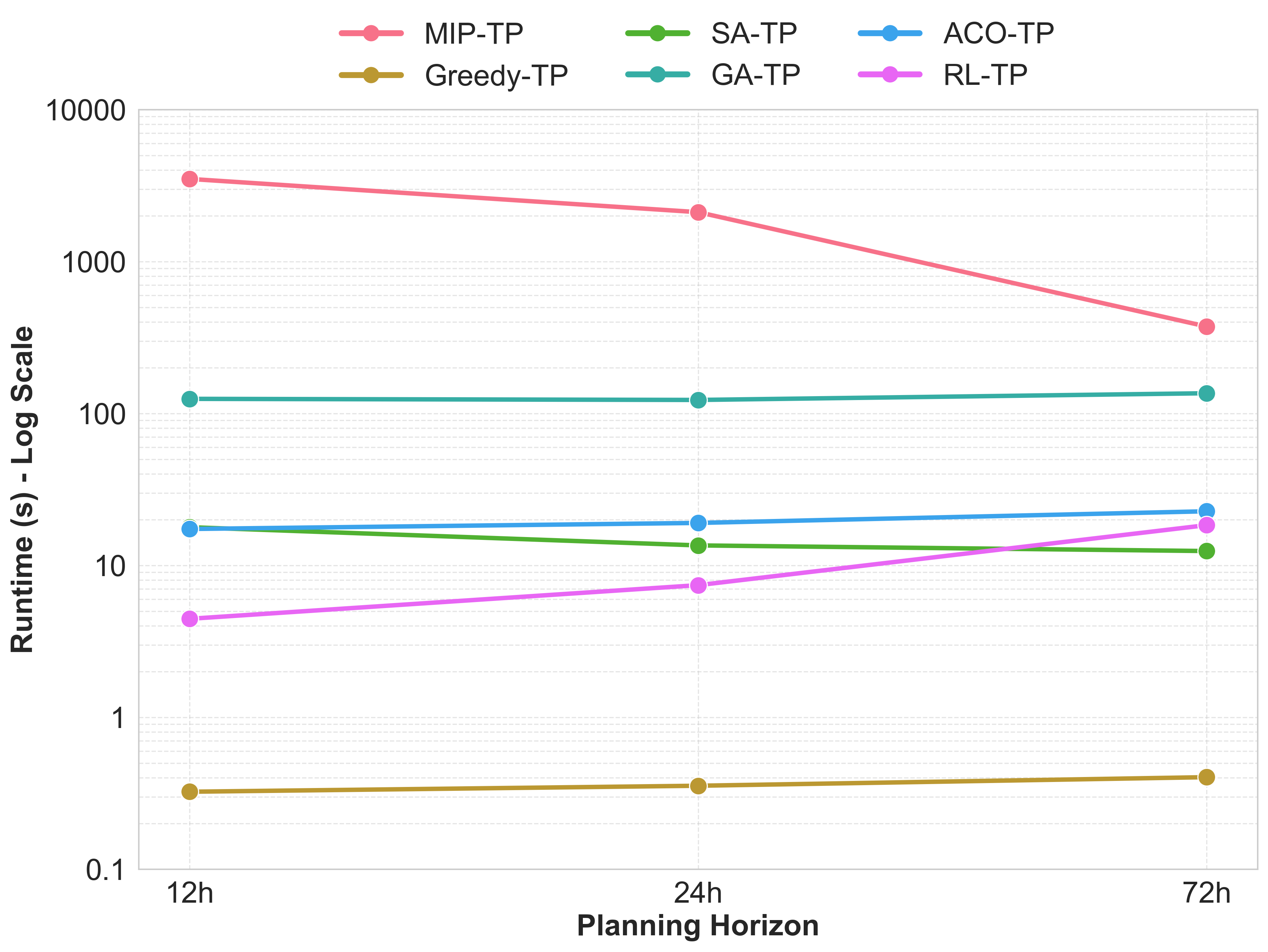}
            \caption{Small-Scale Scenario ($N_S=3, N_T=100, |Dist|=Regional$)}
            \label{fig:runtime_small}
        \end{subfigure}
        \hfill
        \begin{subfigure}[b]{0.49\textwidth}
            \centering
            \includegraphics[width=\textwidth]{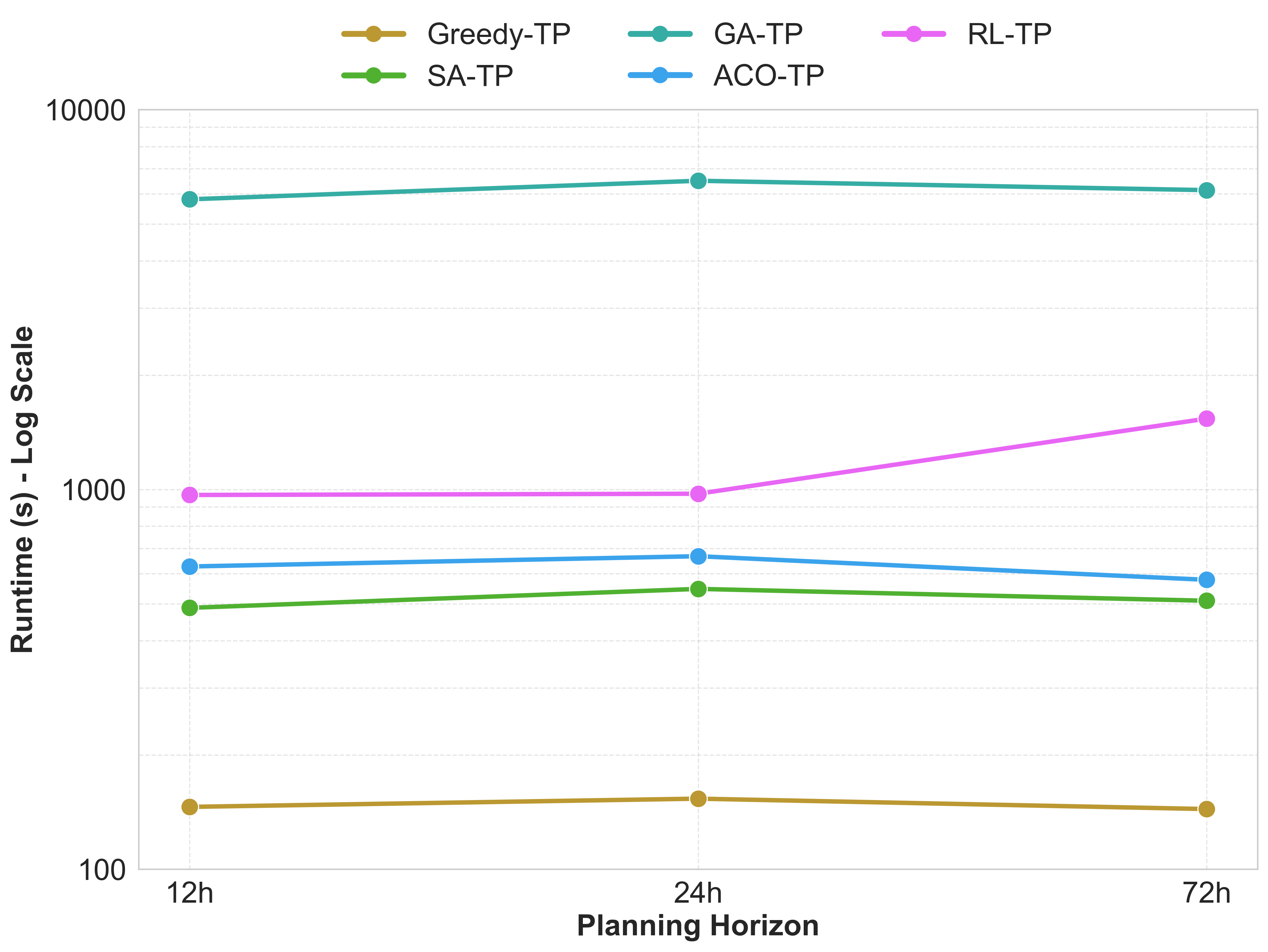}
            \caption{Large-Scale Scenario ($N_S=100, N_T=2000, |Dist|=Regional$)}
            \label{fig:runtime_large}
        \end{subfigure}

        \caption{Computational effort and scalability analysis.}
        \label{fig:runtime}
    \end{figure}

    A robust benchmark must capture the NP-hard nature of the scheduling problem. Figure~\ref{fig:runtime} illustrates the runtime (RT) characteristics across varying planning horizons (12h to 72h) and constellation sizes.

    Figure~\ref{fig:runtime}a (Small-Scale) illustrates the tractability limits of the tested methods.
    The benchmark captures the combinatorial explosion inherent in exact solvers, with MIP runtimes increasing sharply and frequently triggering the predefined timeout limit as the horizon extends.
    In contrast, heuristic and RL-inference approaches show much slower runtime growth.

    Figure~\ref{fig:runtime}b (Large-Scale) serves as a stress test for large-scale scalability.
   With the constellation size increased to 100 satellites, the reported results suggest substantial runtime pressure for population-based meta-heuristics (GA, ACO). A plausible explanation is that evaluating candidate schedules and verifying transition constraints over thousands of tasks becomes increasingly expensive at each iteration. The resulting runtime spread is useful from a benchmarking perspective because it differentiates methods that may be better suited to offline planning from those that may be more compatible with tighter response-time requirements, such as RL inference and simple greedy construction.

    \noindent \textbf{(3) Distributional robustness and sensitivity analysis}

\begin{figure}[htb]
        \centering
        \begin{subfigure}[b]{0.8\textwidth}
            \centering
            \includegraphics[width=\textwidth]{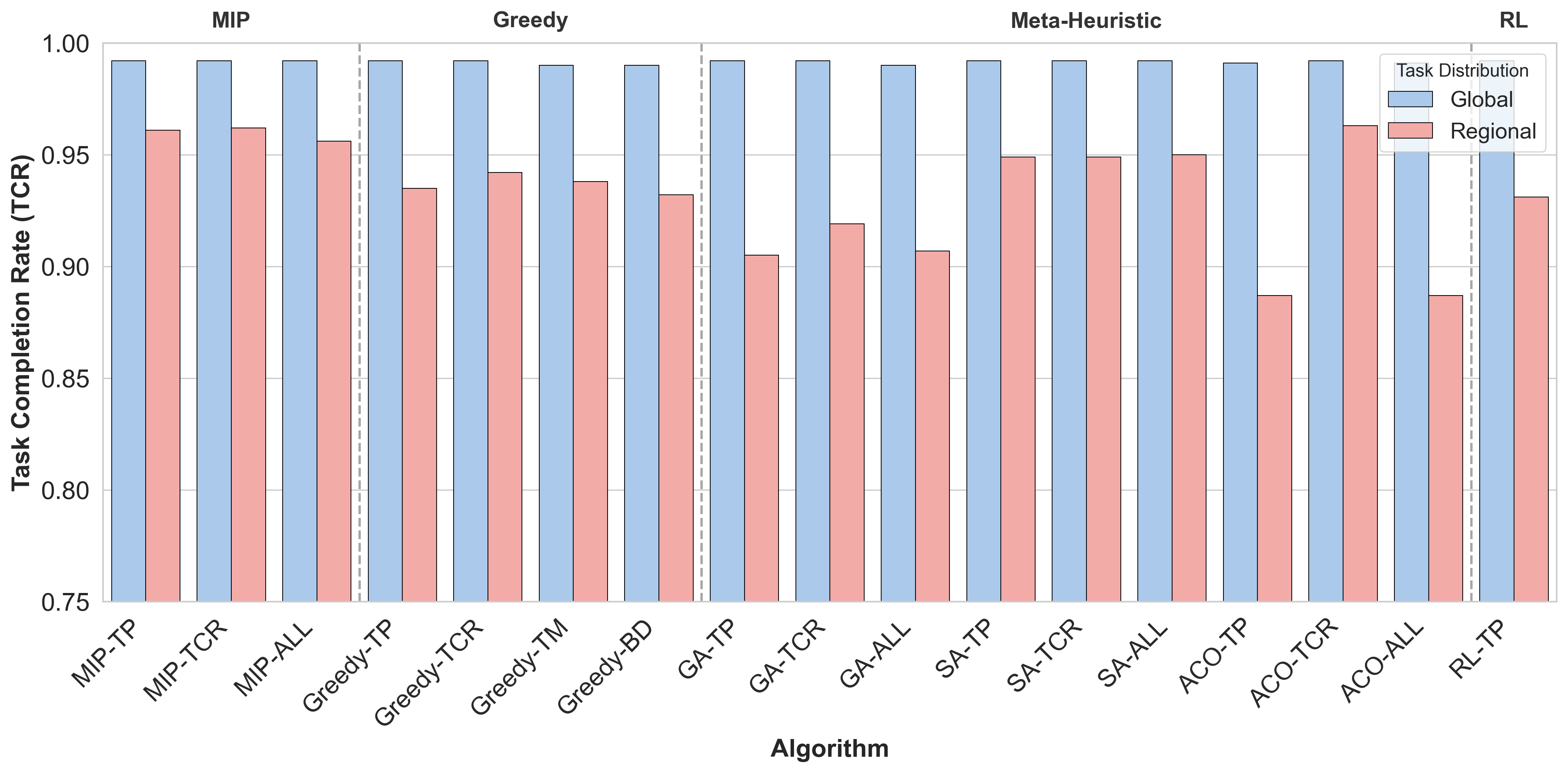}
            \caption{Small-Scale Scenario ($N_S=3, N_T=100, T_{\mathrm{hor}}=24\mathrm{h}$)}
            \label{fig:robustness_small}
        \end{subfigure}

        \vspace{0.05cm}

        \begin{subfigure}[b]{0.8\textwidth}
            \centering
            \includegraphics[width=\textwidth]{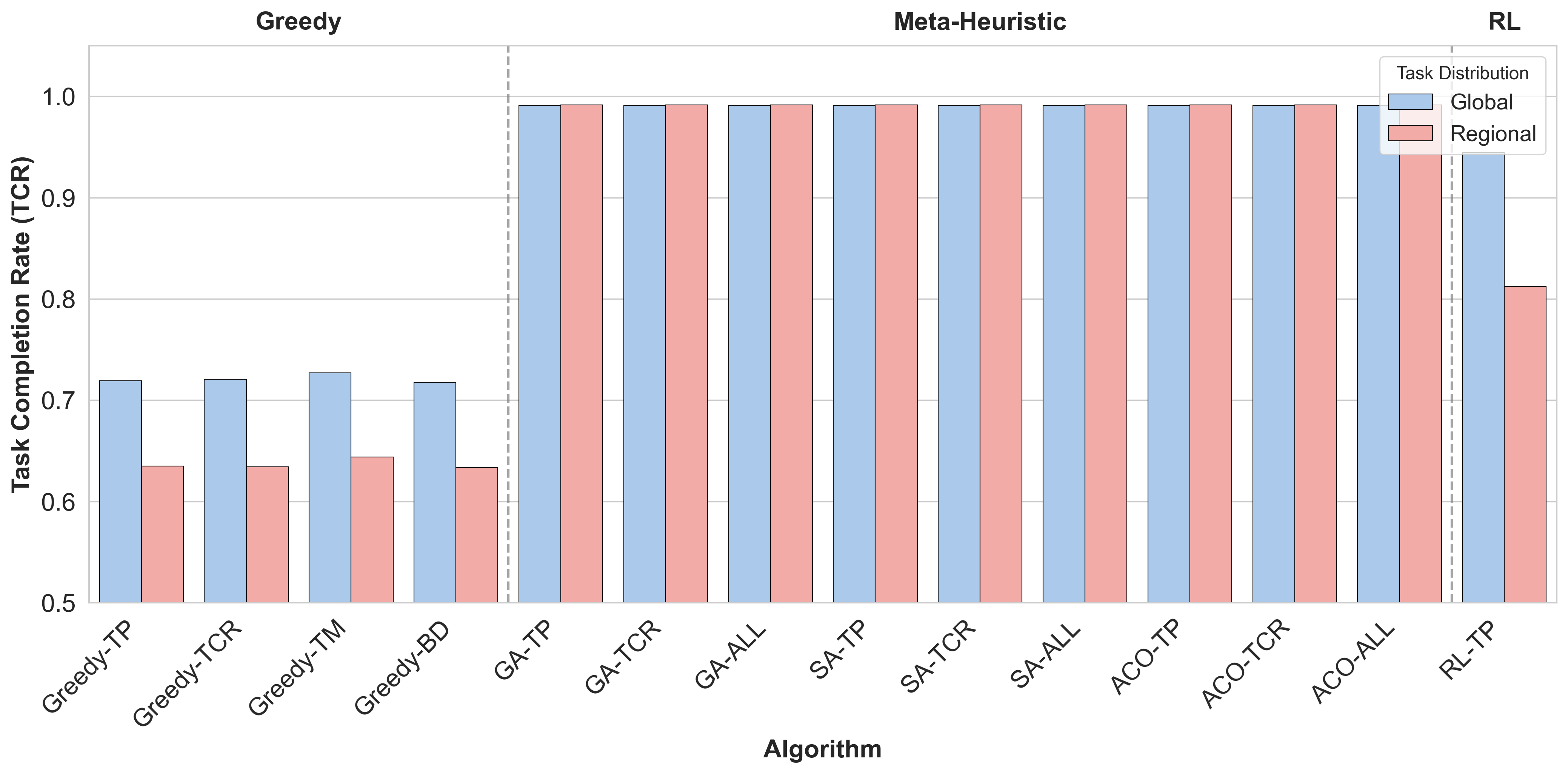}
            \caption{Large-Scale Scenario ($N_S=100, N_T=2000, T_{\mathrm{hor}}=24\mathrm{h}$)}
            \label{fig:robustness_large}
        \end{subfigure}

        \caption{Distributional robustness and sensitivity analysis.}
        \label{fig:robustness}
    \end{figure}

    To evaluate solver performance under varying constraint tightness, Figure~\ref{fig:robustness} compares performance under Global-Random versus Region-Clustered.

    Across both scales, the reported Clustered scenarios are associated with lower performance than the corresponding Global-Random cases. This pattern is consistent with intensified resource contention, where overlapping visible time windows and limited slew feasibility reduce the number of tasks that can be scheduled without conflict.

    Within this comparison, the Greedy methods appear particularly sensitive to clustered demand, plausibly because early assignments reduce flexibility for later high-value tasks. The search-based methods (SA, ACO) and the RL baseline are comparatively less affected in the reported figures, which is consistent with their greater ability to revise choices or account for longer-range structure during decision making. Taken together, these patterns suggest that the benchmark is capable of representing high-conflict operating conditions in which solver behaviour diverges for understandable structural reasons.

    \noindent \textbf{(4) Multi-objective trade-off and Pareto analysis}

\begin{figure}[htbp]
        \centering
        \begin{subfigure}[b]{0.48\textwidth}
            \centering
            \includegraphics[width=\textwidth]{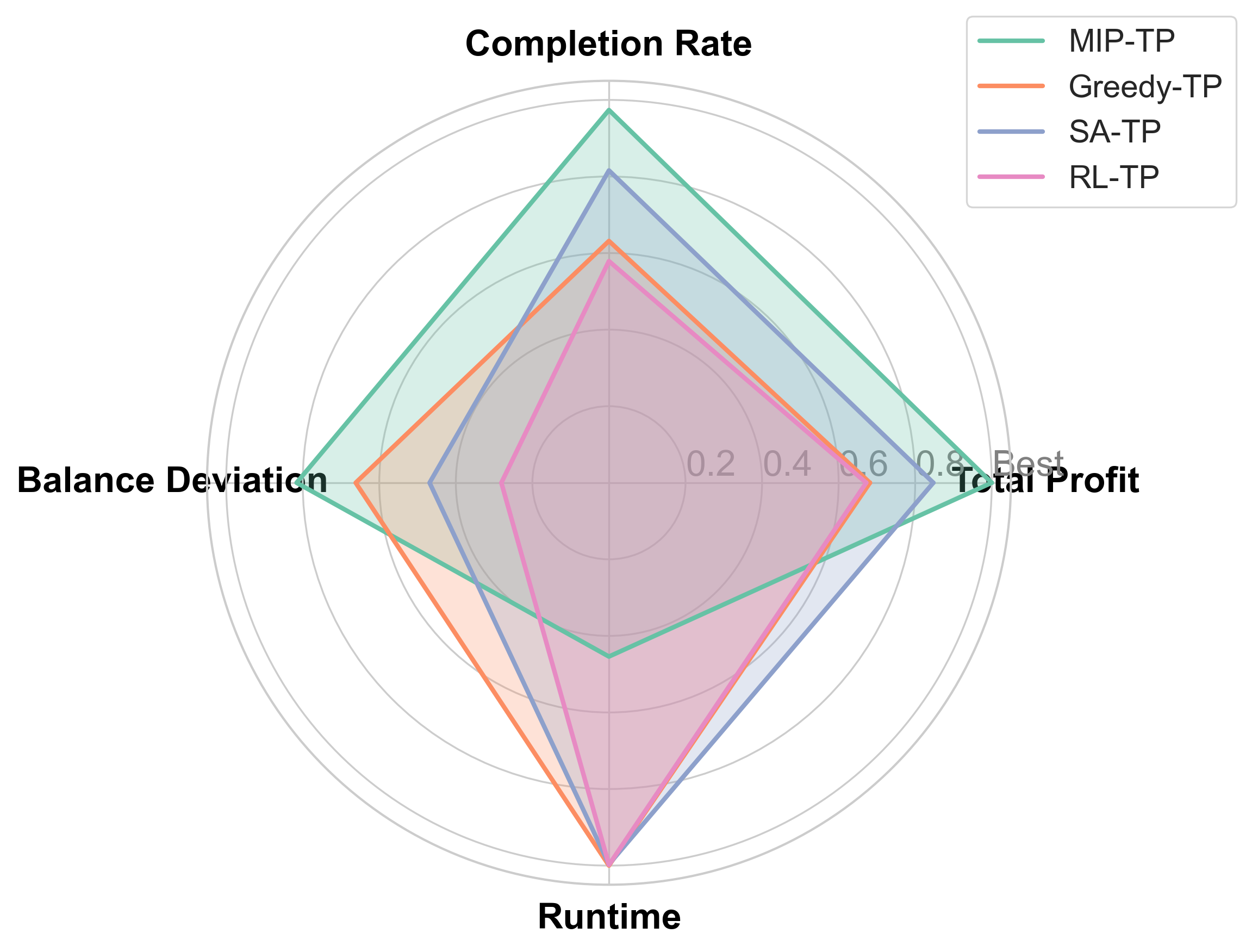}
            \caption{Small-Scale Scenario ($N_S=3, N_T=100, |Dist|=Regional$)}
            \label{fig:radar_small}
        \end{subfigure}
        \hfill
        \begin{subfigure}[b]{0.48\textwidth}
            \centering
            \includegraphics[width=\textwidth]{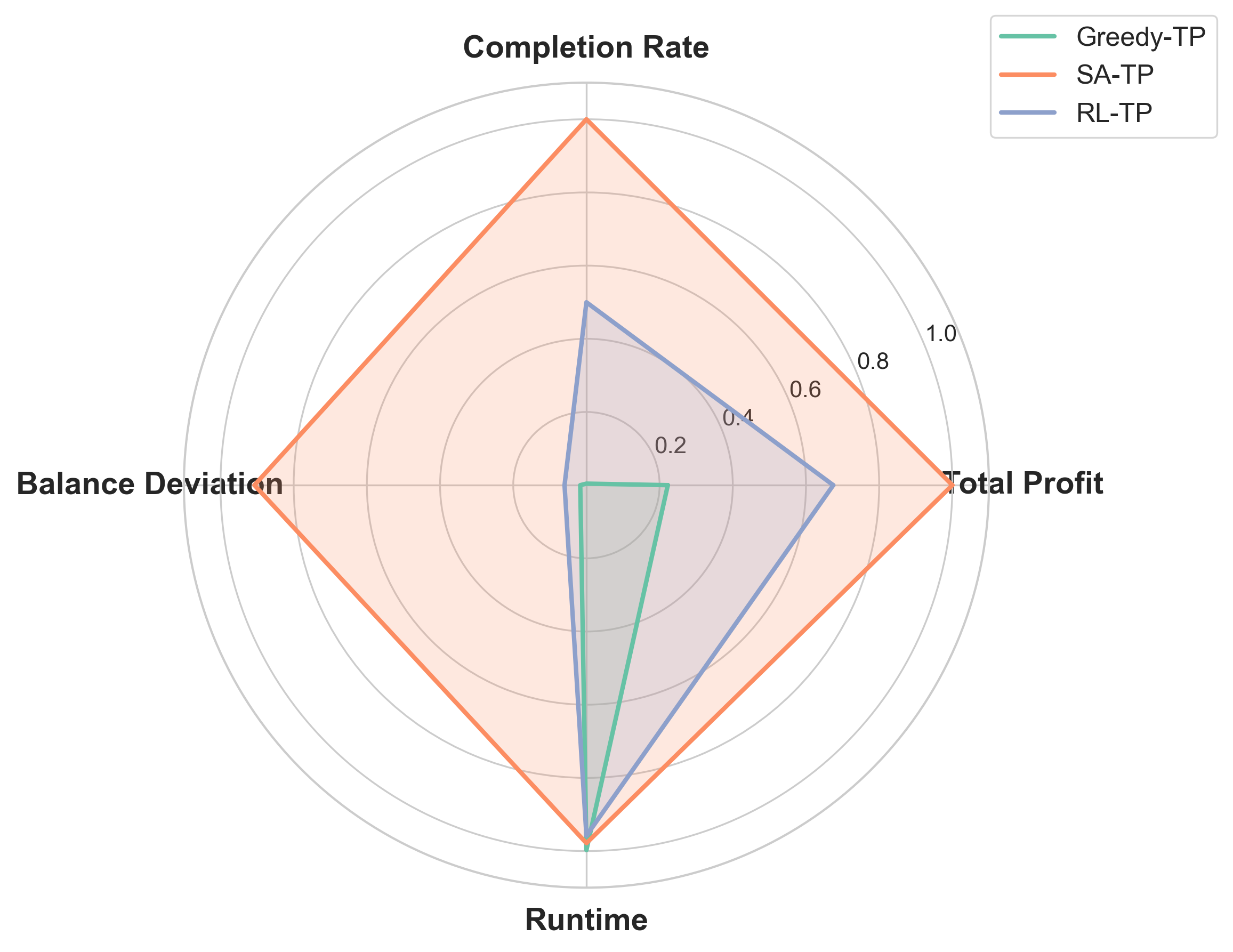}
            \caption{Large-Scale Scenario ($N_S=100, N_T=2000, |Dist|=Regional$)}
            \label{fig:radar_large}
        \end{subfigure}

        \caption{Multi-objective trade-off and Pareto analysis via performance profiles.}
        \label{fig:radar}
    \end{figure}

    EOSSP is inherently a multi-objective optimisation problem requiring a delicate balance between competing operational goals. Figure~\ref{fig:radar} uses normalised radar charts to compare solver performance profiles for each solver class, simultaneously encompassing Task Profit (TP), Task Completion Rate (TCR), Balance Degree (BD), and Runtime (RT).
    To ensure visual consistency where a larger area denotes superior performance, all axes are aligned such that higher values are preferable; specifically, minimisation metrics (such as Runtime) are inverted using a $1 - \text{normalised value}$ transformation.

    The {Small-Scale} profile (Figure~\ref{fig:radar}a) illustrates the cost of optimality in a visual form. In these instances, the Exact Solver (MIP) achieves the strongest values on the objective-oriented metrics (TP, TCR, BD), but it does so with substantially weaker runtime performance, which yields a highly skewed profile.

    The {Large-Scale} profile (Figure~\ref{fig:radar}b) makes the main trade-offs between solver families easier to inspect. In the reported results, the Greedy rules favour speed and timeliness but lag on balance and profit, whereas the meta-heuristics obtain stronger quality-oriented scores at a higher computational cost. The RL models appear comparatively balanced across the plotted dimensions, although the preferred method would still depend on the operational emphasis placed on speed, solution quality, and workload distribution. This kind of multi-metric view is useful because it reduces the risk of evaluating solvers on the basis of a single headline metric alone.

\subsubsection{Specific Scenarios}
    \noindent \textbf{(1) Capacity sensitivity analysis}

    This set of experiments investigates the impact of satellite resource budgets, specifically onboard energy and data storage, on scheduling performance.
    We evaluated five capacity configurations—ranging from homogeneous constraints (Low/High bounds across all nodes) to heterogeneous mixes (Mixed A/B/C, simulating distinct satellite generations)—on the 3-satellite, 200-task scenario.

    \begin{figure}[htbp]
        \centering
        \includegraphics[width=1.0\textwidth]{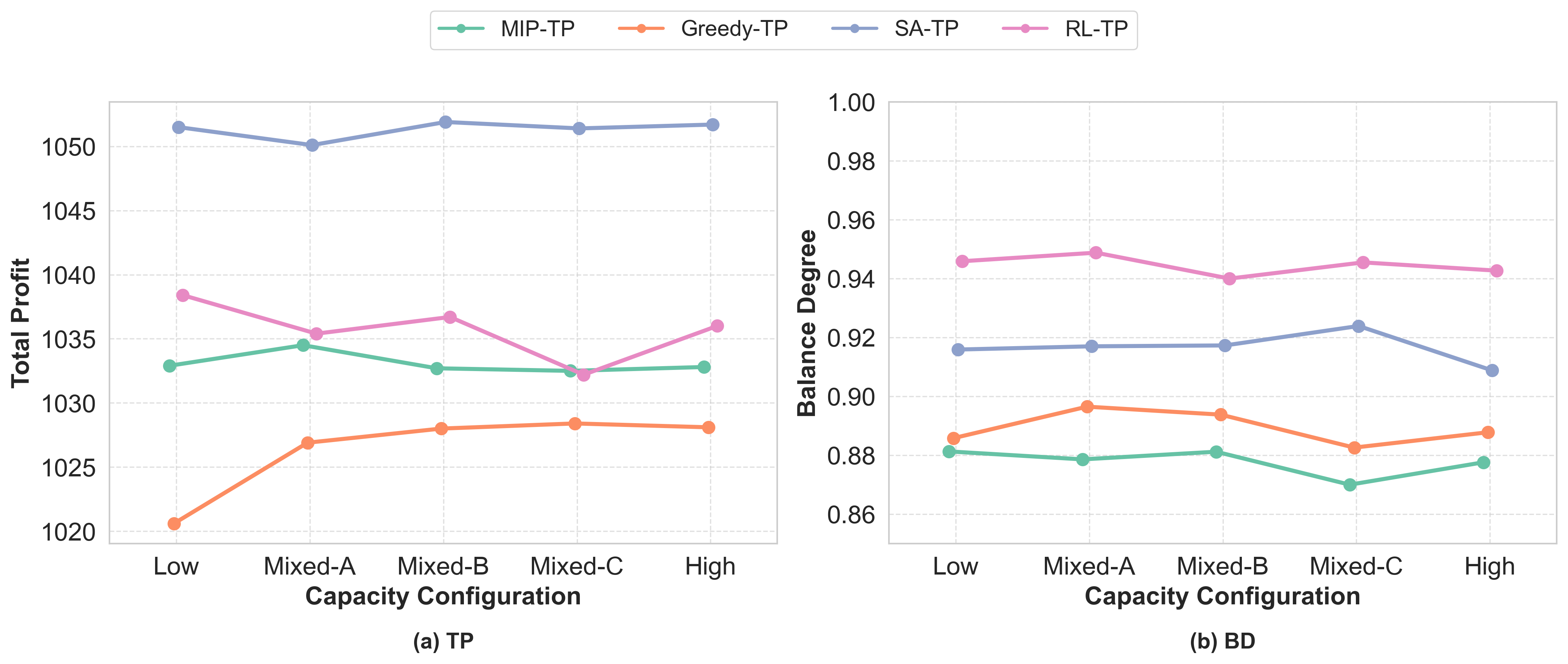}
        \caption{Algorithm sensitivity to resource capacity in Regional Distribution.}
        \label{fig:capacity_analysis}
    \end{figure}

    The results highlight two distinct operational regimes dictated by resource availability. Under the Regional Distribution (Dist 1), where targets are geographically clustered, overlapping visibility windows cause rapid depletion of local satellite buffers, forcing the system into a strictly Resource-Bound state.

    Figure~\ref{fig:capacity_analysis}a illustrates the performance variance in this constrained setting. The Greedy heuristics show a positive response to capacity expansion, with Task Profit (TP) increasing from 1020.6 (Low) to 1028.1 (High), which is consistent with the idea that myopic construction benefits directly from relaxed physical limits. By comparison, the RL baseline appears less sensitive to resource scarcity in the reported experiments, maintaining relatively high solution quality (TCR $\approx 0.87$) even in the Low-Capacity scenarios. One possible interpretation is that the learned policy is more willing to defer early low-value observations in order to preserve resources for denser or higher-value opportunities later in the orbit.

    Furthermore, in Heterogeneous settings (Mixed A/B/C), Figure~\ref{fig:capacity_analysis}b shows that SA and RL obtain higher Load Balancing values (BD $> 0.9$) than the Greedy methods in the reported experiments. This pattern is consistent with the view that more adaptive methods can distribute tasks more evenly across satellites with different capacities, whereas simpler heuristics tend to concentrate work on the most immediately attractive resources.

    \noindent \textbf{(2) Agility sensitivity analysis}

    This experiment evaluates the impact of satellite manoeuvrability on scheduling density by testing three agility profiles (Limited, Low, High), which correspond to varying maximum angular slew velocities (deg/s) and stabilisation times.

    \begin{figure}[htbp]
        \centering
        \includegraphics[width=1.0\textwidth]{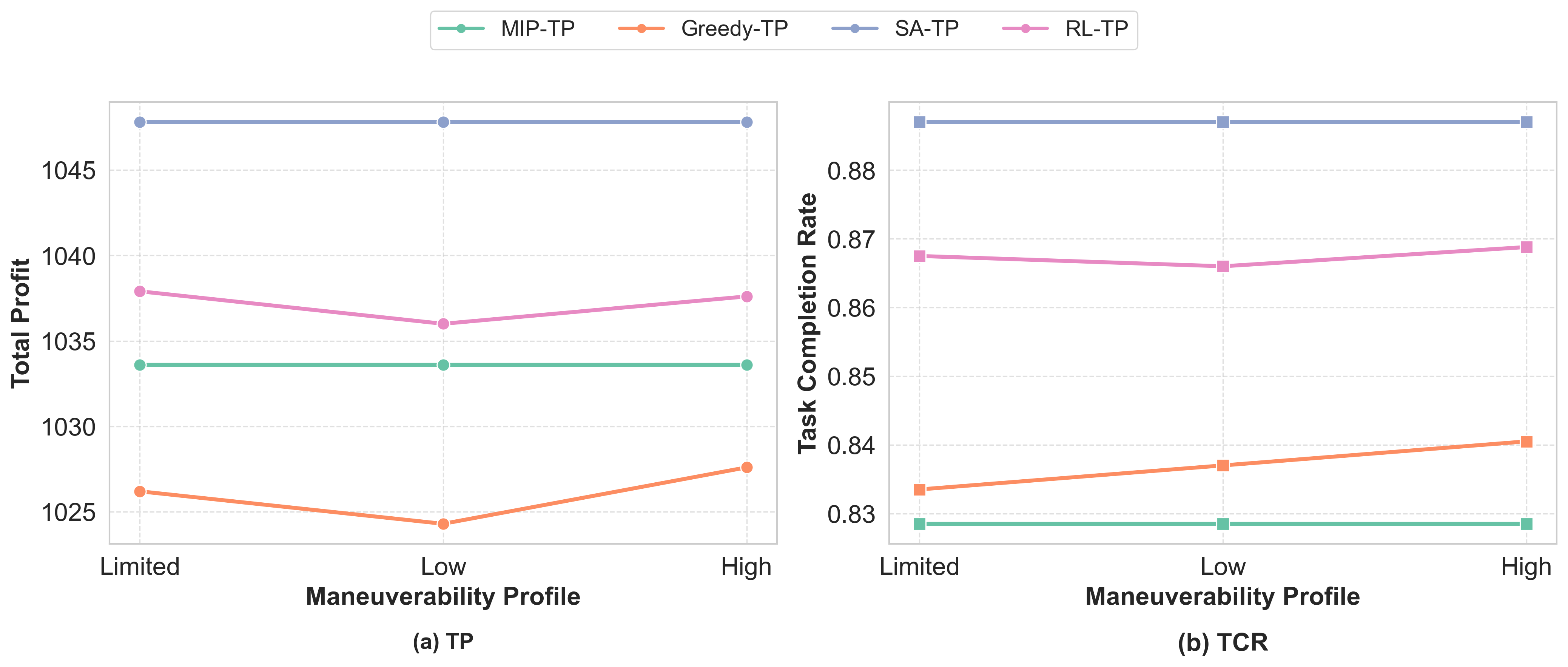}
        \caption{Algorithm sensitivity to satellite agility.}
        \label{fig:agility_analysis}
    \end{figure}

    The results in the Regional Distribution (Dist 1) suggest diminishing returns from further increases in agility in this scenario for the tested scenario size ($N_S=3, N_T=200$). As shown in Figure~\ref{fig:agility_analysis}a, the Task Profit (TP) for the exact solver (MIP) remains effectively constant ($\approx 1033.6$) across all profiles. This indicates that for this specific spatial layout, feasibility is primarily bounded by the geometric distribution of the visibility windows rather than the inter-task transition delays; once a minimum slew threshold is met, further increasing angular velocity provides diminishing returns to the global objective value.

    However, at the algorithmic level, higher agility appears to benefit the time-greedy constructive heuristics. Figure~\ref{fig:agility_analysis}b shows that the Task Completion Rate (TCR) of the Greedy algorithm increases from 0.833 (Limited) to 0.841 (High). This trend is consistent with the fact that improved manoeuvrability reduces the required transition time ($t_{transition}$) between consecutive observations and therefore opens additional short insertion opportunities.
    Across all three profiles, SA and the RL baseline also remain above the Greedy baseline on the reported metrics. In some runs, they even exceed the profit of the time-limited MIP solutions, which should be interpreted cautiously: because the MIP results are truncated by a time budget, these comparisons indicate stronger incumbent solutions within that budget rather than superiority over a proven optimum.

    \noindent \textbf{(3) Orbit-architecture sensitivity analysis}

    This subsection analyses the impact of constellation orbital topology at medium scales ($N_S=100$ and $200$) with moderate task loads ($N_T=300$ and $500$). We compare three distinct Walker constellation configurations: Balanced, Dense (Few-Planes, densely packed satellites), and Sparse (Many-Planes, sparsely populated orbits). 
    {To ensure a consistent baseline for this structural comparison, all scheduling results reported in this analysis are obtained using the GA-ALL.}

    \begin{figure}[htbp]
        \centering
        \includegraphics[width=1.0\textwidth]{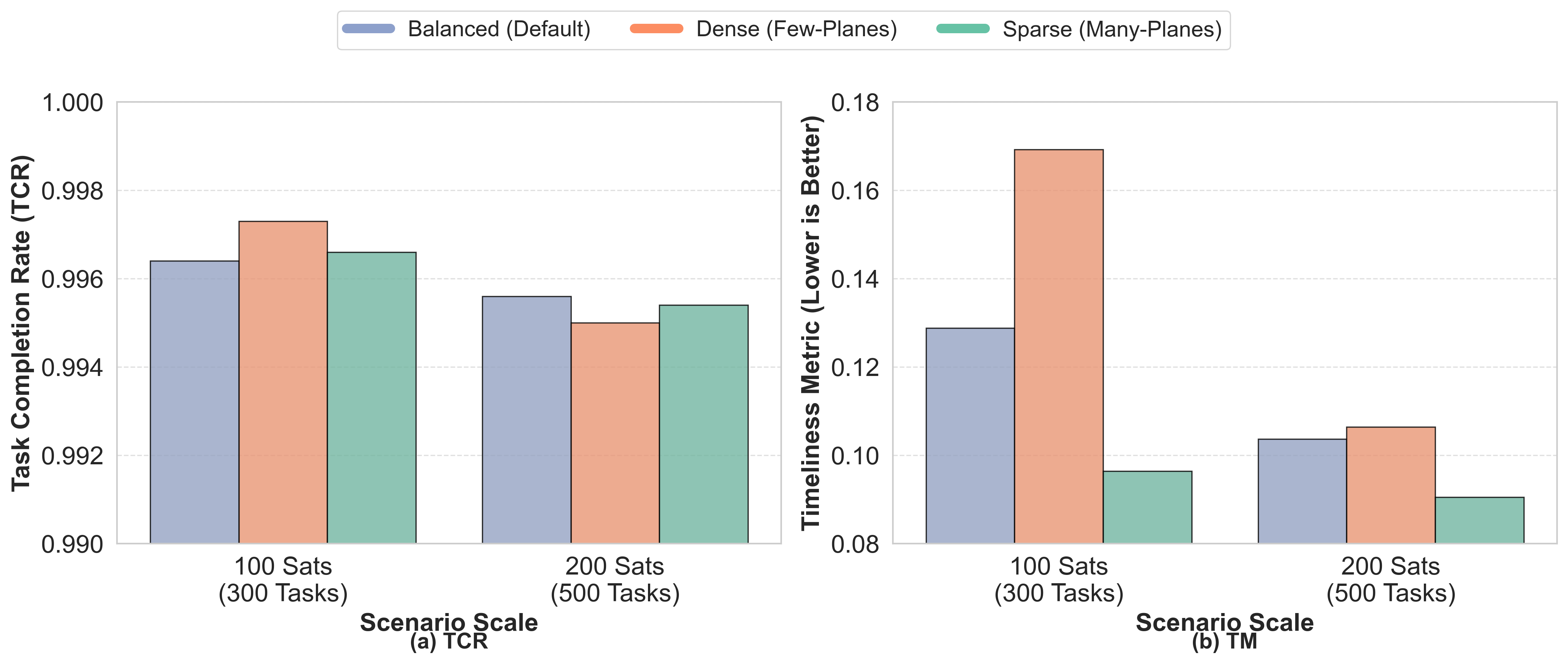}
        \caption{Impact of Orbit Architecture at medium scales.}
        \label{fig:orbit_mid}
    \end{figure}

    As illustrated in Figure~\ref{fig:orbit_mid}a, the Task Completion Rate (TCR) changes very little across the tested orbital configurations at this scale. Across all three topologies, completion remains close to saturation (TCR $> 0.99$), which suggests that the problem is comparatively under-constrained in terms of basic feasibility. When the satellite-to-task ratio is this high ($\approx 1:3$), the aggregate sensor footprint of each evenly distributed Walker constellation appears sufficient to provide substantial redundancy.

    However, while completion rates remain similar, the orbital geometry appears to influence the Timeliness score more noticeably. Figure~\ref{fig:orbit_mid}b shows that the Sparse Configuration (Many-Planes, e.g., 40 planes $\times$ 5 satellites) attains the lowest Timeliness (TM $\approx 0.090$) among the tested cases, ahead of both the Balanced ($20 \times 10$, TM $\approx 0.102$) and Dense ($10 \times 20$, TM $\approx 0.104$) topologies. A plausible explanation is geometric: increasing the number of orbital planes reduces longitudinal gaps between ground tracks, which can shorten waiting time before a target re-enters a satellite's field of view.

    \noindent \textbf{(3) Realistic-target Scenarios}

    This subsection evaluates algorithmic performance using real-world city targets and operational ground tracks. The experiments scale the constellation size from 1 to 20 satellites under a fixed load of 100 tasks and a 24-hour planning horizon. This setup specifically tests Resource Saturation and Coordination Efficiency.

    \begin{figure}[htb]
        \centering
        \begin{subfigure}[b]{0.45\textwidth}
            \centering
            \includegraphics[width=\textwidth]{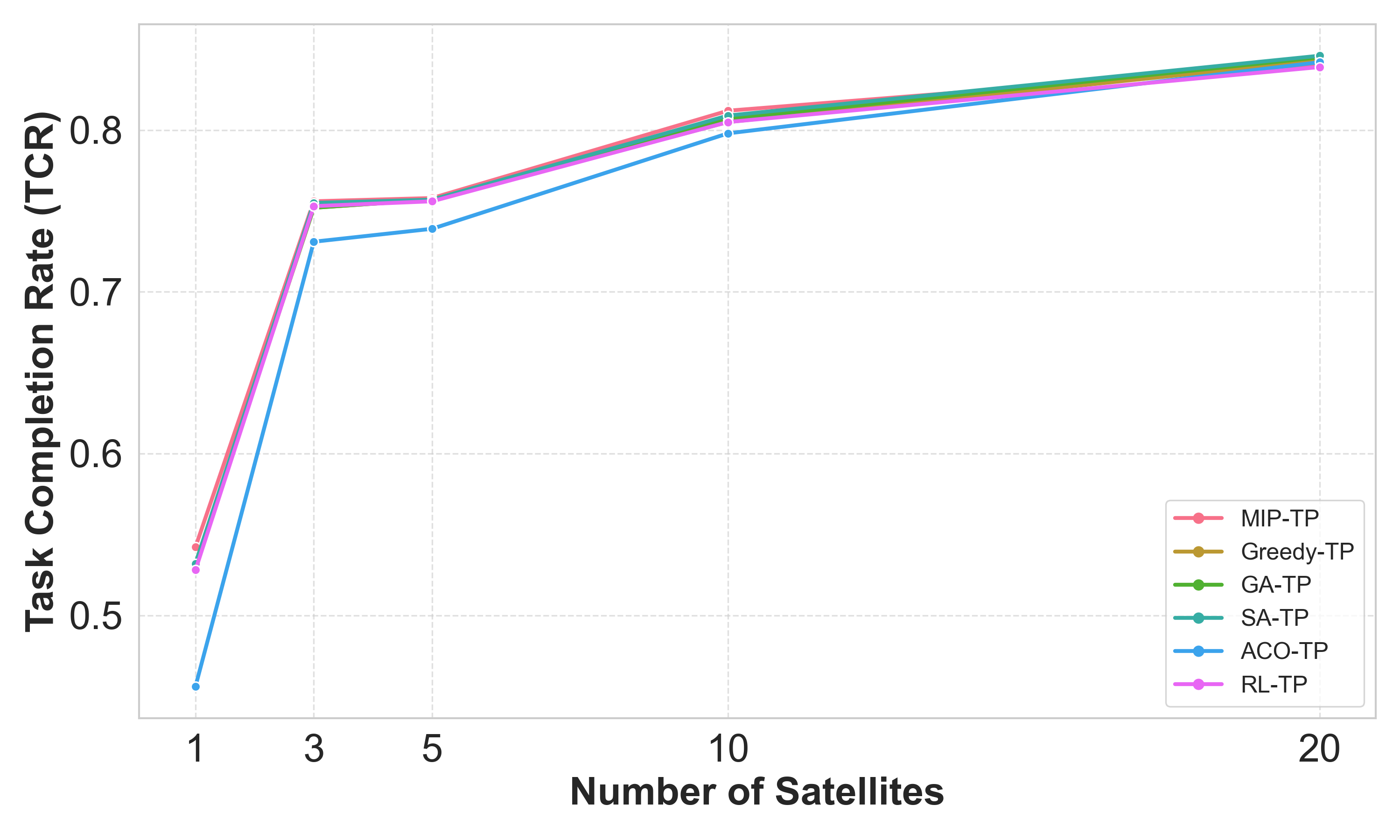}
            \caption{TCR}
            \label{fig:real_saturation}
        \end{subfigure}
        \hfill
        \begin{subfigure}[b]{0.45\textwidth}
            \centering
            \includegraphics[width=\textwidth]{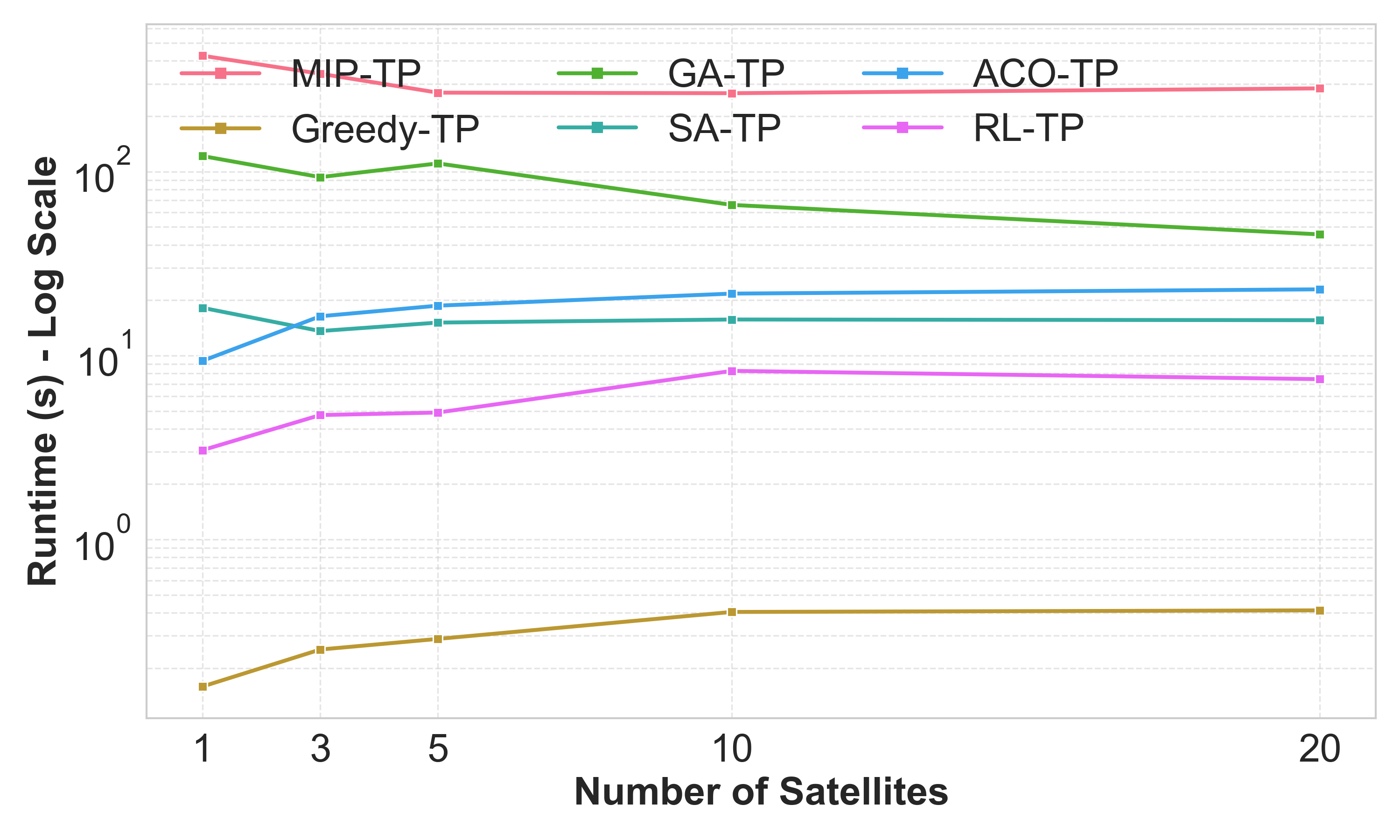}
            \caption{Runtime}
            \label{fig:real_runtime}
        \end{subfigure}

        \vspace{0.05cm} 

        \begin{subfigure}[b]{0.45\textwidth}
            \centering
            \includegraphics[width=\textwidth]{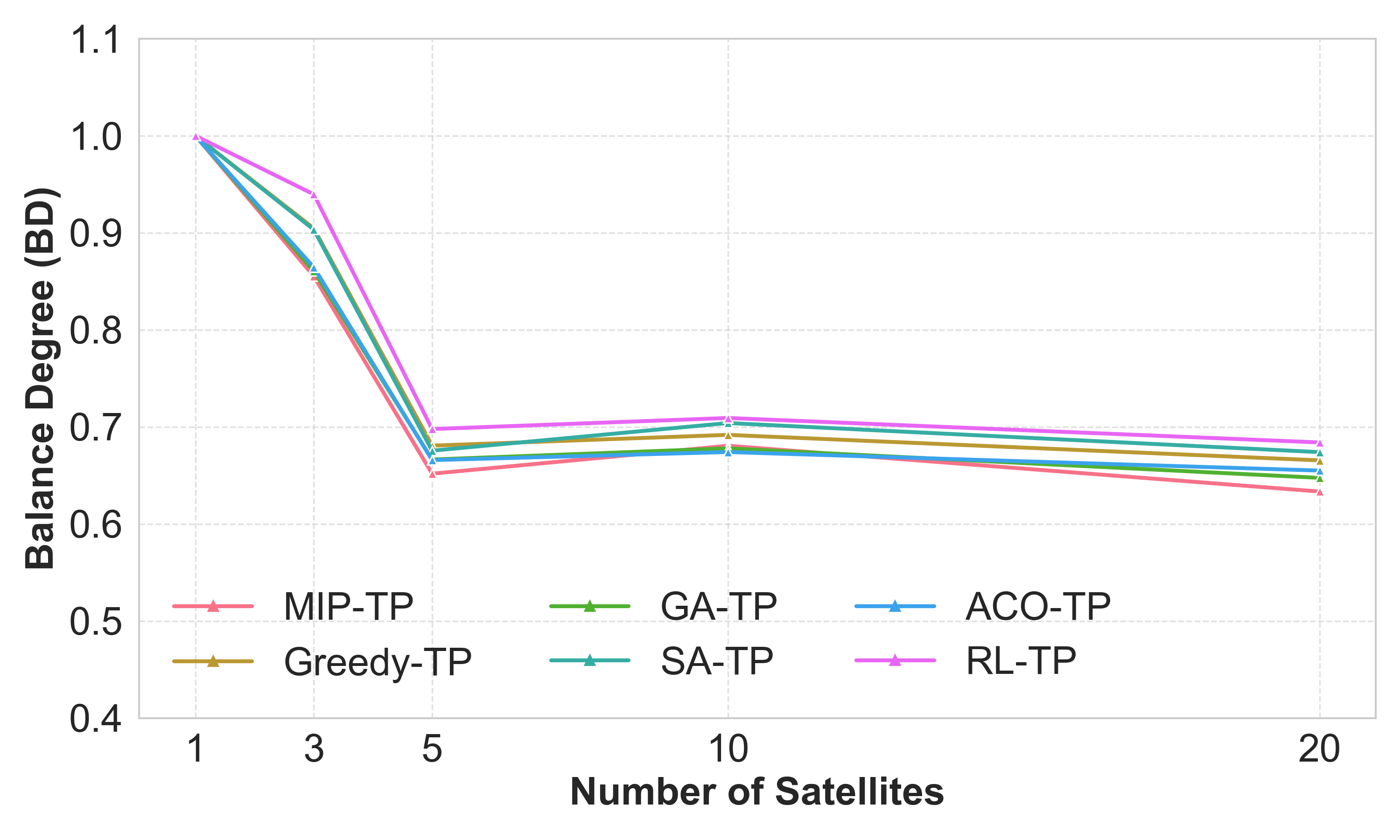}
            \caption{BD}
            \label{fig:real_balance}
        \end{subfigure}

        \caption{Performance analysis in Realistic-target scenarios with scaling constellation size ($N_S=1 to 20$).}
        \label{fig:realistic_analysis}
    \end{figure}


    As illustrated in Figure~\ref{fig:real_saturation}, the Task Completion Rate (TCR) follows a concave growth pattern. In the single-satellite case ($N_S=1$), resources are scarce, and the Exact Solver (MIP) provides a useful reference level (TCR $\approx$ 0.54) that the heuristic methods do not fully match. As the constellation size increases to 20 satellites ($N_S=20$), the system moves toward a more resource-abundant regime, and most algorithms, including RL and the constructive heuristics, approach a common performance ceiling (TCR $\approx$ 0.84). In this sense, the experiment suggests that the benchmark can help identify a practical saturation region beyond which additional assets yield smaller marginal gains in completion.


    Figure~\ref{fig:real_runtime} highlights the computational cost of coordination. In the reported experiments, the meta-heuristics (GA, ACO) show roughly linear or super-linear runtime growth as the search space expands with more satellites, whereas the RL approach remains comparatively flat and low-latency ($<10$s). The Greedy approaches still appear to be the fastest overall, but this speed comes with weaker load-balance scores (Figure~\ref{fig:real_balance}) in the larger constellations. RL, therefore, appears to offer a comparatively balanced compromise in this particular experiment, combining high completion rates with moderate computational overhead.

    \subsection{Visualisation of Results} \label{subsec:visualisation}

    To facilitate a qualitative assessment of the generated scheduling plans, EOS-Bench incorporates an interactive visualisation suite. This module bridges the gap between raw metric data and operational intuition by providing a spatio-temporal representation of the constellation's activities.

    \begin{figure}[htb]
        \centering
        \includegraphics[width=0.7\textwidth]{ 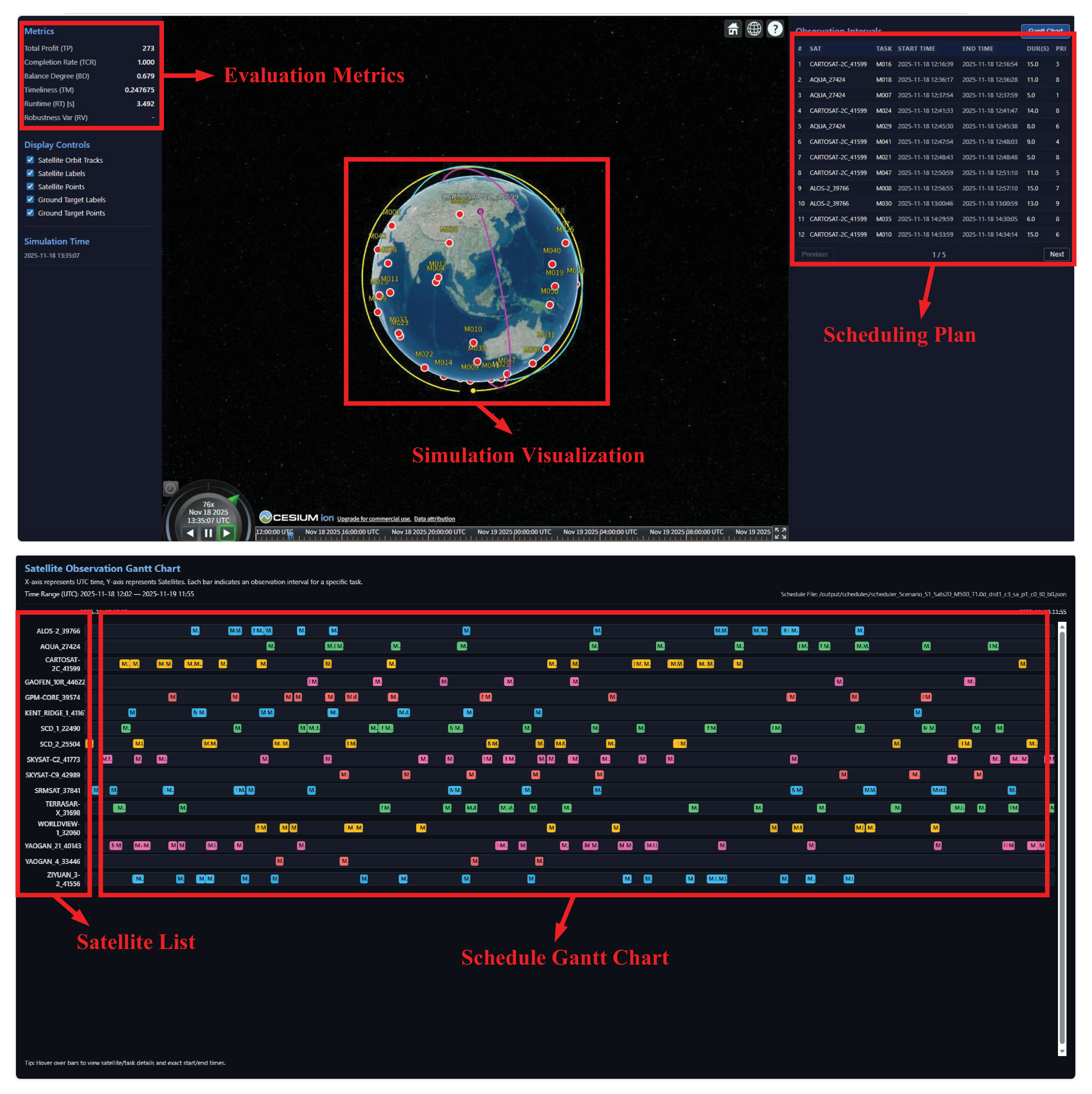 }
        \caption{Scheduling visualisation in EOS-Bench. Interactive visualisation of satellite trajectories, target locations, and observation activities in a benchmark scheduling scenario.}
        \label{fig:viz_globe}
    \end{figure}

    The visualisation engine is built on web-standard technologies (HTML5/CesiumJS), ensuring cross-platform accessibility without local environment dependencies. As shown in Figure~\ref{fig:viz_globe}, the 3D interface renders high-fidelity orbital tracks and highlights active observation links between satellites and ground targets during the simulation. This allows researchers to inspect conflict-dense regions and coordination patterns visually.

    \section{Future Directions}\label{sec:directions}
The development of EOS-Bench establishes a unified and extensible foundation for benchmarking Earth observation satellite scheduling. However, it also opens several important directions for future research. While the current framework provides a controlled and reproducible environment for evaluating diverse solution paradigms, the evolving complexity of real-world space operations calls for continued expansion in both modelling fidelity and evaluation scope. In particular, advancing towards more realistic operational settings, richer task representations, and adaptive decision-making frameworks will be essential to further enhance the relevance and impact of benchmark-driven research in this field.

    \subsection{Regional Area Targets}

    Although the current release of EOS-Bench focuses on point targets, many practical Earth observation requests are specified as continuous geographic regions rather than isolated locations, especially in disaster assessment, environmental monitoring, and wide-area mapping~\cite{pe2014satellite,zhong2019deep}. Extending the benchmark from point targets to regional area targets would therefore improve practical realism and broaden the class of scheduling problems that can be studied.

    Compared with point-target scheduling, regional observation introduces additional geometric and combinatorial structure. Rather than selecting independent observation points, algorithms must construct coordinated coverage plans that account for strip generation, overlap, coverage continuity, and coordination across multiple passes or satellites~\cite{galceran2013survey,gu2022large}. Concrete approaches to regional scheduling already exist in the literature, although they remain less explored than point-target formulations. For example, Perea et al.~\cite{perea2015swath} formulate multi-satellite swath-acquisition planning as a set-covering MIP, while other studies investigate large-area multi-satellite scheduling and image-strip-based planning~\cite{Ze2019multi}. These works show that regional coverage scheduling differs fundamentally from point-target selection because the decision variables and preprocessing steps are closely tied to swath geometry and spatial coverage structure.

    A natural extension of EOS-Bench would therefore be to generate polygonal requests, preprocess them into strips or grid cells, and include coverage-oriented metrics such as covered-area ratio, revisit density, and overlap efficiency. More broadly, integrating regional area targets would make it possible to evaluate scheduling methods under stricter spatial continuity requirements and more realistic observation products, while also opening the door to uncertainty-aware regional observation settings under cloud effects and other dynamic environmental factors~\cite{wangRobust2021}.

    \subsection{Payload Modalities}

    Future extensions of EOS-Bench should broaden the benchmark's scope beyond optical imaging payloads to incorporate additional sensing modalities whose data quality, geometric constraints, and operational flexibilities differ fundamentally from those of passive optical systems. Synthetic aperture radar (SAR), hyperspectral imagers, and spaceborne LiDAR each introduce modality-specific factors that directly influence core benchmark components, including how observation opportunities are modelled and how task profit is computed.

    With regard to observation-opportunity modelling, SAR systems, for example, are inherently side-looking and impose geometry-dependent access constraints that diverge from nadir-centric assumptions common in optical scheduling. Commercial SAR platforms routinely support substantial squint capability through a combination of spacecraft attitude control and electronic beam steering. This expanded steering envelope alters the set of valid line-of-sight conditions for a given target and requires the scheduler to reason about look direction, incidence angle, and the coupling between spacecraft attitude and power or momentum budgets. Some works have developed schedulers that leverage the characteristic viewing geometries of SAR sensors for minimising system response times~\cite{kim2015mission}.

    Task profit is currently modelled generically in EOS-Bench with a static profit assigned to each task and a binary variable indicating whether an acquisition occurred. While this definition is sufficient for comparing solution approaches across simplified mission settings, it does not fully capture practical differences in data quality. In time-critical missions, such as disaster response, task value may vary with acquisition time, viewing geometry, or image quality. Furthermore, a single acquisition may not fully satisfy a task because of data-quality gaps. For example, recent work has developed a data-quality-aware SAR observation scheduler that considers how geometric distortions in SAR imagery obscure portions of a region of interest~\cite{kramer2025scheduling}. Future iterations of EOS-Bench should therefore introduce modular payload-specific opportunity and reward models so that different sensing modalities can be represented within a common benchmark architecture.

    \subsection{Heterogeneous, Reconfigurable, and Very-Large Constellations}

    At present, the benchmark uses symmetric Walker--Delta patterns for medium- and large-scale scenarios. This is useful for controlled comparison, but operational Earth observation fleets increasingly involve mixed inclinations, altitudes, launch epochs, and platform capabilities. Allowing user-defined heterogeneous constellations would therefore make the benchmark more representative of contemporary and future missions.

    This extension is also important from the perspective of constellation geometry. Non-standard regional coverage studies have shown that irregular or optimised constellation layouts can outperform classical symmetric patterns in specific coverage settings~\cite{lee2020satellite}. A natural next step is therefore to allow fully user-defined constellation configurations, namely arbitrary collections of orbital elements drawn from TLE catalogues, mission design tools, or trade studies, so that scheduling algorithms can be evaluated on mixed-inclination, multi-altitude, and federated fleets. This extension would also make it possible to study how irregular coverage patterns and asymmetric platform capabilities interact with different scheduling paradigms, broadening the benchmark from an idealised symmetric setting to one that reflects the full diversity of current and future Earth observation systems.

    Beyond heterogeneous constellation structure, future benchmark development could also move from attitude agility to orbital reconfiguration. Recent work on the reconfigurable Earth observation satellite scheduling problem has shown that satellites may perform active orbital transfers, such as phasing or plane changes, to manipulate future visibility opportunities and improve mission performance~\cite{pearl2025REOSSP,pearl2024Benchmarking}. Extending EOS-Bench to support such manoeuvrable assets would require modelling propellant-limited $\Delta v$ budgets, the power requirements of electric or chemical propulsion, and the coupling between multi-revolution orbital dynamics and combinatorial task scheduling.

    A further challenge lies in scale. Existing research mainly focuses on scenarios involving hundreds of satellites, while scenarios with thousands or tens of thousands of satellites remain far less explored. When satellite attitude agility, payload diversity, large task volumes, heterogeneous request types, complex couplings among planning elements, and strict timeliness requirements are all considered simultaneously, the resulting scheduling problem becomes extremely difficult to formulate and solve. Extending EOS-Bench towards heterogeneous, reconfigurable, and very-large constellations would therefore provide a more demanding testbed for collaborative planning under complex operational constraints.

    \subsection{Integrated Planning and Onboard Autonomy under Uncertainty}

    The current EOS-Bench pipeline mainly evaluates observation planning. In real missions, however, observation, storage, downlink, and ground-station availability are tightly coupled. Ignoring this coupling can produce schedules that are feasible from the imaging perspective but difficult to execute end-to-end~\cite{barbulescu2004scheduling,barbulescu2006afscn,zufferey2008graph}. The literature on antenna and ground-station scheduling already provides a strong basis for this extension, including heuristic, graph-based, Lagrangian, and exact optimisation approaches, as well as more recent integrated observation--downlink methods~\cite{marinelli2011lagrangian,vazquez2014antenna,linares2024milp,sabol2021dsn}. Adding downlink resources and ground-station windows would move EOS-Bench closer to end-to-end mission planning and provide a testbed for algorithms that jointly optimise observation acquisition and data delivery.

    At the same time, increased platform agility and larger constellations are driving demand for stronger onboard and distributed autonomy. As satellite manoeuvrability and payload capabilities improve, platforms exhibit increasing heterogeneity in performance, payload configuration, and onboard energy reserves, which in turn makes dynamic resource allocation substantially more difficult~\cite{claudet2022dsn,zhangj2019svm,zhangj2022genetic}. Moreover, the range of autonomous satellite applications is expanding, creating greater demand for intelligent decision-making and mission-planning capabilities in dynamic and uncertain environments.

    Recent studies have examined onboard mission generation based on user preferences~\cite{liUserPreference2024}, distributed scheduling~\cite{liDistributedDRL2024,CHEN2026Autonomous}, dynamic scheduling~\cite{yao2026meta}, and distributed constraint optimisation for Earth observation scheduling~\cite{krigmanDCOP2024}. Together with uncertainty-aware scheduling under cloud effects~\cite{wangRobust2021}, these studies point towards a benchmark setting in which centralised, distributed, and onboard-autonomous planners can be compared under rolling re-planning, partial information, and dynamic resource variation. Future versions of EOS-Bench could therefore support resource perception, online state updates, and adaptive re-planning based on satellite status and environmental changes, thereby broadening its relevance for intelligent autonomous operations.

    \subsection{Explainable and Auditable Intelligent Scheduling}

    As learning-based and hybrid scheduling methods become more capable, their interpretability becomes increasingly important. In operational Earth observation systems, users may need to understand why a task was accepted, delayed, reassigned, or rejected, particularly when schedules are generated automatically under multiple competing objectives and constraints.

    The benchmark proposed in this paper is well-suited to support this line of work because task attributes, access windows, conflicts, and resource limits are already represented explicitly. This makes it possible to trace the decision context behind task acceptance and rejection, and to study explanation mechanisms at the level of constraints, conflicts, and trade-offs. The structure of the benchmark scenarios also enables the behaviour of scheduling algorithms to be examined across different workload levels, constellation configurations, and resource capacities, making it possible to assess whether decisions remain consistent with the defined constraints and objectives.

    More broadly, recent work in explainable AI emphasises the need for explanations that are not only human-readable, but also trustworthy, actionable, and auditable~\cite{longoXAI2024}. Reviews in Earth observation likewise show growing interest in explainability, accountability, and regulatory alignment for intelligent EO systems~\cite{gevaertXAIEO2022}. Future releases of EOS-Bench could therefore include explanation-oriented outputs and evaluation criteria, such as the faithfulness of reported reasons, consistency across similar scenarios, and the auditability of scheduling traces produced by optimisation and learning-based methods. In this way, the benchmark could support not only solution-quality assessment, but also the development and validation of interpretable and auditable intelligent scheduling methods.

\section{Conclusions}\label{sec:conclusion}

This paper introduced EOS-Bench, a unified, open-source, and extensible benchmark framework for Earth observation satellite scheduling.
The goal of EOS-Bench is to provide a standardised and reproducible experimental platform rather than to propose a new scheduling algorithm.
By unifying scenario modelling, constraint definitions, and evaluation metrics, the benchmark establishes a common ground for fair comparison across diverse solution paradigms.

We constructed a benchmark library comprising 1,390 scenarios and 13,900 instances that span a wide range of constellation scales, task volumes, planning horizons, and resource-tightness conditions across both agile and non-agile platforms. To bridge the gap between nominal problem size and actual scheduling complexity, we introduced a pre-optimisation scenario characterisation scheme that mathematically quantifies intrinsic structural bottlenecks, such as opportunity sparsity and timeline congestion. Furthermore, a consistent evaluation protocol was designed to assess performance from multiple perspectives, including solution quality, computational efficiency, load balancing, and timeliness, thereby supporting multidimensional and cross-method comparison.

The comprehensive experiments, illustrated in the main text via a representative agile-satellite subset, indicate that EOS-Bench can support systematic evaluation across optimisation-based, heuristic, meta-heuristic, and learning-based approaches.
The benchmark enables scalability analysis across varying problem scales, sensitivity studies with respect to satellite agility and resource budgets, and structured comparison under different orbital configurations.
The reported results suggest that the benchmark can reveal performance trends, scalability limits, and trade-offs among competing scheduling paradigms under consistent modelling assumptions. Crucially, the empirical findings also confirm that our scenario-level descriptors successfully explain why nominally similar instances can induce vastly different levels of algorithmic difficulty.

Overall, EOS-Bench provides a reproducible, extensible, and scalable testbed for the Earth observation scheduling community.
By offering a shared experimental foundation, complete with full cross-platform data in the supplementary repository, it is intended to support more transparent benchmarking, more rigorous comparative analysis, and continued methodological progress. Building upon this initial release, future development of the framework will pursue several key directions: extending task representations to regional area targets, integrating diverse payload modalities (e.g., SAR), and supporting heterogeneous or orbitally reconfigurable constellations. Ultimately, incorporating end-to-end mission couplings, onboard autonomy under uncertainty, and metrics for explainable scheduling decisions will further align the benchmark with the evolving complexities of next-generation space operations.

\section*{Acknowledgements}
R.~Vazquez acknowledges support of grant PID2023-147623OB-I00 funded by MICIU/AEI/ 10.13039/501100011033 and by “ERDF A way of making Europe”. 
M.~Mavrovouniotis is supported by the `EXCELSIOR’ project which
has received funding from the European Union’s Horizon 2020 research and
innovation programme under Grant Agreement No 857510, from the Government
of the Republic of Cyprus through the Directorate General for the European
Programmes, Coordination and Development and the Cyprus University of
Technology.
 X.~Chen's work was supported by the Natural Science Foundation of Hubei Province under Grant No.~JCZRMS202600542. Y.~Song's work was supported by the National Natural Science Foundation of China No.~72501042. Q. Luo and X. Wang's work was supported by the European Union MSCA Grant (101278720).

	\bibliographystyle{elsarticle-num-names}
	\bibliography{benchmark}

\appendix

\section{Illustrative Example of Descriptor Calculation} \label{app:scenario_characterisation_example}

To provide a concrete interpretation of the scenario characterisation descriptors introduced in Section~\ref{sec:scenario_characterisation}, this appendix presents a small example instance.
The aim is to show, in a compact and transparent setting, how the adopted descriptors can be computed from task opportunities, pairwise interference relations, and satellite-timeline congestion.

\subsection{Example instance}
\label{app:example_scenario}

\begin{figure}[htb]
    \centering
    \includegraphics[width=0.75\textwidth]{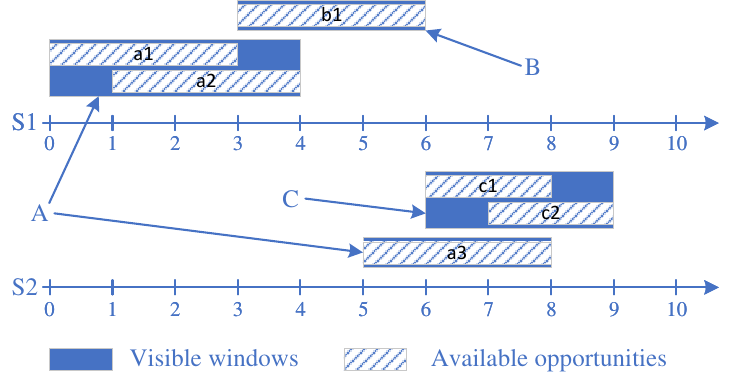}
    \caption{Illustrative visible-window and available-window structure on two satellite timelines. Solid bars denote visible windows, and hatched bars denote available opportunities derived from them.}
    \label{fig:toy_windows}
\end{figure}

Consider an example instance with two satellites ($S1$ and $S2$), four tasks ($A$, $B$, $C$, and $D$), a planning horizon of 10 time units, and an analysis step of 1 time unit.
The visible windows and the corresponding available opportunities are illustrated in Figure~\ref{fig:toy_windows}.

Task $A$ has two visible windows, namely $[0,4]$ on $S1$ and $[5,8]$ on $S2$, from which three available opportunities are generated:
$a_1=[0,3]$, $a_2=[1,4]$, and $a_3=[5,8]$.
Task $B$ has one visible window, $[3,6]$ on $S1$, which generates one available window $b_1=[3,6]$.
Task $C$ has one visible window, $[6,9]$ on $S2$, which generates two available opportunities, $c_1=[6,8]$ and $c_2=[7,9]$.
Task $D$ has no visible window within the current horizon and therefore no available window.

The resulting task-level structural information is summarised in Table~\ref{tab:toy_task_info}.
Here, the elasticity of each task is defined as the product of its number of optional satellites and its number of visible windows.

\begin{table}[htb]
    \centering
    \footnotesize
    \caption{Task-level structural information for the example instance.}
    \label{tab:toy_task_info}
    \begin{tabular}{ccccc}
        \toprule
        Task & Optional satellites & Visible windows & Available opportunities & Elasticity \\
        \midrule
        $A$ & 2 & 2 & 3 & 4 \\
        $B$ & 1 & 1 & 1 & 1 \\
        $C$ & 1 & 1 & 2 & 1 \\
        $D$ & 0 & 0 & 0 & 0 \\
        \bottomrule
    \end{tabular}
\end{table}

\subsection{Pairwise interference and satellite-timeline congestion}
\label{app:example_conflicts}

The four tasks form six task pairs in total:
$(A,B)$, $(A,C)$, $(A,D)$, $(B,C)$, $(B,D)$, and $(C,D)$.
Among them, only $(A,B)$ and $(A,C)$ are comparable, because only these two pairs contain available opportunities on a common satellite timeline.

For $(A,B)$, task $A$ provides two available opportunities on $S1$, namely $a_1=[0,3]$ and $a_2=[1,4]$, while task $B$ provides one available window on $S1$, namely $b_1=[3,6]$.
Hence, the comparable available-window pairs are $(a_1,b_1)$ and $(a_2,b_1)$.
Under the present convention, both pairs are conflicting, so $(A,B)$ has two comparable pairs and two conflicting pairs.

For $(A,C)$, task $A$ provides one available window on $S2$, namely $a_3=[5,8]$, while task $C$ provides two available opportunities on $S2$, namely $c_1=[6,8]$ and $c_2=[7,9]$.
The comparable pairs are therefore $(a_3,c_1)$ and $(a_3,c_2)$, and both are conflicting.
All remaining task pairs are non-comparable because they do not share any common satellite timeline with available opportunities.
These pairwise relations are summarised in Table~\ref{tab:toy_pair_info}.

\begin{table}[htb]
    \centering
    \footnotesize
    \caption{Comparable and conflicting available-window relations in the example instance.}
    \label{tab:toy_pair_info}
    \begin{tabular}{cccc}
        \toprule
        Task pair & Comparable pairs $C_{ij}$ & Conflicting pairs $F_{ij}$ & Conflict ratio $F_{ij}/C_{ij}$ \\
        \midrule
        $(A,B)$ & 2 & 2 & 1.00 \\
        $(A,C)$ & 2 & 2 & 1.00 \\
        $(A,D)$ & 0 & 0 & -- \\
        $(B,C)$ & 0 & 0 & -- \\
        $(B,D)$ & 0 & 0 & -- \\
        $(C,D)$ & 0 & 0 & -- \\
        \bottomrule
    \end{tabular}
\end{table}

The same structure can be read from the satellite timelines.
On $S1$, task $A$ is active over $[0,4]$ after merging its available opportunities on that satellite, whereas task $B$ is active over $[3,6]$.
The resulting conflict segment is therefore $[3,4]$.
On $S2$, task $A$ is active over $[5,8]$, whereas task $C$ is active over $[6,9]$ after merging its available opportunities on that satellite, producing a conflict segment $[6,8]$.
Thus, the instance contains one conflict segment on each satellite.
The corresponding timeline-level conflict structure is given in Table~\ref{tab:toy_sat_info}.

\begin{table}[htb]
    \centering
    \footnotesize
    \caption{Satellite-timeline conflict structure in the example instance.}
    \label{tab:toy_sat_info}
    \begin{tabular}{ccccc}
        \toprule
        Satellite & Conflict segment & Conflict steps & Conflict duration & Conflicting tasks \\
        \midrule
        $S1$ & $[3,4]$ & 1 & 1 & 2 \\
        $S2$ & $[6,8]$ & 2 & 2 & 2 \\
        \bottomrule
    \end{tabular}
\end{table}

\subsection{Descriptor calculation for the example instance}
\label{app:example_metric_values}

Using the definitions in Section~\ref{sec:scenario_characterisation}, the task-oriented descriptors can be obtained directly from Tables~\ref{tab:toy_task_info} and~\ref{tab:toy_pair_info}.
The average available opportunities per task is
\[
\Gamma_{\mathrm{ao}}
=
\frac{3+1+2+0}{4}
=
1.5.
\]
Among the four tasks, three tasks ($B$, $C$, and $D$) have at most two available opportunities, so the opportunity constrained task ratio is
\[
\Gamma_{\mathrm{oc}}
=
\frac{3}{4}
=
0.75.
\]
Among the six task pairs, two pairs, namely $(A,B)$ and $(A,C)$, contain at least one conflicting available-window pair, which gives
\[
\Gamma_{\mathrm{ti}}
=
\frac{2}{6}
=
0.333.
\]
Restricting attention to comparable task pairs, the pairwise conflict ratios are 1.00 for $(A,B)$ and 1.00 for $(A,C)$.
Their average task pair conflict ratio is therefore
\[
\Gamma_{\mathrm{at}}
=
\frac{1.00+1.00}{2}
=
1.0.
\]
Finally, averaging the task elasticities in Table~\ref{tab:toy_task_info} yields
\[
\Gamma_{\mathrm{te}}
=
\frac{4+1+1+0}{4}
=
1.5.
\]

The satellite-oriented descriptors follow from the timeline-conflict structure in Table~\ref{tab:toy_sat_info}.
For the observation contention ratio, the conflict-load area equals 3, because the overlap on $S1$ lasts for 1 time unit and the overlap on $S2$ lasts for 2 time units.
The total active observation load equals 13.
More specifically, the active load on $S1$ is
\[
1\times 3 + 2\times 1 + 1\times 2 = 7,
\]
corresponding to the intervals $[0,3]$, $[3,4]$, and $[4,6]$,
and the active load on $S2$ is
\[
1\times 1 + 2\times 2 + 1\times 1 = 6,
\]
corresponding to the intervals $[5,6]$, $[6,8]$, and $[8,9]$.
Hence,
\[
\Lambda_{\mathrm{oc}}
=
\frac{3}{7+6}
=
\frac{3}{13}
\approx
0.231.
\]

Both satellites contain at least one conflict segment, so the conflict satellite ratio is
\[
\Lambda_{\mathrm{cs}}
=
\frac{2}{2}
=
1.0.
\]
The total conflict duration is $1+2=3$, and the aggregate satellite timeline is $2\times 10=20$, which gives
\[
\Lambda_{\mathrm{to}}
=
\frac{3}{20}
=
0.15.
\]
In all conflict instances, exactly two tasks are simultaneously in conflict, so
\[
\Lambda_{\mathrm{ac}}
=
\frac{2\times 1 + 2\times 2}{1+2}
=
2.
\]
Finally, for each conflict segment, the excess demand equals $q_{\ell}-1$.
In this example, both conflict segments contain two conflicting tasks, so the excess demand is always $2-1=1$.
The duration-weighted excess demand is therefore
\[
(2-1)\times 1 + (2-1)\times 2 = 3.
\]
The peak number of conflicting tasks is $q_{\max}=2$, and the total conflict duration is $1+2=3$.
Hence, the Excess demand ratio is
\[
\Lambda_{\mathrm{ed}}
=
\frac{3}{(2-1)\times 3}
=
1.0.
\]

The resulting descriptor values are summarised in Table~\ref{tab:toy_metric_summary}.

\begin{table}[htb]
    \centering
    \footnotesize
    \caption{Characterisation descriptor values for the example instance.}
    \label{tab:toy_metric_summary}
    \begin{tabular}{lll}
        \toprule
        Perspective & Descriptor & Value \\
        \midrule
        \multirow{5}{*}{Task-oriented}
        & Average Available Opportunities $(\Gamma_{\mathrm{ao}})$ & 1.5 \\
        & Opportunity Constrained Task Ratio $(\Gamma_{\mathrm{oc}})$ & 0.75 \\
        & Task Interference Ratio $(\Gamma_{\mathrm{ti}})$ & 0.333 \\
        & Average Task Pair Conflict Ratio $(\Gamma_{\mathrm{at}})$ & 1.0 \\
        & Task Elasticity Ratio $(\Gamma_{\mathrm{te}})$ & 1.5 \\
        \midrule
        \multirow{5}{*}{Satellite-oriented}
        & Observation Contention Ratio $(\Lambda_{\mathrm{oc}})$ & 0.231 \\
        & Conflict Satellite Ratio $(\Lambda_{\mathrm{cs}})$ & 1.0 \\
        & Timeline Overload Ratio $(\Lambda_{\mathrm{to}})$ & 0.15 \\
        & Average Conflict tasks $(\Lambda_{\mathrm{ac}})$ & 2 \\
        & Excess Demand Ratio $(\Lambda_{\mathrm{ed}})$ & 1.0 \\
        \bottomrule
    \end{tabular}
\end{table}

\section{Additional Experimental Results Available in the Public Repository}\label{app:first}

Due to space limitations, the main text reports only a representative subset of the experimental results generated with EOS-Bench.
To improve transparency and reproducibility, the complete numerical results, logs, and related visual outputs for both agile and non-agile satellites (in total 1,370 + 1,130 = 2,500 instances) are publicly available in the GitHub repository associated with this paper via
\url{https://github.com/Ethan19YQ/EOS-Bench}.

In particular, the repository provides a workbook, \texttt{Scenario\_Level\_Results.xlsx}, which contains the complete numerical outputs for the 2,500 experimental instances considered in Section~\ref{sec:experiments}.
These materials include the full solver-level results for the five evaluation metrics, namely Task Profit (TP), Task Completion Rate (TCR), Balance Degree (BD), Timeliness (TM), and Runtime (RT), across the standard scenarios and the specific sensitivity-analysis scenarios. 
Importantly, while the main text discusses only a representative subset focusing on agile platforms to highlight key performance trends, the repository provides the complete dataset encompassing both agile and non-agile satellite configurations.
Table~\ref{tab:repo_results} summarises the main categories of experimental results included in \texttt{Scenario\_Level\_Results.xlsx}, organised into two primary sections for agile and non-agile platforms, respectively. Note that the manoeuvrability sensitivity analysis is exclusively applicable to agile platforms.

\begin{table}[htb]
\centering
\footnotesize
\setlength{\tabcolsep}{4pt}
\renewcommand{\arraystretch}{1.2}
\caption{The complete experimental results available in \texttt{Scenario\_Level\_Results.xlsx} in the public repository.}
\label{tab:repo_results}
\begin{tabular}{p{0.22\textwidth} p{0.38\textwidth} p{0.34\textwidth}}
\toprule
\textbf{Category} & \textbf{Description} & \textbf{Worksheets in \texttt{data.xlsx}} \\

\midrule
\multicolumn{3}{c}{\textbf{Part I: Agile Platforms}} \\
\midrule

\makecell[l]{Overall summary}
& \makecell[l]{Consolidated overview of the experimen-\\tal groups and their parameter settings.}
& \texttt{Total} \\

\makecell[l]{Standard scenarios}
& \makecell[l]{Results for the main benchmark experi-\\ments under different constellation sizes, \\task loads, planning horizons, and target \\distributions.}
& \makecell[l]{
\texttt{st-s3-t100}, \texttt{st-s3-t200},\\
\texttt{st-s10-t200}, \texttt{st-s10-t500},\\
\texttt{st-s100-t500}, \texttt{st-s100-t2000},\\
\texttt{st-s500-t2000}, \texttt{st-s500-t5000}
} \\

\makecell[l]{Resource-capacity \\scenarios}
& \makecell[l]{Sensitivity analysis under different \\homogeneous and heterogeneous satellite \\resource-capacity configurations.}
& \makecell[l]{
\texttt{sp-re-s3-t200}, \texttt{sp-re-s10-t500},\\
\texttt{sp-re-s100-t500}, \texttt{sp-re-s500-t2000}
} \\

\makecell[l]{Manoeuvrability \\scenarios}
& \makecell[l]{Sensitivity analysis under different agility \\and transition-time profiles.}
& \makecell[l]{
\texttt{sp-ma-s3-t200}, \texttt{sp-ma-s10-t500},\\
\texttt{sp-ma-s100-t500}, \texttt{sp-ma-s500-t2000}
} \\

\makecell[l]{Constellation- \\configuration scenarios}
& \makecell[l]{Results for alternative Walker--Delta \\orbital architectures with different \\plane/satellite allocations.}
& \makecell[l]{
\texttt{sp-co-s50}, \texttt{sp-co-s100},\\
\texttt{sp-co-s200}, \texttt{sp-co-s500},\\
\texttt{sp-co-s1000}
} \\

\makecell[l]{Realistic-target \\and runtime}
& \makecell[l]{Additional results for realistic-target \\settings and runtime-oriented experiments.}
& \texttt{sp-rt-t100} \\

\midrule
\multicolumn{3}{c}{\textbf{Part II: Non-agile Platforms}} \\
\midrule

\makecell[l]{Standard scenarios}
& \makecell[l]{Results for the main benchmark experi-\\ments under different constellation sizes, \\task loads, planning horizons, and target \\distributions.}
& \makecell[l]{
\texttt{n-st-s3-t100}, \texttt{n-st-s3-t200},\\
\texttt{n-st-s10-t200}, \texttt{n-st-s10-t500},\\
\texttt{n-st-s100-t500}, \texttt{n-st-s100-t2000},\\
\texttt{n-st-s500-t2000}, \texttt{n-st-s500-t5000}
} \\

\makecell[l]{Resource-capacity \\scenarios}
& \makecell[l]{Sensitivity analysis under different \\homogeneous and heterogeneous satellite \\resource-capacity configurations.}
& \makecell[l]{
\texttt{n-sp-re-s3-t200}, \texttt{n-sp-re-s10-t500},\\
\texttt{n-sp-re-s100-t500}, \texttt{n-sp-re-s500-t2000}
} \\

\makecell[l]{Constellation- \\configuration scenarios}
& \makecell[l]{Results for alternative Walker--Delta \\orbital architectures with different \\plane/satellite allocations.}
& \makecell[l]{
\texttt{n-sp-co-s50}, \texttt{n-sp-co-s100},\\
\texttt{n-sp-co-s200}, \texttt{n-sp-co-s500},\\
\texttt{n-sp-co-s1000}
} \\

\makecell[l]{Realistic-target \\and runtime}
& \makecell[l]{Additional results for realistic-target \\settings and runtime-oriented experiments.}
& \texttt{n-sp-rt-t100} \\

\bottomrule
\end{tabular}
\end{table}

\end{document}